\documentclass[final,3p,times,authoryear]{elsarticle}

\usepackage[utf8]{inputenc}
\usepackage{amsmath}
\usepackage{amsfonts}
\usepackage{amssymb}
\usepackage[english]{babel}
\usepackage{tikz}
\usepackage{graphicx}
\usepackage{caption}
\usepackage{subcaption}
\usepackage{float}
\usepackage{bm}
\usepackage{booktabs}
\usepackage[linktocpage, linkcolor=black]{hyperref}

\setcitestyle{square}

\hypersetup{
    colorlinks=true,
    linkcolor=red,
    citecolor=blue
}

\newcommand{\rcol}[1]{%
  \begingroup
    \textcolor{black}{#1}%
  \endgroup
}

\usepackage{tikz}
\usetikzlibrary{math,calc}
\usetikzlibrary{shapes.geometric,shapes.arrows,arrows.meta}
\usetikzlibrary{quotes,angles}
\usetikzlibrary{angles}
\usetikzlibrary{shapes}
\usetikzlibrary{shapes.geometric}

\newcommand\poly[4]{
    \begin{tikzpicture}[auto=right,>={Triangle[open,fill=white,length=4pt,angle=60:1pt 2,inset=0pt]}]
    \def\rps{#1}
    
    \def\rayonext{#2 cm}
    \def\couleur{blue!15}
    \pgfmathsetmacro\nbrotation{int(\rps-1)}
    \pgfmathsetmacro\rayonint{multiply(\rayonext,cos(180/\rps))}
   
    \pgfmathsetmacro\startarrow{45}

	\edef\rpss{\the\numexpr\rps-1\relax}
    \node [name=S, draw,blue!0!black,rotate=0,minimum size=2*\rayonext,regular polygon,fill=\couleur,regular polygon sides=\rps] at (0, 0) {}; 
    \node[name=cc] at (0,0) {};
    \path foreach \X in {1,...,\rps} {(S.corner \X) coordinate (corner \X)};
    \path foreach \X in {1,...,\rps} {(S.side \X) coordinate (side \X)};
    \foreach \x in {1,...,\rps}{
        \draw [black,dashed, shorten >=0cm,shorten <=0cm](S.center) -- (S.corner \x);
    }
    \pic [draw, "$\alpha$", angle eccentricity=1.2, angle radius=1/3*\rayonext] {angle = corner \rpss--cc--corner \rps};
    \draw [red] (S.side 1)--(S.center) -- (S.corner 1) node[midway, right]{$#3$};
    \pic [draw, red,  "$\alpha/2$" ,angle eccentricity=1.2, angle radius=1/3*\rayonext] {angle = corner 1--cc--side 1};

	\node[name=cross] at ($0.5*(S.side 1)+0.5*(S.side \rps)$) {};    
    \draw [blue] (S.side 1)--($0.5*(S.side 1)+0.5*(S.side \rps)$) -- (S.corner 1) node[midway, right]{$#4$};
    \pic [draw, red ,angle eccentricity=1.2, angle radius=1/6*\rayonext] {angle = cross--side 1--corner 1};

\end{tikzpicture}
}

\usepackage{lineno}

\journal{Journal of Computational Science}

\begin{document}

\begin{frontmatter}

\title{PalaCell2D: A framework for detailed tissue morphogenesis}

\author{Rapha\"{e}l Conradin\corref{cor1}%
}
\ead{raphael.conradin@unige.ch%
}

\author{Christophe Coreixas}
\author{Jonas Latt}
\author{Bastien Chopard}
\address{Computer Science Department, University of Geneva, Carouge, Switzerland}

\cortext[cor1]{Corresponding author}

\begin{abstract}
In silico, cell based approaches for modeling biological morphogenesis are used to test and validate our understanding of the biological and mechanical process that are at work during the growth and the organization of multi-cell tissues. As compared to in vivo experiments, computer based frameworks dedicated to tissue modeling allow us to easily test different hypotheses, and to quantify the impact of various biophysically relevant parameters.

Here, we propose a formalism based on a detailed, yet simple, description of cells that accounts for intra-, inter- and extra-cellular mechanisms. More precisely, the cell growth and division is described through the space and time evolution of the membrane vertices. These vertices follow a Newtonian dynamics, meaning that their evolution is controlled by different types of forces: a membrane force (spring and bending), an adherence force (inter-cellular spring), external and internal pressure forces. Different evolution laws can be applied on the internal pressure, depending on the intra-cellular mechanism of interest. In addition to the cells dynamics, our formalism further relies on a lattice Boltzmann method, using the Palabos library, to simulate the diffusion of chemical signals. The latter aims at driving the growth and migration of a tissue by simply changing the state of the cells. 

All of this leads to an accurate description of the growth and division of cells, with realistic cell shapes and where membranes can have different properties. While this work is mainly of methodological nature, we also propose to validate our framework through simple, yet biologically relevant benchmark tests at both single-cell and full tissue scales. This includes free and chemically controlled cell tissue growth in an unbounded domain. The ability of our framework to simulate cell migration, cell compression and morphogenesis under external constraints is also investigated in a qualitative manner.
\end{abstract}




\end{frontmatter}


\section{Introduction}

Morphogenesis is the biological mechanism that drives the development of cells, tissues, and organs, by directly impacting their shape and function. 
This mechanism is tightly related to the various states cells undergo: proliferation, differentiation, apoptosis, migration and adhesion. 
Despite recent advances in developmental biology (molecular biology, live-imaging and \textit{ex vivo} methodologies~\citep{STJOHNSTON_PLOS_13_2015}), we are mainly able to explain key mechanisms of morphogenesis in an isolated manner. In addition, obtaining quantitative results in a reproducible manner remains a tedious task, hence, the growing interest in computational methods~\citep{FLETCHER_BJ_106_2014,VANLIEDEKERKE_CPM_2_2015,TANAKA_PhD_2016,SHARPE_DEVELOPMENT_144_2017,MERZOUKI_PhD_2018,BUTTENSCHON_PLOSCB_16_2020}. \rcol{These methods obviously rely on empirical observations and ad-hoc parameters in order to compensate for our reduced understanding of active and mechanical properties of living cells. Yet, having access to these parameters allows researchers to (1) accurately quantify the impact of each mechanism of interest (elasticity, bending, internal pressure, chemical signaling, etc), and (2) simulate the morphogenesis of various types of living cells.} 

Depending on the level of abstraction and the scale considered, several methodologies are available. At the largest scales, 
cell tissues and organs can be approximated by a continuum material whose properties obey a given partial differential equation (PDE). This PDE is solved using standard numerical discretizations such as finite-element or finite-volume. More precisely, that approach requires the tissue/organ to be discretized in a collection of elements, each of them representing a group of cells. Continuum based models were applied to study various morphogenetic problems, such as bone~\citep{RODRIGUEZ_JoB_27_1994} and brain growth~\citep{TALLINEN_NP_12_2016}. 

\rcol{Lattice-based approaches are another type of computational model that can simulate cell tissue dynamics over a wide range of scales~\citep{VANLIEDEKERKE_CPM_2_2015,METZCAR_JCOCCI_2_2019}. These methods rely on a fixed space discretization, whose lattice sites (or nodes) correspond to the position of cells. The number of cells represented by a lattice site can vary greatly depending on the considered approach. For cellular automata (CA), a lattice site can contain either a single cell~\citep{BLOCK_PRL_99_2007}, or many of them~\citep{RADSZUWEIT_PRE_79_2009}. 
On the contrary, cellular Pott models (CPM) assume that a living cell occupies several lattice sites, hence, properly representing the evolution of the cell shape~\citep{GRANER_PRL_69_1992,GLAZIER_PRE_47_1993}. 
Both CA and CPM simulate cellular processes (growth, mitosis, apoptosis, migration, etc) in a stochastic manner, e.g., Metropolis algorithm to minimize the energy given by a Hamiltonian, where the latter contains all biophysical constraints of interest. 
Interestingly, these methods were shown to accurately simulate various biophysical problems. At the largest scales, CA were shown to be of particular interest for cancer modeling~\citep{ANDERSON_CELL_127_2006,ENDERLING_PBMB_106_2011}. At smaller scales, CPM
were used for cells sorting and cells migration~\citep{Scianna_Preziosi_Wolf_2013}, or even bio-tube morphogenesis~\citep{Hirashima_Rens_Merks_2017}. Nevertheless, due to the stochastic nature of lattice-based approaches, it is not always clear how to derive straightforward relationships between the model parameters and biophysical properties of living cells.}

At the scale of individual cells, several lattice-free approaches have also been proposed over the years, and each of them describes cells in a different way. 
The centroid method models cells through their geometrical center (centroid), hence its name. The latter centroids obey Langevin equation of motion, and they can interact with one another as soon as the distance between the two cell centroids is smaller than a given radius of influence --in compliance with Johnson-Kendall-Roberts theory for adhesive spheres~\citep{CHU_PRL_94_2005}. This approach was used to simulate, e.g., cell tissue rheology and motility~\citep{PALSSON_PNAS_97_2000}, as well as, cell spreading/proliferation~\citep{DRASDO_PB_2_2005}. Interestingly, centroid-based methods were also hybridized with continuum approaches to create a multi-scale model that was used to investigate breast cancer~\citep{KIM_BMB_75_2013}.
Nevertheless, these models cannot provide any information on how cell deformation and morphology impact pattern formation and growth processes.

Vertex models are an alternative computational modeling of cells which represent cells through their membrane instead of their geometrical center~\citep{HONDA_JTB_84_1980,FLETCHER_PBMB_113_2013,FLETCHER_BJ_106_2014,MERZOUKI_PhD_2018}. More precisely, cell membranes are approximated with polygons, and their edges are in common with adjacent cells. Each membrane vertex follows Newton mechanics for which internal and external forces are derived through energy potentials. This approach is of particular interest for the simulation of 2D epithelial tissue, from either an apical or lateral viewpoint, and can even be extended to 3D simulations of single monolayers of cells in a straightforward manner~\citep{FLETCHER_BJ_106_2014}. Other application examples include the study of cell mechanical properties~\citep{MERZOUKI_SM_12_2016}, wound healing~\citep{FLETCHER_PBMB_113_2013}, and tissue buckling~\citep{MERZOUKI_NC_17_2018}.
Nevertheless, by sharing common vertices/edges with adjacent cells in order to avoid interpenetration issues, topological rearrangements do not naturally emerge from the interaction between cells due to the absence of interstitial gaps. Consequently, ad-hoc mechanisms must be added to the vertex model to account for these topological transitions.

A way to circumvent this issue is to move towards more accurate cell descriptions. The most detailed computational model~\citep{JAMALI_PO_5_2010} relies on a large collection of vertices that discretize both the cell membrane and nucleus. Contrarily to vertex models, this approach does not require vertices to be shared with adjacents cells, hence, properly modeling interstitial gaps between cells. Simple Voigt models (damper and spring coupled in parallel) are used to account for viscoeslastic constraints on the membrane, the cytoskeleton, and the nucleus. In addition, several cellular processes (growth, mitosis, motility, apoptosis, polarization) can be simulated by adapting spring and damper constants in either a static or dynamic manner. By accounting for external forces, this model can also include environmental effects such as cell-cell and cell-subtract interactions, or even external force fields (e.g., pushing/pulling and electromagnetic forces). Obviously, this model can reproduce most of cell developmental mechanisms, but it is extremely computationally demanding. Furthermore, mechanisms related to cell proliferation and death are rather complicated to implement. 
One way to simplify the above approach is to give up on cell growth, mitosis and apoptosis mechanisms. Doing so, (1) only the cell membrane needs to be discretized, (2) the total amount of vertices remains constant over time, and (3) no dynamic recomputing of the model constants is required. Such an approach was proposed to study gastrulation~\citep{TAMULONIS_DB_351_2011,VANDERSANDE_DB_460_2020}. In that context, only springs are used to model interactions between the membrane vertices of q given cell. Springs between consecutive vertices model the membrane elasticity and tension, whereas those between opposite vertices represent the elastic behavior of the cytoskeleton. The inner structure of the cell is not accounted for. Only the cytoplasm is mimicked through a force depending on the cell area variations. Eventually, repulsive/adhesive contact fores are used to model the interaction between vertices belonging to adjacent cells.

If one still wants to simulate cell proliferation and apoptosis in a detailed and rather simple manner, 
the immersed boundary methodology is an interesting alternative~\citep{REJNIAK_BMB_66_2004,DILLON_CM_466_2008,TANAKA_PhD_2016}. 
The latter originates from Computational Fluid Dynamics, and it was originally introduced to study fluid-structure interactions~\citep{PESKIN_AN_11_2002}. 
Contrarily to the simpler spring or Voigt models, forces are computed in an iterative manner depending on interactions with inner and outer fluids (cytoplasm and extra-cellular matrix, respectively). Similarly to the above simplified formulation, it only represents vertices on the cell membrane.
Interestingly, even if it does not explicitly represent the cell nucleus, this approach can be used to simulate cell proliferation and apoptosis in a simple manner. Indeed, when the distance between two consecutive vertices is greater/smaller than a predefined threshold, then vertices are added/removed to ensure an accurate and uniform discretization of the cell membrane.
Once the cell is large enough (usually twice its rest size), mitosis can be triggered by simply cutting the mother cell (along its minor axis) into two daughter cells, and adding vertices to close their membrane. 
The main advantage of immersed-boundary formulations is that they self-adjust to the number of vertices, hence, no constant modification is required during cell proliferation and apoptosis. This particular approach was notably used to simulate, tissue bending~\citep{REJNIAK_BMB_66_2004}, pattern formation and necrosis~\citep{DILLON_CM_466_2008}, and morphogenic signaling~\citep{TANAKA_Bioinformatics_31_2015}.
Nevertheless, inner and outer fluids must be simulated to compute forces that apply on the membrane vertices, and these fluids generate spurious velocity currents at the cell membrane interface. In addition, the iterative computation of forces is expensive as compared to spring and Voigt models.

In order to simulate most cellular processes in a detailed, yet simple manner, we propose a general formulation that keeps most advantages of previous works~\citep{JAMALI_PO_5_2010,TAMULONIS_DB_351_2011,TANAKA_Bioinformatics_31_2015} while getting rid of most of their drawbacks. More precisely, we dynamically adjust the number of vertices used for the evolution of the cell membrane (as in~\citep{JAMALI_PO_5_2010}), but, similarly to ~\citep{TAMULONIS_DB_351_2011}, the cell nucleus, cytoskeleton and cytoplasm do not explicitly appear in our formulation. This is possible assuming the same cell division mechanism as in immersed boundary cell-based approaches~\citep{TANAKA_Bioinformatics_31_2015}. Discarding the cell inner structure further allows us to propose simple rules to adjust the cell mechanical properties, which are modelled through linear and torsion springs (elasticity and bending, respectively). Eventually, the present model does not require the simulation of inner and outer fluids. Nevertheless, the impact of the cytoskeleton, cytoplasm and extra-cellular matrix \rcol{is accounted for through internal and external forces.} 

As a first study, we propose to present in details our new formalism, and to validate it in a qualitative manner through various benchmark tests. More precisely, choices made regarding vertex dynamics, single-cell proliferation mechanisms and chemical signaling are detailed in Section~\ref{sec:Model}. The minimal number of vertices required to accurately discretize the cell membrane is further identified through a simple convergence study. The incompressible, visco-elastic behavior of cells is investigated in Section~\ref{sec:Preliminary}, alongside with a first attempt to model cell motility inside a tissue. Section~\ref{sec:Validation} contains more complex validation cases, such as free and chemically piloted tissue growth in an unbounded domain, as well as, cell proliferation under external constraints. Eventually, conclusions are drawn in Section~\ref{sec:Conclusion}.

\section{Model description\label{sec:Model}}

\subsection{Cell description\label{subsec:cell_description}}
Vertex models are commonly used to describe the evolution of cell tissues in the context of morphogenesis~\citep{FLETCHER_BJ_106_2014}. These models describe cells as polygons composed of vertices, and which share edges with neighboring cells. This allows for an efficient modeling of phenomena occurring at the scale of tissues, whereas those emerging at the cellular level must be imposed through an ad-hoc mechanism -- the so called T1 and T2 transitions. 

\begin{figure}[h]
    \centering
    \begin{subfigure}[b]{0.45\textwidth}
    \includegraphics[width=0.49\textwidth]{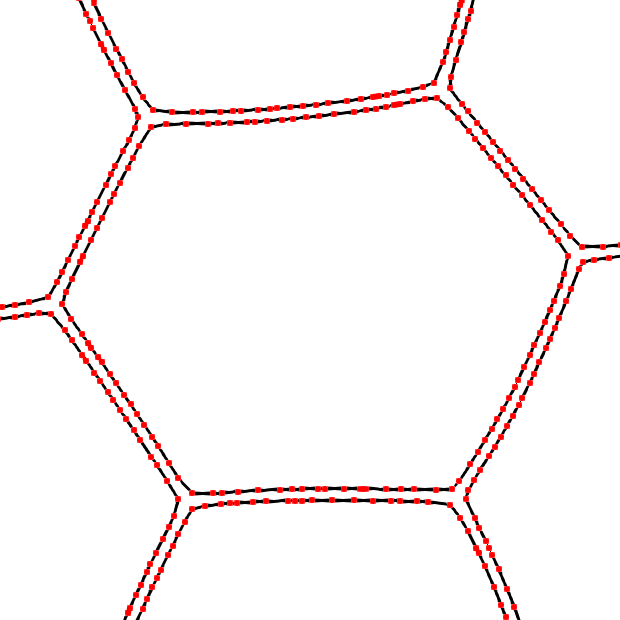}
    \hfill
    \includegraphics[width=0.49\textwidth]{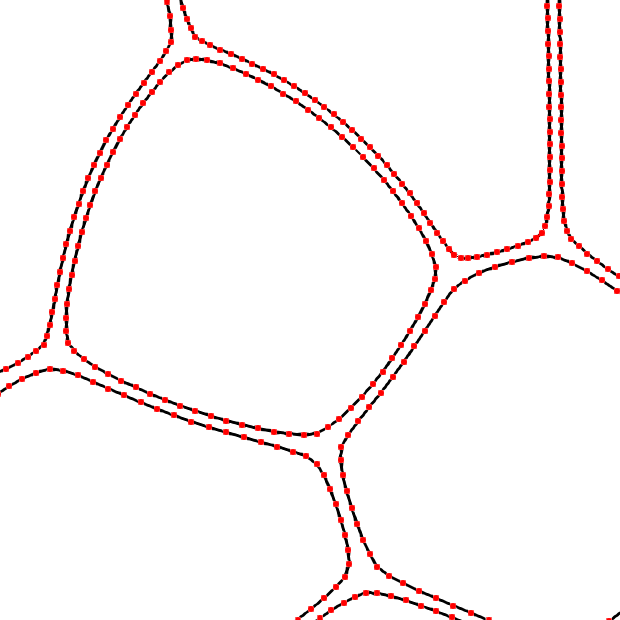}
    \caption{PalaCell}
    \end{subfigure}
    \hspace{0.05\textwidth}
    \begin{subfigure}[b]{0.45\textwidth}
    \includegraphics[width=0.49\textwidth]{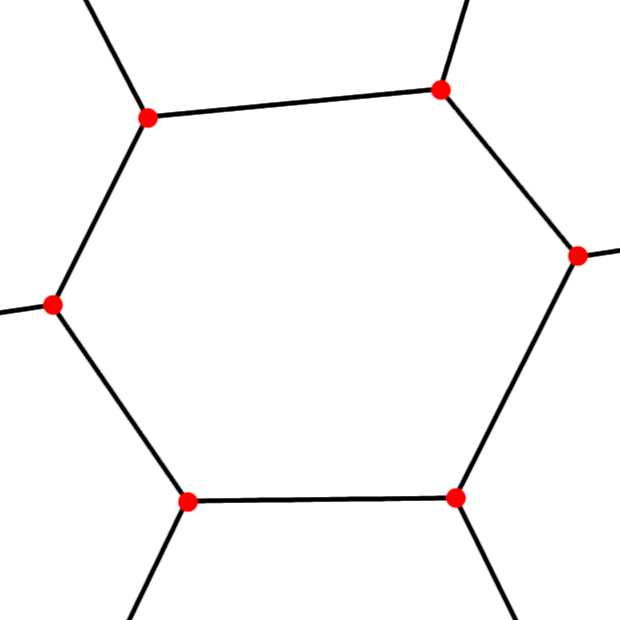}
    \hfill
    \includegraphics[width=0.49\textwidth]{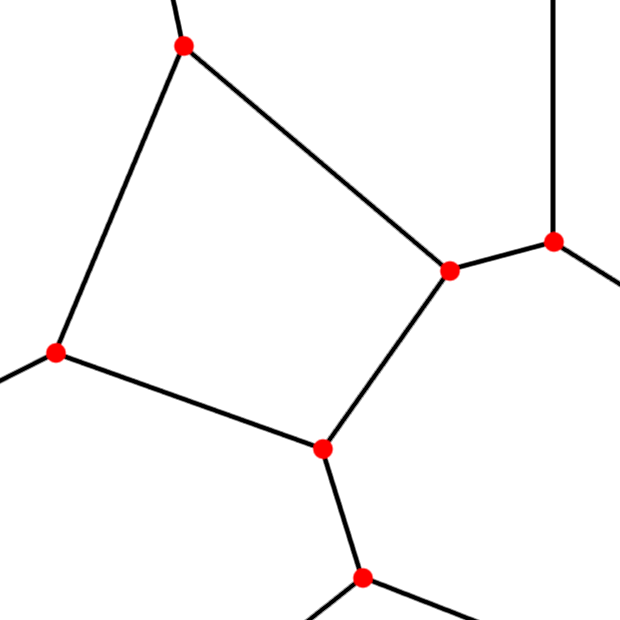}
    \caption{Vertex model}
    \end{subfigure}
    \hfill
    \caption{Cells resolution, with the vertices in red and the membrane in black.}
    \label{fig:resolution}
\end{figure}

On the contrary, we propose hereafter to model a cell as a collection of vertices. In that context, the membrane of a cell is composed of (approximately one hundred) vertices whose \rcol{dynamic properties are} governed by mechanical and biological mechanisms (see Fig.~\ref{fig:resolution}). 
As each cell is totally independent, more complex cell shapes and behaviors naturally emerge from the model without any ad-hoc mechanism. 
As an example, cell migration within a tissue can be simulated through simple mechanisms (see Sec.~\ref{subsec:motility}), whereas, in approaches available in the literature, it is commonly imposed through consecutive T1 transitions within vertex models~\citep{FLETCHER_BJ_106_2014,MERZOUKI_PhD_2018}.

		 
        
\begin{figure}[H]
    \centering
    \begin{subfigure}{0.3\textwidth}
       \includegraphics[width=0.9\textwidth]{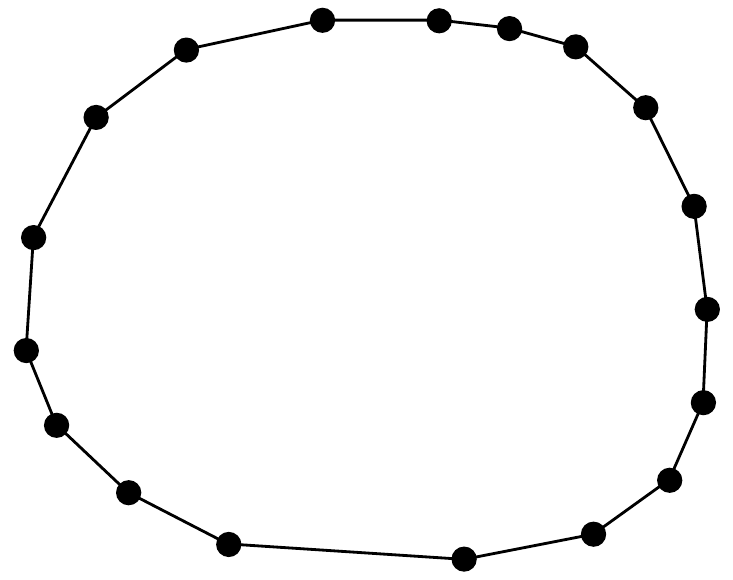}
       \caption{Initial state: inhomogeneous discretization}
    \end{subfigure}
    \begin{subfigure}{0.3\textwidth}
       \includegraphics[width=0.9\textwidth]{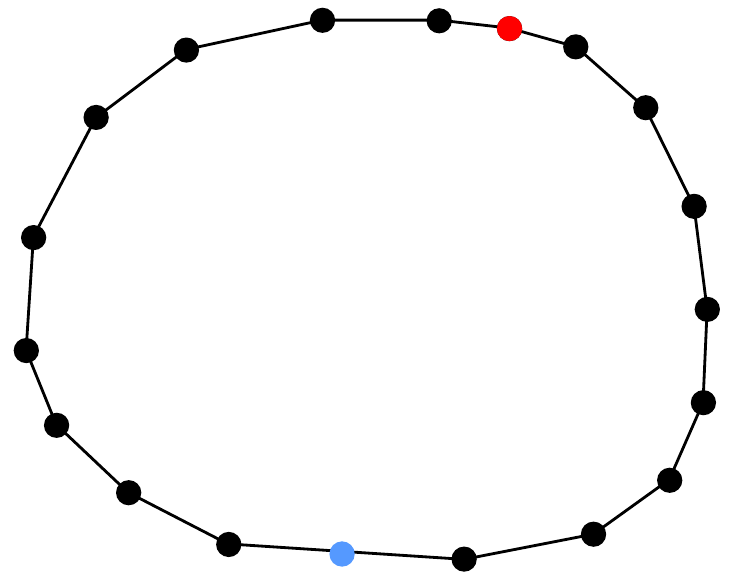}
       \caption{Vertices are added (blue) and removed (red).}
    \end{subfigure}
    \begin{subfigure}{0.3\textwidth}
       \includegraphics[width=0.9\textwidth]{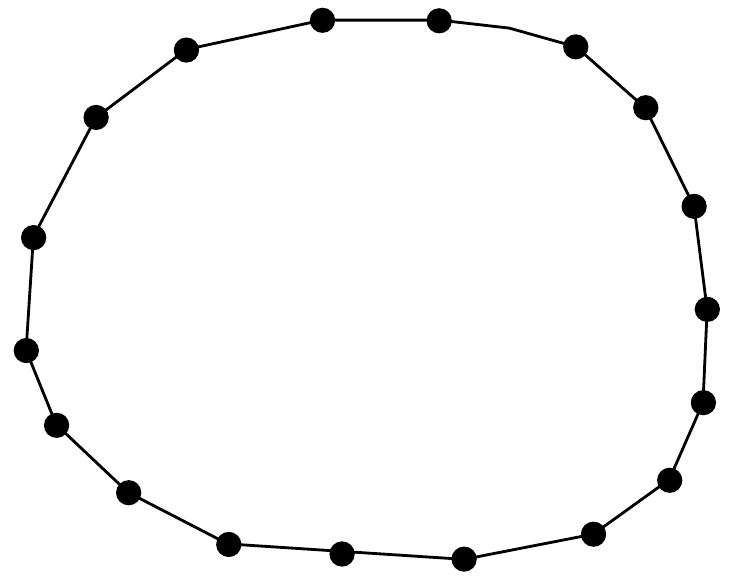}
       \caption{Final state: homogeneous discretization.}
    \end{subfigure}
    \caption{Example of addition/suppression mechanisms to keep a uniformly discretized membrane. In practice, the red vertex is removed because both its neighbors are at a distance lower than $l_0/2$.}
    \label{fig:addrm_vertex}
\end{figure}

In order to simulate the growth and division of cells, while keeping the membrane resolution constant, new vertices must be added. Similarly, when the perimeter of cells is reduced (e.g., apoptosis, external constraints, etc) extra vertices must be removed to avoid local vertex clusterings. Concretely speaking, the addition and the suppression of a vertex depend on the distance $d_{i,i+1}$ between two vertices $i$ and $i+1$, and a characteristic length $l_0$ that corresponds to the cell perimeter divided by a predefined number of vertices $n$ (see \ref{app:appendix} for its expression). In our formulation, both mechanisms are imposed through the following rules~\citep{TANAKA_PhD_2016}: 
\begin{align*}
\text{adding vertex:} &\text{if } d_{i,i+1} > 2l_0, \\
\text{removing vertex:} &\text{if } d_{i,i+1} < l_0/2,
\end{align*}
and further illustrated in Fig.~\ref{fig:addrm_vertex}.

\subsection{Mechanical dynamics}

\begin{figure}[!ht]
    \begin{subfigure}{.45\textwidth}
        \centering
        \includegraphics[width=.9\linewidth]{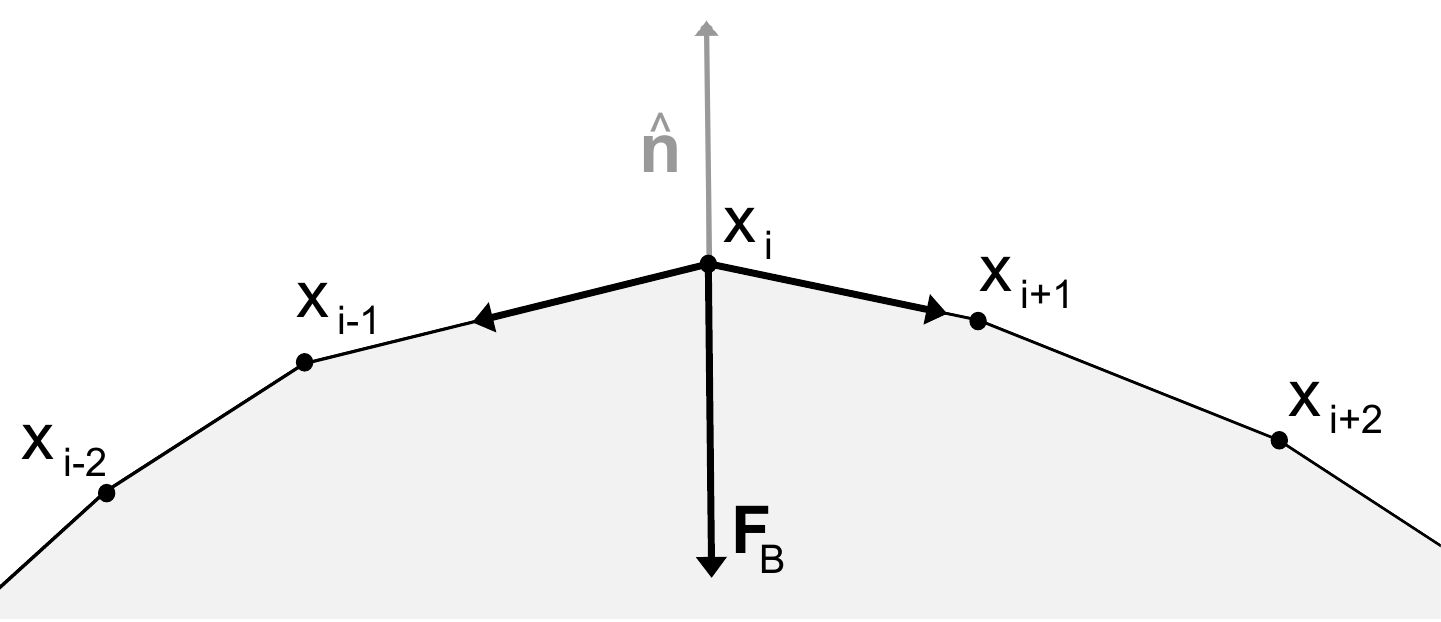}
        \caption{Bending force}
    \end{subfigure}
    \begin{subfigure}{.45\textwidth}
        \centering
        \includegraphics[width=.9\linewidth]{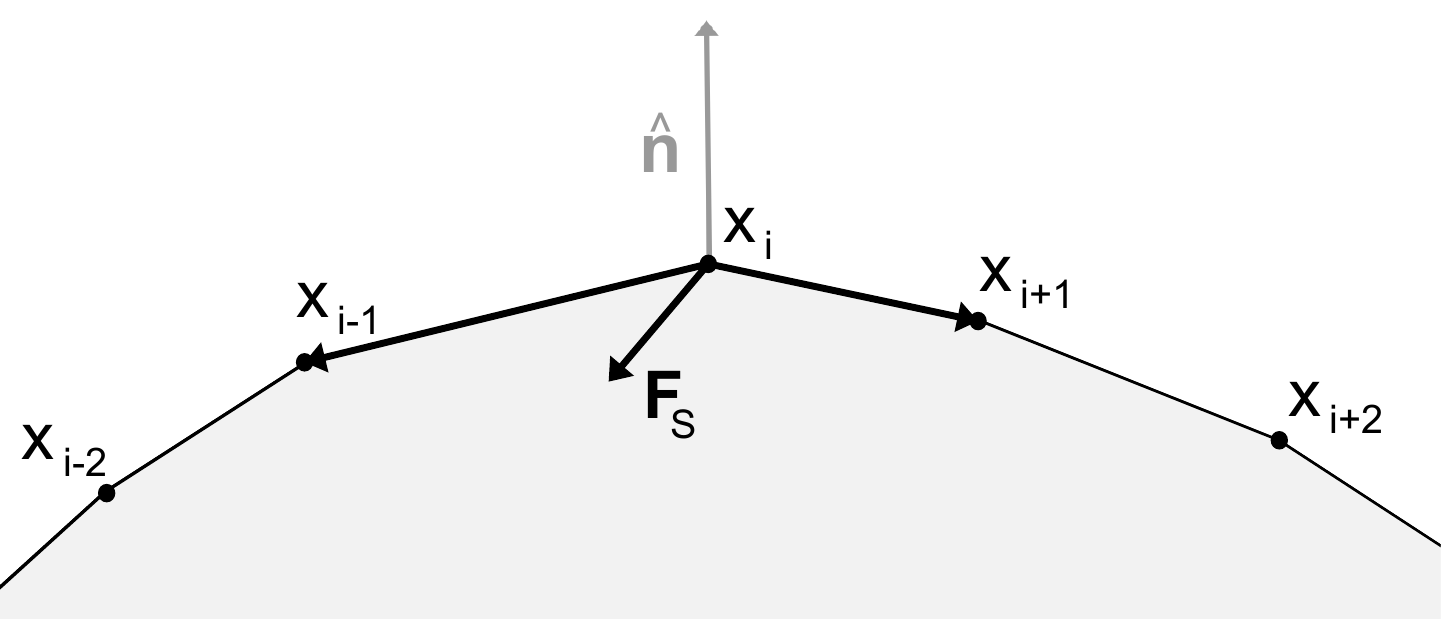}
        \caption{Spring force}
    \end{subfigure}
    \caption{Illustration of the membrane force impact on the vertex dynamics. The normal vector $\bm{\hat{n}}$ is taken at vertex $i$.}
    \label{fig:memf}
\end{figure}

The evolution of vertices follows Newton's law of motion: 
\begin{equation}
\bm{F} = \bm{F}_{P} + \bm{F}_{M} + \bm{F}_{A} + \bm{F}_{ext}.
\label{eq:all}
\end{equation}
The resulting force $\bm{F}$ (\ref{eq:all}) acts on a vertex through various mechanisms: pressure $\bm{F}_P$, membrane resistance to deformation $\bm{F}_M$, adherence between cells $\bm{F}_A$, and external force $\bm{F}_{ext}$. The latter force can either be local (e.g., interaction with a wall, migration) or global (e.g., gravity, interaction with the extra-cellular matrix), and is generally used to model an external constraint. 
The force $\bm{F}_{M}$ accounts for both resistance to elongation and torsion, which allows us to control the membrane tension:
\begin{equation}
\bm{F}_{M} = \bm{F}_{B} + \bm{F}_{S}.
\label{eq:memb}
\end{equation}
$\bm{F}_B$ stands for the bending force which naturally appears when consecutive vertices are non-aligned.
Regarding the spring force $\bm{F}_S$, it connects two vertices and aims at balancing the distance between vertices, hence, ensuring a uniform discretization of the cell membrane. 
While $\bm{F}_B$ is purely perpendicular to the membrane, $\bm{F}_S$ has both a tangential and a normal component. These membrane forces are illustrated in Fig.~{\ref{fig:memf}}.

\begin{figure}[!ht]
    \centering
    \begin{subfigure}{.3\textwidth}
        \centering
        \includegraphics[width=\linewidth]{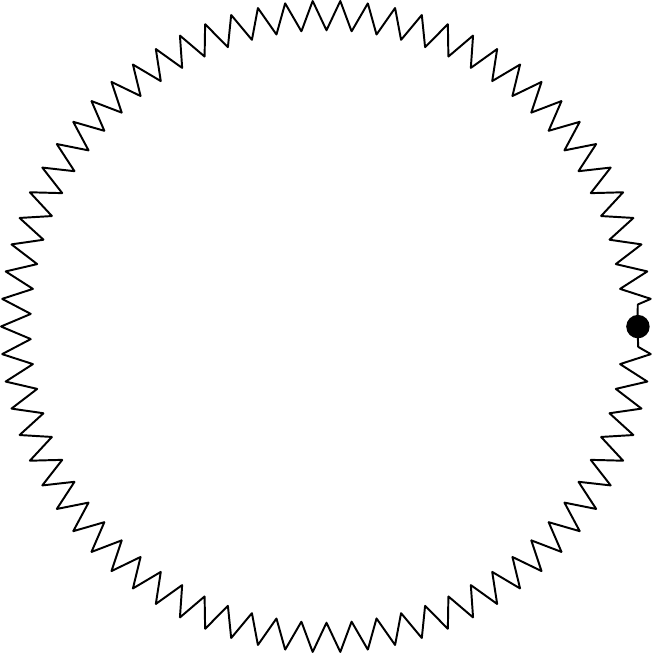}
        \caption{Global spring}
        \label{fig:membrane-1}
    \end{subfigure}
    \hspace{2cm}
    \begin{subfigure}{.3\textwidth}
        \centering
        \includegraphics[width=\linewidth]{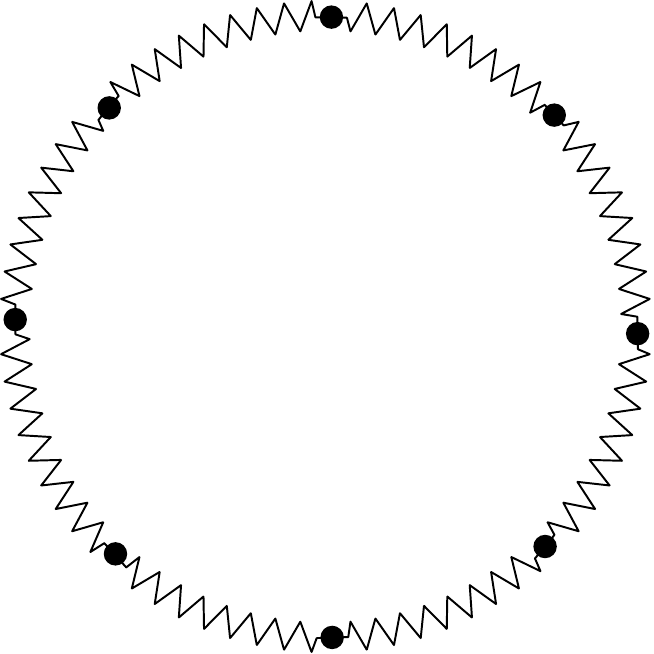}
        \caption{Local springs}
        \label{fig:membrane-8}
    \end{subfigure}
    \caption{Dual representation of the membrane springs. The global spring is used to ensure the cell elasticity will not depend on the number of vertices. Regarding local springs, they are dynamically adjusted when the number of vertices change, hence, ensuring a homogeneous representation of the membrane.}
    \label{fig:membrane}
\end{figure}

Going into more details, the bending force $\bm{F}_{B}$ depends on the local curvature, i.e.,
when the cell is locally convex/concave, it points inward/outward the cell. At vertex $i$, it reads as~\citep{TANAKA_PhD_2016}
\begin{equation}
\bm{F}_B(\bm{x}_{i}) = \rcol{k_b} \left( \frac{\bm{x}_{i+1} - \bm{x}_i}{\lVert \bm{x}_{i+1} - \bm{x}_i \rVert} + \frac{\bm{x}_{i-1} - \bm{x}_i}{\lVert \bm{x}_{i-1} - \bm{x}_i \rVert} \right), 
\label{eq:bend}
\end{equation}
with $\rcol{k_b}$ being the bending constant.
The spring force $\bm{F}_S$ depends on the distance with respect to neighbors vertices. In order to avoid local discrepancies due to, e.g., adding vertices to simulate proliferation, a global spring is used to model the spring force over the whole membrane (Fig.~\ref{fig:membrane-1}). A global spring constant $K_s$ is then assigned to this spring. Once the number of vertices $n$ is known, this global spring is divided into $n$ smaller springs (Fig.~\ref{fig:membrane-8}), whose spring constant is derived assuming springs are in series, i.e., $k_s = n K_s$. The latter condition also has the very nice property of leading to a cell radius that is independent of the membrane discretization (see \ref{app:appendix}). For a vertex $i$, the spring force reads as  
\begin{equation}
\bm{F}_S(\bm{x}_i) = k_s \left[ (\bm{x}_{i+1} - \bm{x}_i) + (\bm{x}_{i-1} - \bm{x}_i) \right]
\label{eq:spring}
\end{equation}
with $k_s$ the local spring constant.

\begin{figure}[!ht]
    \begin{subfigure}{.5\textwidth}
        \centering
        \includegraphics[width=.9\linewidth]{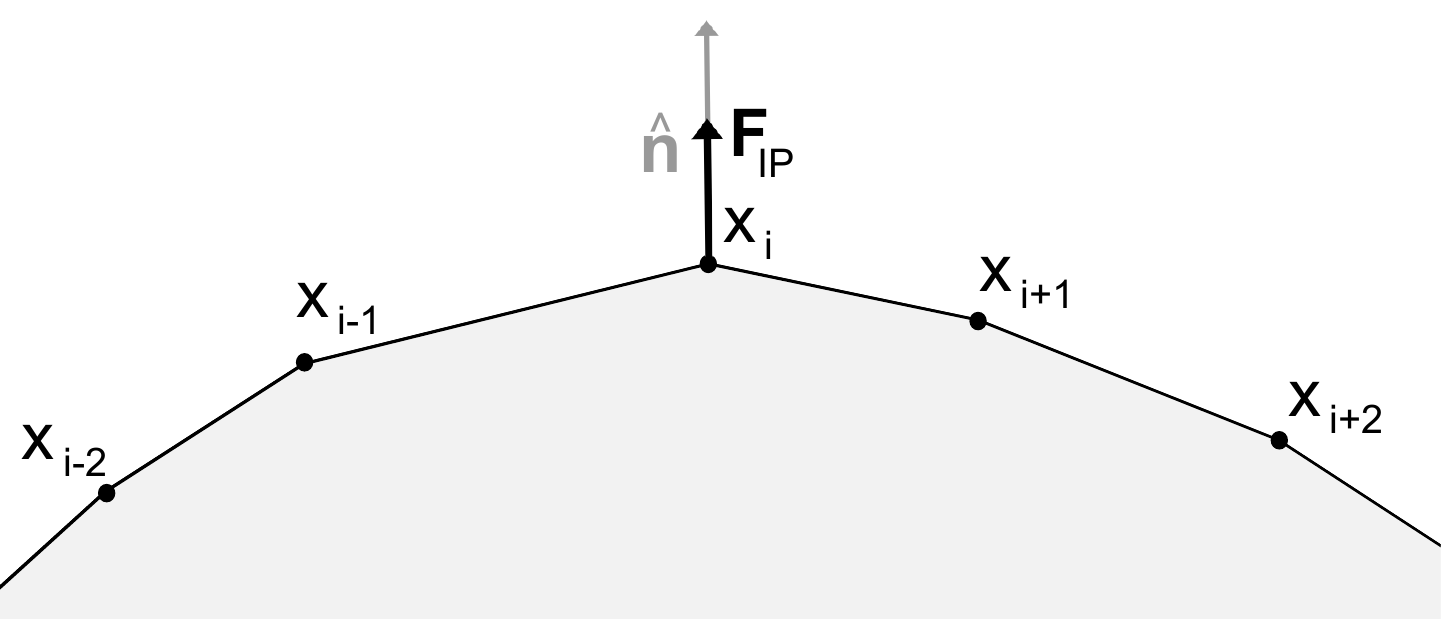}
        \caption{Internal pressure force}
    \end{subfigure}
    \begin{subfigure}{.45\textwidth}
        \centering
        \includegraphics[width=.9\linewidth]{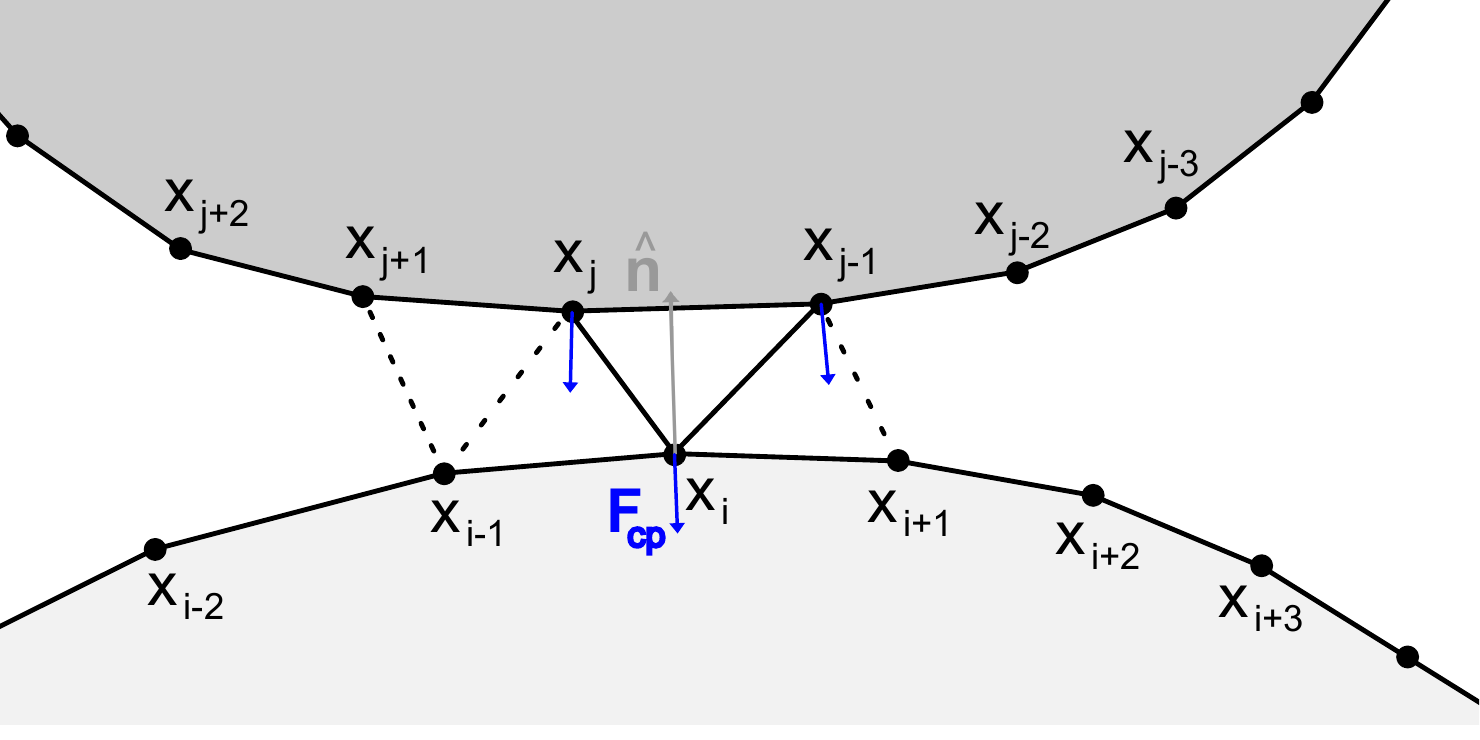}
        \caption{Contact pressure force}
        \label{subfig:5b}
    \end{subfigure}
    \caption{Representation of pressure forces. For the contact pressure force, solid lines corresponds to connections between the current vertex $i$ and vertices belonging to the other cell $(j,j-1)$, whereas dashed lines correspond to connections with adjacent vertices $(i-1,i+1)$.}
    \label{fig:presf}
\end{figure}

The pressure force $\bm{F}_P$ can be decomposed into two parts
\begin{equation}
\bm{F}_{P} = \bm{F}_{IP} + \bm{F}_{CP}
\label{eq:press}
\end{equation}
where the internal pressure force $\bm{F}_{IP}$ allows us to discard the inner structure of the cell (cytoplasm, cytoskeleton and nucleus), by accounting for internal mechanisms, e.g., mitosis and apoptosis (see Section~\ref{subsec:growth}). On the contrary, the contact pressure force $\bm{F}_{CP}$ is purely mechanical, and it corresponds to the force imposed by vertices belonging to neighboring cell. The pressure force $\bm{F}_P$ (\ref{eq:press}) is used to balance the membrane force $\bm{F}_M$ (\ref{eq:memb}), and consequently, it can be used to trigger cell proliferation or death. More precisely, the internal pressure force is applied on a portion of the membrane centered around vertex $i$, i.e., $\mathrm{force} = \mathrm{pressure} \times \mathrm{surface}$. In our case, it is expressed as
\begin{equation}
\bm{F}_{{IP}}(\bm{x}_i) = p_{int} \frac{\left\lVert \bm{x}_{i+1} - \bm{x}_{i} \right\rVert + \left\lVert \bm{x}_{i-1} - \bm{x}_{i} \right\rVert}{2} \bm{\hat{n}},
\label{eq:ipress}
\end{equation}
with $p_{int}$ the pressure inside the cell (whose evolution is described in Eq.~(\ref{eq:internal_pressure})), and $\bm{\hat{n}}$ the local normal to the membrane, which reads as 

\begin{equation*}
    \bm{\hat{n}} = \frac{\bm{\hat{n}}_{i+1,i}+\bm{\hat{n}}_{i-1,i}}{\vert\vert\bm{\hat{n}}_{i+1,i}+\bm{\hat{n}}_{i-1,i}\vert\vert},
\end{equation*}
with
\begin{equation*}
\bm{\hat{n}}_{i+1,i}=\frac{\bm{x}_{i+1}-\bm{x}_{i}}{\vert\vert\bm{x}_{i+1} - \bm{x}_{i}\vert\vert}
\begin{pmatrix}
\phantom{-}0 & -1\phantom{-}\\
\phantom{-}1 & \phantom{-}0\phantom{-}
\end{pmatrix}, \quad
\bm{\hat{n}}_{i-1,i}= \frac{\bm{x}_{i-1}-\bm{x}_{i}}{\vert\vert\bm{x}_{i-1}- \bm{x}_{i}\vert\vert}
\begin{pmatrix}
\phantom{-}0 & \phantom{-}1\phantom{-}\\
-1 & \phantom{-}0\phantom{-}
\end{pmatrix},    
\end{equation*}
and where matrices correspond to rotations of $\pi/2$ and $-\pi/2$ respectively.
Regarding the contact pressure, it only impacts the vertex dynamics when (at least) two cells are in contact. Vertices belonging to different cells are considered to be in contact, if their distance is smaller than a predefined maximal range of interaction $d_{max}$. In that case, the contact pressure force $\bm{F}_{CP}$ is the sum of internal pressure forces that belong to each cell in contact with the current one, weighted by the number of connections $w_j$: 
\begin{equation}
    \bm{F}_{CP}(\bm{x}_i) = \sum_{j \in CV_i} \frac{\bm{F}_{IP}(\bm{x}_j)}{w_{j}}.
    \label{eq:cpress}
\end{equation}
with $CV_i$ the set of vertices belonging to other cells, and that are inside the range of interactions of vertex $i$. For the particular case illustrated in Fig.~\ref{subfig:5b}, we have $CV_i=\{j-1,j\}$, where the corresponding weights read are $w_{j}=w_{j-1}=2$ because $j$ and $j-1$ are connected to two vertices: $(i-1,i)$ and $(i,i+1)$ respectively.

\begin{figure}[!ht]
    \centering
    \begin{subfigure}{0.45\textwidth}
        \centering
        \includegraphics[width=0.9\linewidth]{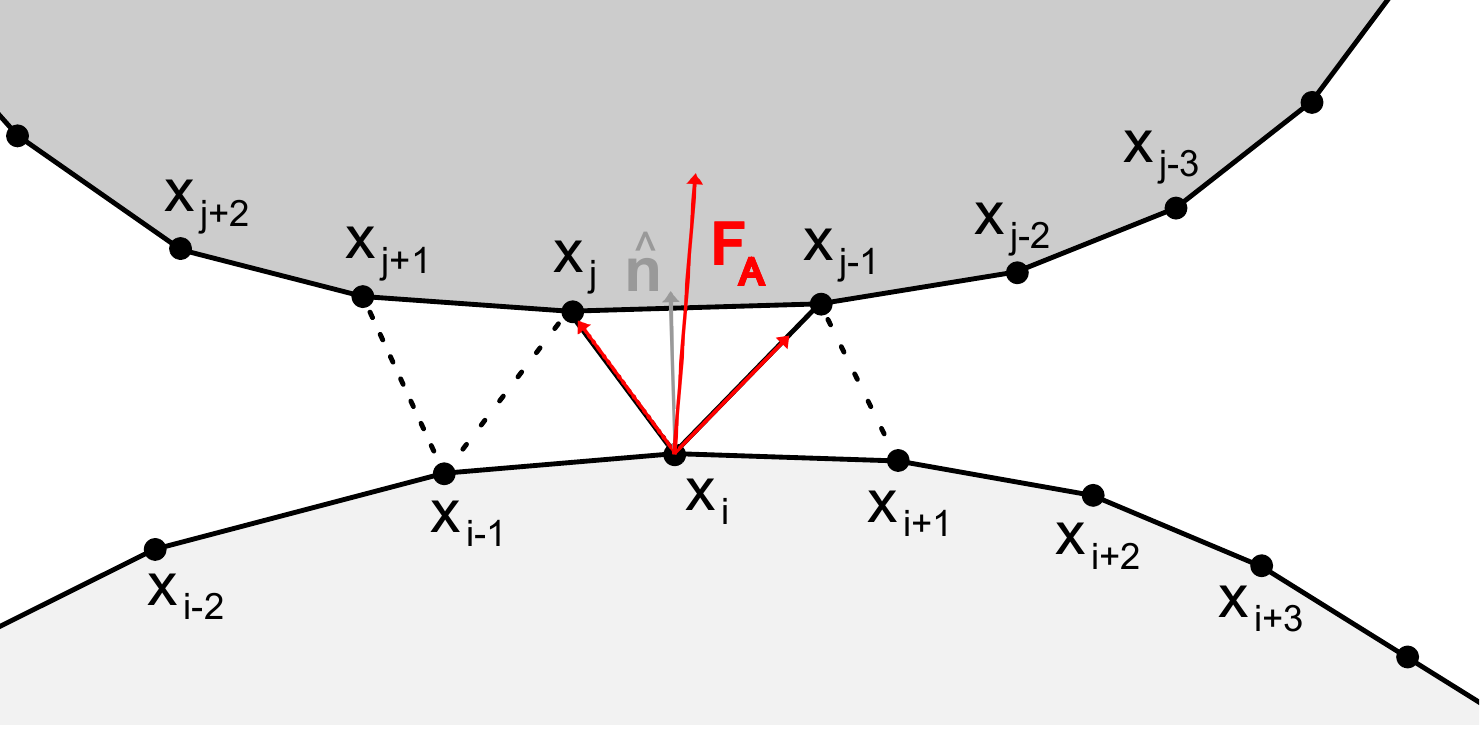}
        \caption{Sketch of the adherence force}
        \label{fig:adhf}
    \end{subfigure}
    \begin{subfigure}{0.45\textwidth}
        \centering
        \includegraphics[width=0.9\linewidth]{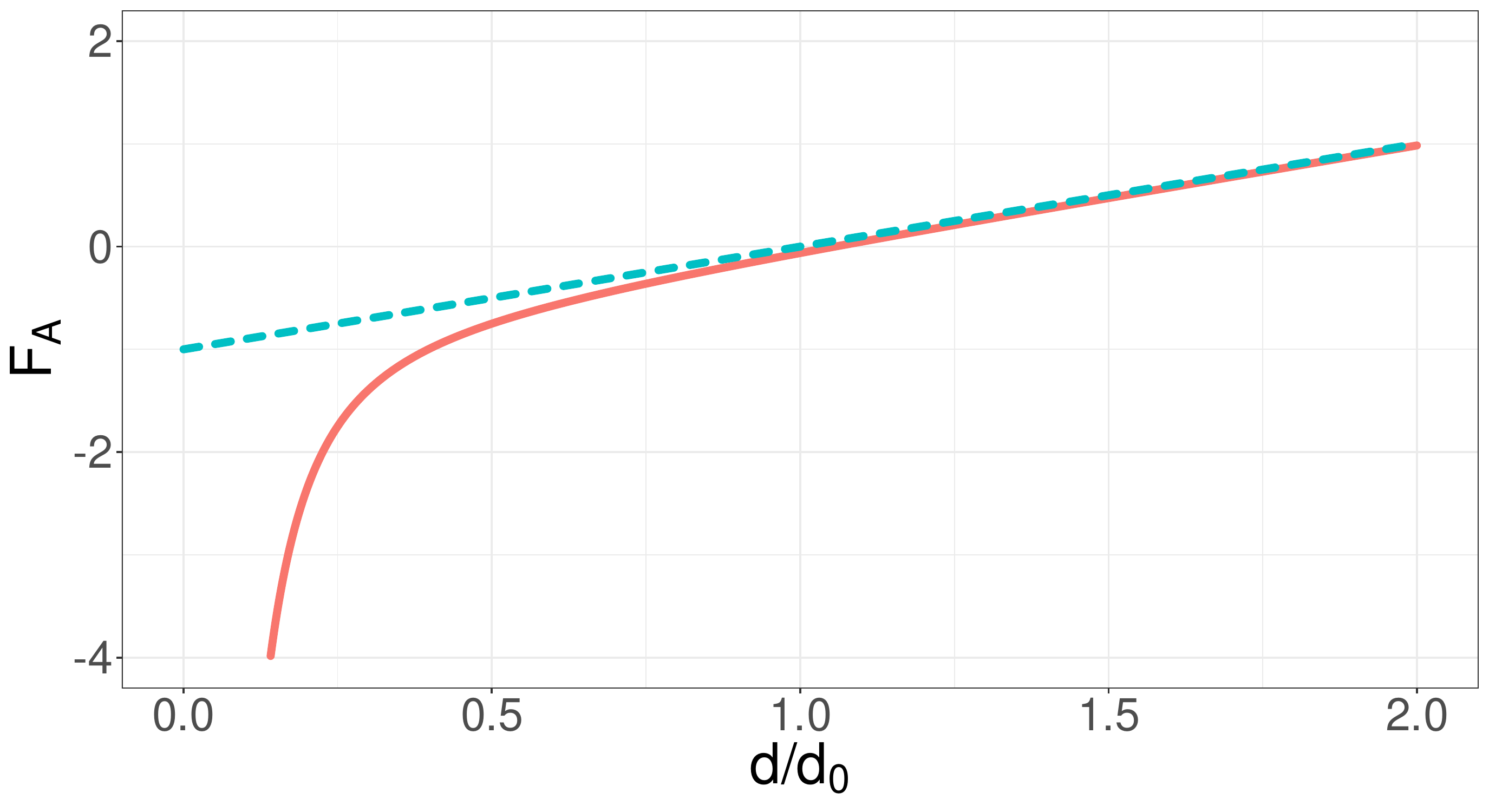}
        \caption{Adherence force modelling: \citep{TANAKA_PhD_2016} (linear) and our model (\ref{eq:adh})}
        \label{fig:adh_comp}
    \end{subfigure}
    \caption{Sketch and comparison of adherence force modelling}
\end{figure}

Finally, the adherence force $\bm{F}_A$ is the force that keeps cells connected to each other. It also activates when vertices, belonging to different and adjacent cells, are within a given range of interaction $d_{max}$. The force used in our model is a adapted from the adherence model proposed in \citep{TANAKA_PhD_2016}. In our case, we replaced the linear approximation by a potential in order to impose a rest distance $d_0=d_{max}/2$. Doing so, the repulsive behavior of $\bm{F}_A$ is strengthen, and it prevents interpenetration issues.
This force is defined as follows
\begin{equation}
\bm{F}_{A}(\bm{x}_i) = k_a \sum_{j \in CV_i} \frac{\bm{x}_i - \bm{x}_j}{\lVert \bm{x}_i - \bm{x}_j \rVert} \left[ \lVert \bm{x}_i - \bm{x}_j \rVert - d_0\left(1 - \frac{d_0^2}{16\lVert \bm{x}_i - \bm{x}_j \rVert^2}\right) \right]
\label{eq:adh}   
\end{equation}
with $k_a$ the adherence constant. The behavior of the adherence force $\bm{F}_A$ is sketched in Fig.~\ref{fig:adhf}, and further compared to~\citep{TANAKA_PhD_2016} in Fig.~\ref{fig:adh_comp}.

\subsection{Biological dynamics}

\subsubsection{Cell growth, relaxation and apoptosis\label{subsec:growth}}
In our work, biological mechanisms are accounted for through the internal pressure $p_{int}$ that controls the cell size via
\begin{equation}\label{eq:internal_pressure}
    p_{int} = \eta \left(\frac{m}{A} - \rho_0\right)
\end{equation}
with $m$ and $A$ the cell mass and area respectively. This biological mechanism corresponds to a relaxation of cells towards their target density $\rho_0$, and it is controlled by the pressure sensitivity $\eta$.  

\begin{figure}[!ht]
    \centering
    \begin{subfigure}{.3\textwidth}
        \centering
        \includegraphics[width=0.9\textwidth]{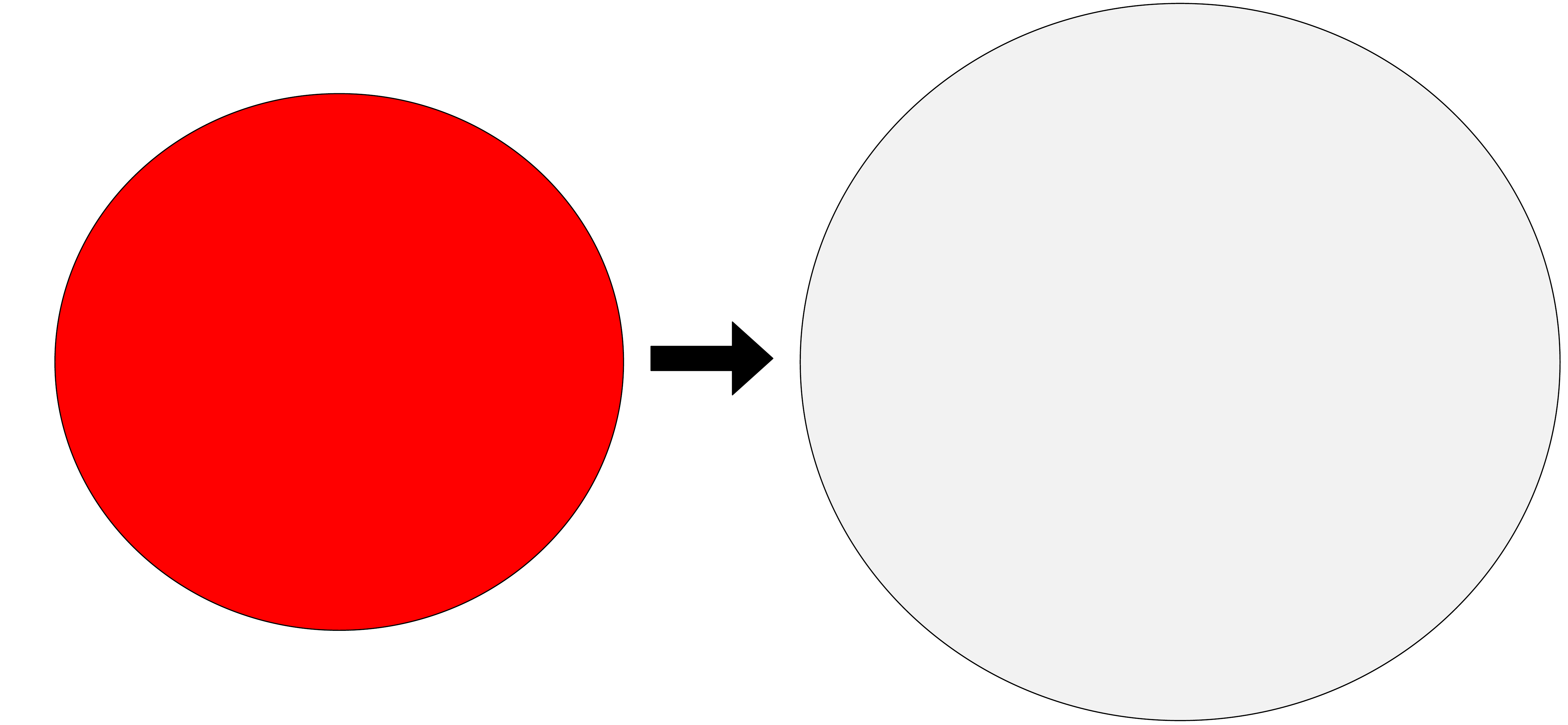}
        \caption{Cell with high density (red)}
        \label{fig:rho_high}
    \end{subfigure}
    \begin{subfigure}{.3\textwidth}
        \centering
        \includegraphics[width=0.9\textwidth]{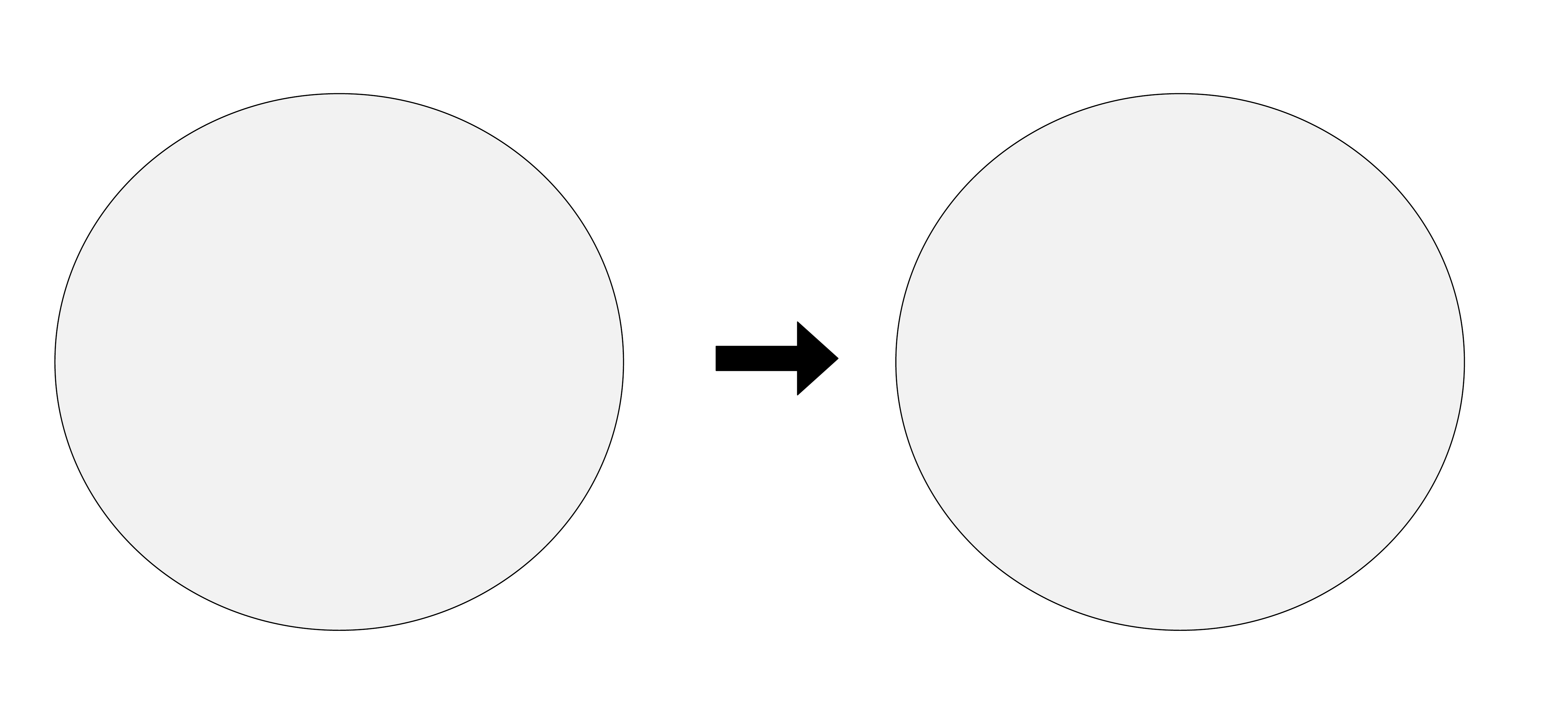}
        \caption{Cell with target density (grey)}
    \end{subfigure}
    \begin{subfigure}{.3\textwidth}
        \centering
        \includegraphics[width=0.9\textwidth]{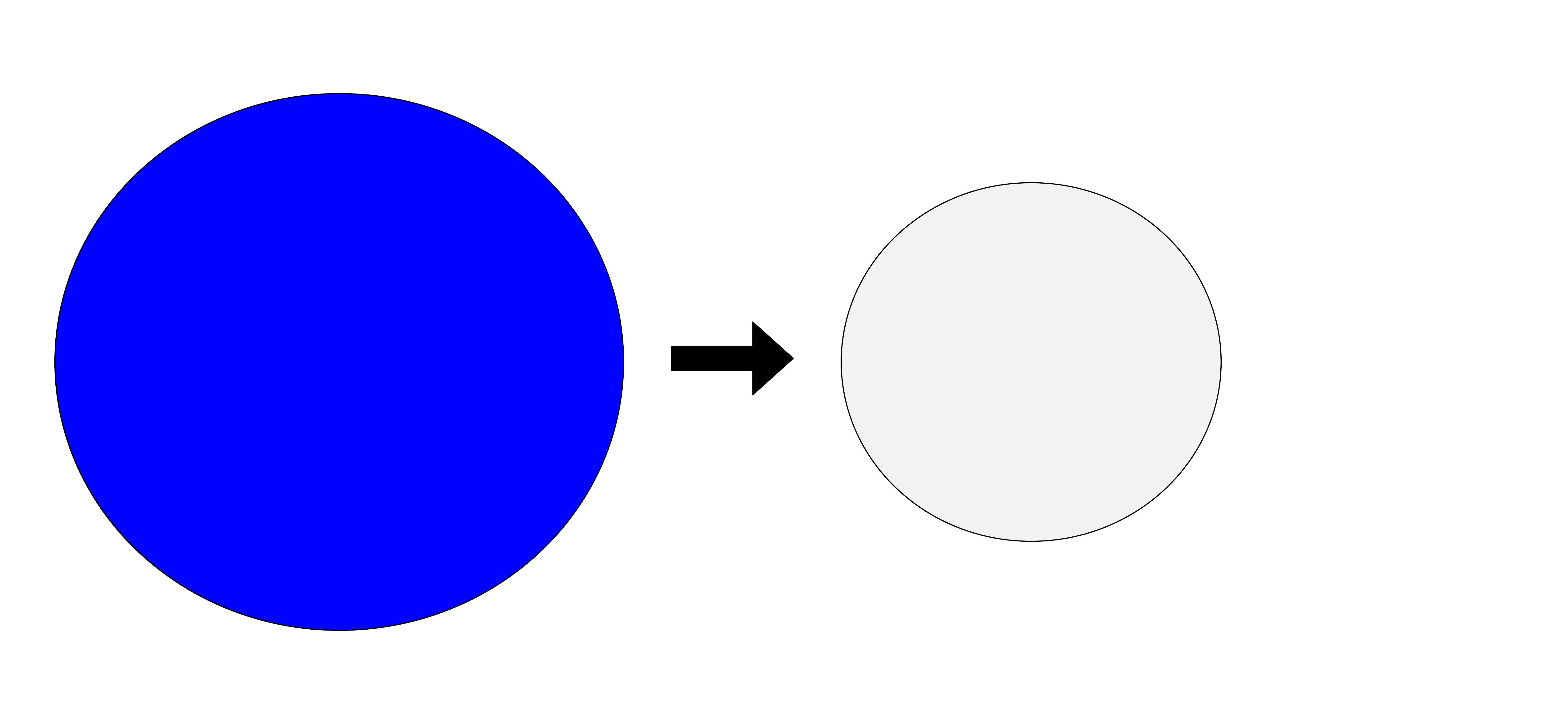}
        \caption{Cell with low density (blue)}
        \label{fig:rho_low}
    \end{subfigure}
    \caption{Cell internal mechanisms: growth, rest and relaxation (from left to right).}
    \label{fig:cell_density}
\end{figure}

Cells have different states during their life cycle. To account for this natural cycle, the mass of a cell is modified depending on the cell state one wants to simulate (proliferation, rest or apoptosis).
As an example, by increasing the mass of the cell, the density of the cell increases accordingly. In turns, the cell starts to grow until the target area/density of the cell is reached (Fig.~\ref{fig:rho_high}). Alternatively, it is possible to reduce the cell size by reducing its mass (Fig.~\ref{fig:rho_low}).

\noindent In addition, the cell state is also directly impacted by forces imposed by surrounding cells. 
For instance, when the pressure on the cell membrane increases, the cell reduces its growth speed~\citep{ROUX_PRIVATE_2020}. 
In our model, as the growth is controlled by the internal pressure, it is used to adapt the growth rate by adjusting the mass rate $dm$, which in turn increases the cell size according to the pressure exerted on the cell membrane.
Hereafter, these mechanisms are modelled through the following rules:
\begin{equation}\label{eq:growthRelax}
dm = \begin{cases}
 \nu A \left( 1 - \frac{p_{ext}}{p_{max}}\right) dt &\text{ for proliferating cells}\\
- \nu_{relax} \left(A - A_0\right) \left(1-\frac{p_{ext}}{p_{max}} \right) dt &\text{ otherwise.}
\end{cases} 
\end{equation}
where $\nu$ and $\nu_{relax}$ are mass growth rates related to proliferation and relaxation respectively state. $A_0$ is the target area, $dt$ the time step of the simulation, and $p_{max}$ a pressure threshold above which the cell proliferation is stopped. $p_{ext}$ is the external pressure acting on the cell, and is defined through
\begin{equation}
p_{ext} =  \frac{2[\bm{F}_{{CP}}(\bm{x}_i) + \bm{F}_{ext}(\bm{x}_i)]\cdot\bm{\hat{n}}}{\left\lVert \bm{x}_{i+1} - \bm{x}_{i} \right\rVert + \left\lVert \bm{x}_{i-1} - \bm{x}_{i} \right\rVert},
\label{eq:pext}
\end{equation}
where ``$\cdot$'' is the scalar product.
All these mechanisms (growth, rest and relaxation) are illustrated in Fig.~(\ref{fig:cell_density}). It is worth noting that rules~(\ref{eq:growthRelax}) are one possibility among many others, and consequently, one could also choose, e.g., a power law of the type $(1-p_{ext}/p_{max})^{\alpha}$ with $\alpha>0$. The impact of such a choice, on the biophysical properties of cells, will be investigated in a future work.

Eventually, apoptosis is triggered when the number of vertices $n$, used to discretized the cell membrane, is below a predefined threshold $n_{a}=30$. This corresponds to $A\lesssim A_0/16$, assuming that $n=120$ (see Section~\ref{subsec:convergence}). In practice, the remaining points are simply removed from the simulation domain, and connections with vertices belonging to other cells are removed as well. Obviously, one could impose a minimal area $A_{min}$, instead of $n_a$, as a criterion to trigger apoptosis.

\subsubsection{Cell mitosis}
\begin{figure}[!ht]
    \centering
    \begin{subfigure}{0.4\textwidth}
        \centering
        \includegraphics[width=0.8\textwidth]{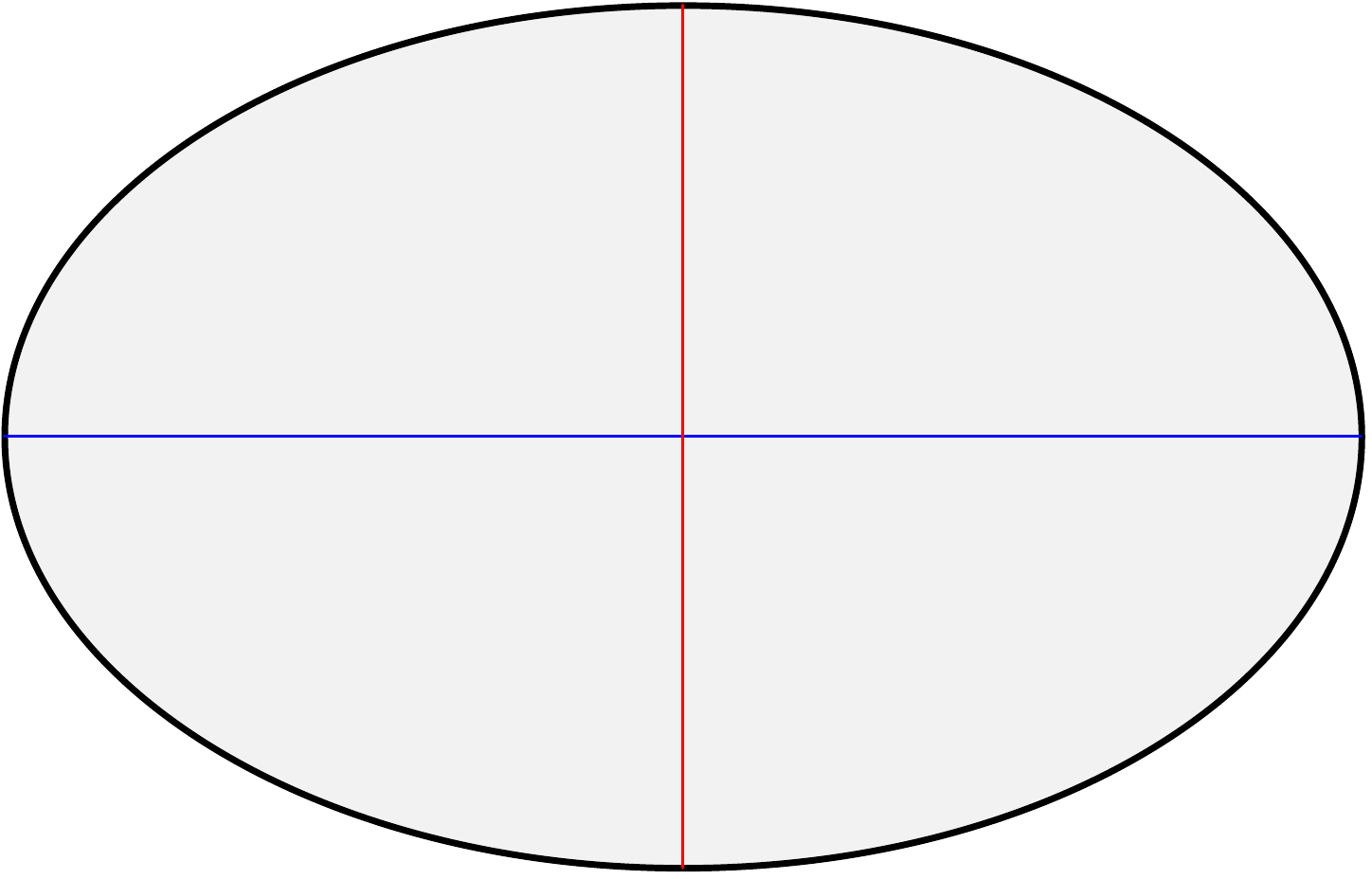}
        \caption{Before division}
        \label{fig:cell_split_before}
    \end{subfigure}
    ~
    \begin{subfigure}{0.4\textwidth}
        \centering
        \includegraphics[width=0.8\textwidth]{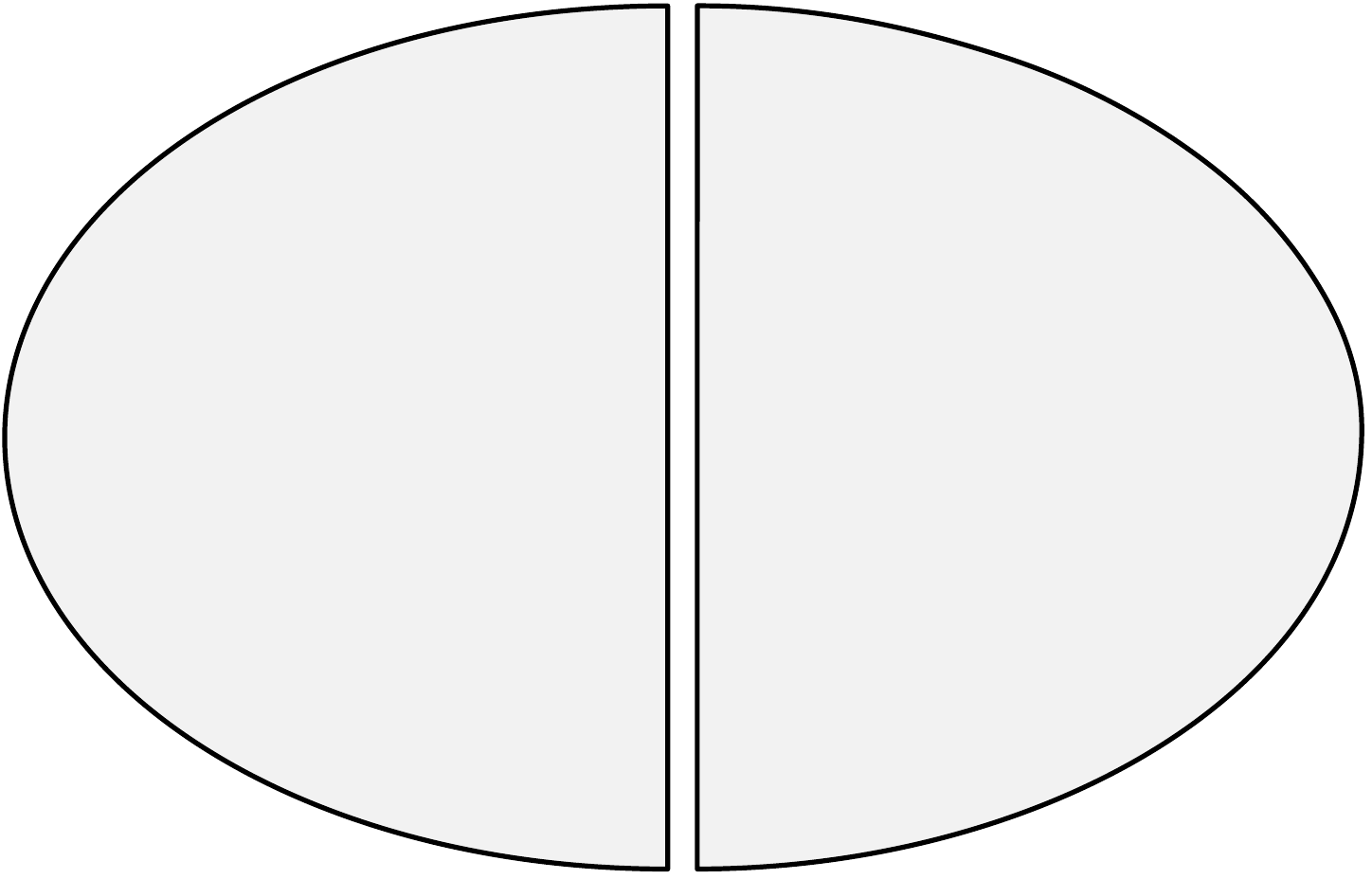}
        \caption{After division}
        \label{fig:cell_split_after}
    \end{subfigure}
    \caption{Simplified cell division, with the major axis represented in blue, and the cutting axis in red. Only lines connecting adjacent vertices are shown here.}
    \label{fig:cell_split}
\end{figure}

When a mother cell reaches a given size (here, $A\geq 2 A_0$), it divides into two daughter cells, according to the mitosis mechanism.
Similarly to \citep{TANAKA_PhD_2016}, this is numerically done by cutting the mother cell in the direction perpendicular to its major axis \citep{Thery_Racine_Pepin_Piel_Chen_Sibarita_Bornens_2005}, as sketched in Fig.~\ref{fig:cell_split_before}. Two new, independent membranes are then created along the cutting axis in order to obtain two distinct daughter cells. Vertices are finally added, to the newly created membrane, to keep the vertices density constant over the perimeter (see Fig.~\ref{fig:cell_split_after}). 

\subsection{Chemical signaling\label{subsec:signaling}}

\rcol{A simple way to move from one cell state to another one is to rely on chemical signaling. This type of signal is usually simulated as a scalar whose evolution is given by a reaction-diffusion equation. Hereafter, this equation is solved thanks to lattice Boltzmann (LB) methods, as proposed by~\citep{TANAKA_Bioinformatics_31_2015,TANAKA_PhD_2016}. 
Contrarily to the cell-based approach dedicated to the evolution of the cell membrane, the LB scheme requires a Cartesian space discretization of the simulation domain to compute the evolution of the chemical signal. In the LB context, solving the reaction-diffusion equation is done through the D2Q5 lattice following instructions proposed, e.g.,} in Sec 5.3.5 of~\citep{KRUGER_Book_2017}. Concretely speaking, the evolution of population $f_k$ ($1\leq k \leq 5$) is given by the \rcol{LB} equation:   
\begin{equation}
    f_k(\bm{x}+\bm{c}_k dt, t + dt) - f_k(\bm{x},t) = -\frac{dt}{\tau_f}\left(f_k(\bm{x},t)-f_k^{eq}(\bm{x},t)\right) + \left(1-\dfrac{dt}{2\tau_f}\right) Q_k(\bm{x},t).
\end{equation}
The first right-hand-side (rhs) term \rcol{is the BGK collision term~\citep{BHATNAGAR_PR_94_1954} that} corresponds to a relaxation towards the equilibrium state $f_k^{eq}(\bm{x},t)=w_k \rho_{LB}(\bm{x},t)$, with $\rho_{LB}$ being the signal density assigned to one voxel. $w_k=(2/6,1/6,1/6,1/6,1/6)$ are the lattice weights, and the lattice velocities read as  
\begin{equation}
    \bm{c}_k = 
\begin{pmatrix}
0 & \phantom{-}1 & \phantom{-}0 & -1 & \phantom{-}0 \\
0 & \phantom{-}0 & \phantom{-}1 & \phantom{-}0 &-1 
\end{pmatrix}.
\end{equation}
The second rhs term is a source term defined as
\begin{equation}
    Q_k(\bm{x},t) = w_k R(\bm{x},t),
\end{equation}
with $R(\bm{x},t)$ being the reaction coefficient. The prefactor $(1-dt/2\tau_f)$ is used to reduce discrete effects as proposed by~\citep{GUO_PRE_65_2002}.

In practice, the signal is produced inside a given cell at a rate $\tau_{p}/dt$, but its evolution is not bounded inside the cell membrane. On the contrary, it can freely propagate in all the simulation domain, even though its amplitude decays according to a predefined decay rate $\tau_{d}/dt$ outside the emitting cell. This decaying mechanism can be seen as a simplification of more realistic signaling processes, for which diffusive compounds are partially bloked by the cell membrane. 
In any case, the reaction term locally modifies the lattice population $f_k$ through the reaction coefficient
\begin{equation}
    R(\bm{x},t) = \begin{cases}
    \tau_{p}/dt& \text{in the cell emitting the signal},\\
    -(\tau_{d}/dt)\rho_{LB}(\bm{x},t) &\text{otherwise}.
    \end{cases}
\end{equation}
Here, $\rho_{LB}$ is computed at each voxel via
\begin{equation}
\rho_{LB}(\bm{x},t) = \sum_{k=1}^5 \left(f_k + \dfrac{dt}{2}Q_k\right)(\bm{x},t).
\end{equation}

\begin{figure}[!ht]
    \centering
    \begin{subfigure}[B]{0.4\textwidth}
        \centering
        \includegraphics[width=0.9\textwidth]{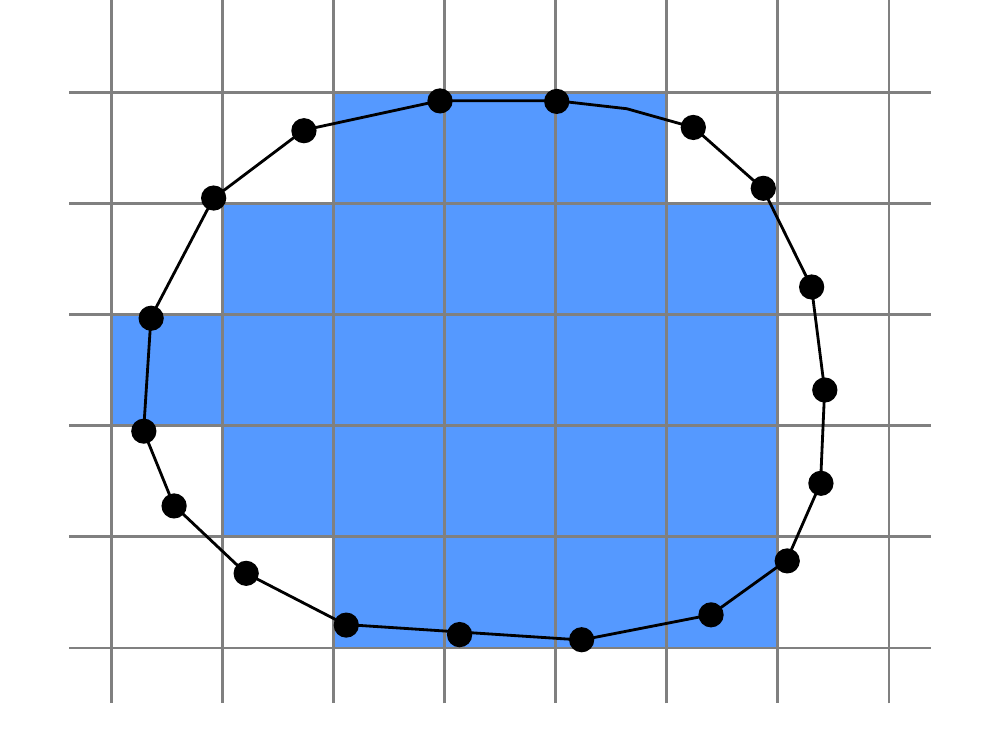}
         \caption{In blue, the LBM lattice elements inside the cell.}
         \label{fig:coupling_det}
    \end{subfigure}
    \begin{subfigure}[B]{0.4\textwidth}
        \centering
        \includegraphics[width=0.99\textwidth]{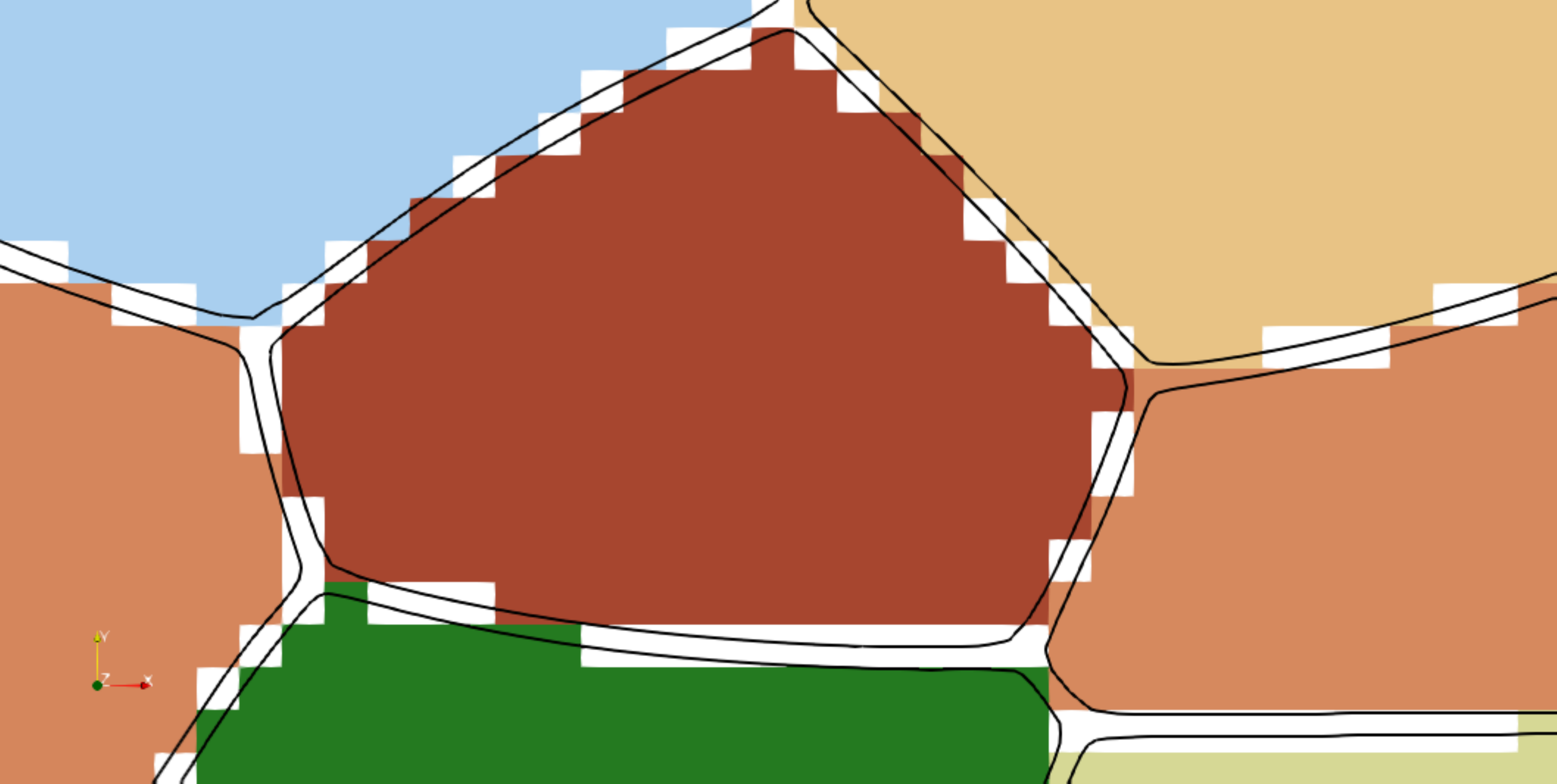}
        \caption{Example of voxelization. Cell membranes are in black, and colored voxels are computed via Palabos library~\citep{LATT_CMA_81_2021}.}
        \label{fig:coupling_voxel}
    \end{subfigure}
    \caption{Illustration of the coupling between the Eulerian lattice grid (LBM) and the Lagrangian representation of cells.}
    \label{fig:coupling}
\end{figure}

To switch from one cell state $s$ to another, the present model requires the signal density inside a cell to be large enough. In other words,
\begin{equation}
s = \begin{cases}
1 &\text{ if } \rho_{s} > \theta_{sig}\\
0 &\text{ otherwise }
\end{cases}
\end{equation}
with $s$ the state of the cell, $\rho_{s}$ the signal density inside the cell, and $\theta_{sig}$ the switching threshold.
The signal density $\rho_s$ is computed by integrating the density provided by the LBM $\rho_{LB}$ over the voxelized area of the cell $A_{LB}$:
\begin{equation}
    \rho_{s} = \frac{1}{A_{LB}}\int_{\bm{x}\in A_{LB}}\rho_{LB}(\bm{x},t)\, \mathrm{d}A_{LB}.
\end{equation}
In our model, we assume that a lattice voxel belongs to $A_{LB}$ if at least half of its area is inside the membrane of the cell (see Fig.~\ref{fig:coupling_det} for an example). In order to account for topology changes of cells during a simulation, we adapted the lattice voxelizer that is available in Palabos library~\citep{LATT_CMA_81_2021}. This is illustraed in Fig.~\ref{fig:coupling_voxel}, where each cell has a flag corresponding to a given color, and white voxels represent interstitial gaps. 

\rcol{Before moving on to cell differentiation, it is worth noting a few shortcomings, as well as possible improvements, of the proposed model for chemical signaling simulation. 
Firstly, the LB scheme can encounter stability issues when simulating steep gradients of chemical signaling with low values of $\tau_f$, i.e., low diffusion coefficient. A common way to tackle that problem is to adapt the numerical scheme by either replacing the BGK operator with a more sophisticated collision model, or by employing flux limiters. 
A detailed discussion about these extensions is out of the scope of this work, and the interested reader may refer to, e.g.,~\citep{COREIXAS_PRE_100_2019} for a review about collision models,~\citep{COREIXAS_RSTA_378_2020} for a stability comparison in the context of fluid flow simulations, and~\citep{HOSSEINI_IJMPC_28_2017} for the use of total variation diminishing schemes to improve the accuracy and stability of LBM for the simulation of advection-diffusion equations with low diffusion coefficients. 
Secondly, while LB schemes are particularly well-suited for high performance computing, they also require more memory storage than standard solvers for reaction-diffusion equations. 
Even if several strategies have been proposed to reduce memory consumption while keeping good performance (see, e.g.,~\citep{LATT_ARXIV_11751_2020}), finite-difference or finite-volume methods might be a better choice when a large number of coupled reaction-diffusion equations are to be solved, as it is the case, e.g., for pattern formation induced by gene expression~\citep{LANDGE_DB_460_2020}. 
Hereafter, a single equation is solved, hence, it is not necessary to rely on more advanced techniques than LBM to simulate chemical signaling.}

\subsection{Cell differentiation}
\begin{figure}[!ht]
    \centering
    \begin{subfigure}{0.3\textwidth}
    \centering
    \includegraphics[width=0.9\textwidth]{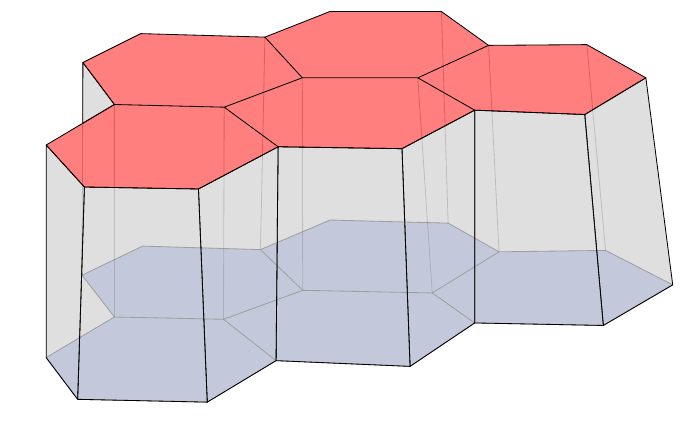}
    \caption{Global 3D view}
    \label{fig:tissue_glob}
    \end{subfigure}
    \begin{subfigure}{0.3\textwidth}
    \centering
    \includegraphics[width=0.9\textwidth]{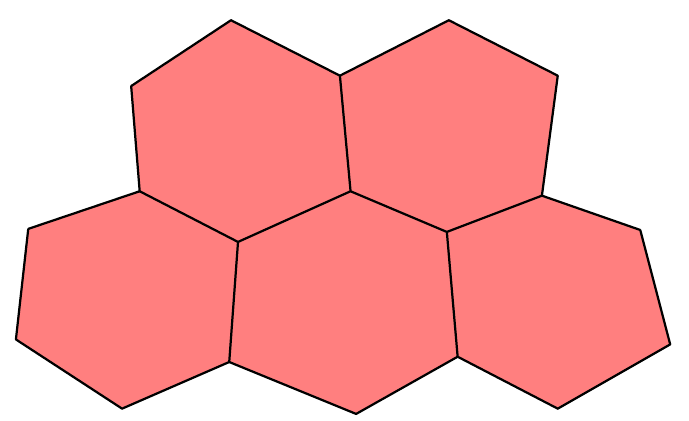}
    \caption{2D apical view}
    \label{fig:tissue_api}
    \end{subfigure}
    \begin{subfigure}{0.3\textwidth}
    \centering
    \includegraphics[width=0.9\textwidth]{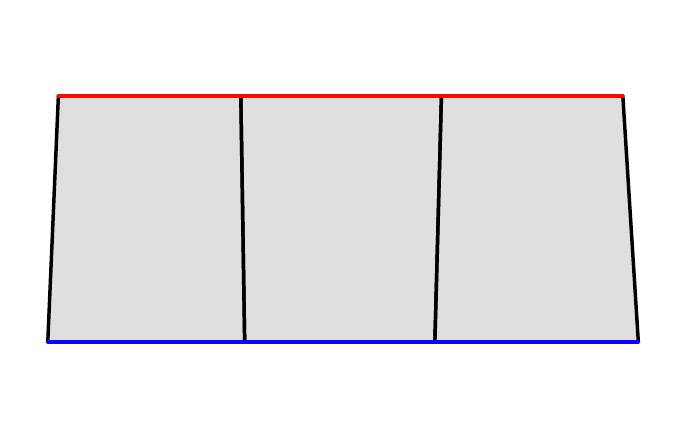}
    \caption{2D lateral view}
    \label{fig:tissue_lat}
    \end{subfigure}
    \caption{2D representations of a simplified 3D tissue. The apical, basal and lateral sides are represented in red, blue and grey respectively.}
    \label{fig:my_label}
\end{figure}

Starting from a 3D monolayer of cells (Fig.~\ref{fig:tissue_glob}), there are several ways to investigate its behavior in 2D. As an example, when simulating cell monolayers from an apical view, all vertices that compose the membrane share the very same properties (Fig.~\ref{fig:tissue_api}). On the contrary, adopting a lateral representation of cells means that membrane properties should be adjusted according to the apical, basal and lateral visco-elastic properties of the cell.

As done in the context of vertex models~\citep{MERZOUKI_PhD_2018,TRUSHKO_DC_54_2020}, this differentiation mechanism is included in our framework by simply tagging vertices (basal, lateral, apical), and fixing their properties (elasticity, bending, etc) depending on their tag. 

Henceforth, the bio-physical properties of the proposed model will be investigated. All parameters, and their value are reported in~\ref{app:param}, and $dt=1$. Eventually, $20$ voxels are used to discretize the initial cell in the case of chemically-driven proliferation (Sec.~\ref{subsec:growth_w_wo_sig}).

\section{Single cell behavior\label{sec:Preliminary}}

\subsection{Convergence study: single cell relaxation \label{subsec:convergence}}

Before investigating the cell properties that naturally emerge from our framework, it is important to identify the minimal number of vertices required to accurately discretize the cell membrane. Doing so, the computational cost of our model can be reduced without decreasing the accuracy of the solver.

By definition, the cell membrane description should be independent of the number of vertices considered for its discretization. Hence, changing the number of vertices should not have any impact on the cell properties. To investigate that point, we consider the relaxation of a single cell, in an unbounded domain, toward its equilibrium state. In that particular case, vertices are only subjected to membrane and internal pressure forces, $\bm{F}_M$ and $\bm{F}_{IP}$ respectively. \rcol{Knowing that our model flows from Newton's law of motion, an analytical formulation for the equilibrium state can be derived by balancing both forces}. In the limit of small angle (high number of vertices), it is reached when the cell radius reads as
\begin{equation}
    r_{eq} =  \rcol{k_b} \frac{\sqrt{1 + \frac{2\eta m}{\pi \rcol{k_b}^2}\left(4\pi K_s + 2 \eta \rho_0 \right)}-1}{4\pi K_s + 2 \eta \rho_0}
    \label{eq:radius}
\end{equation}
with $\rcol{k_b}$ the bending constant, $K_s$ the global spring constant, $\eta$ the pressure sensitivity, $\rho_0$ the target density, and $m$ the cell mass. The full derivation of the cell radius at equilibrium can be found in~\ref{app:appendix}.

\begin{figure}[!btp]
    \centering
    \includegraphics[width=0.5\textwidth]{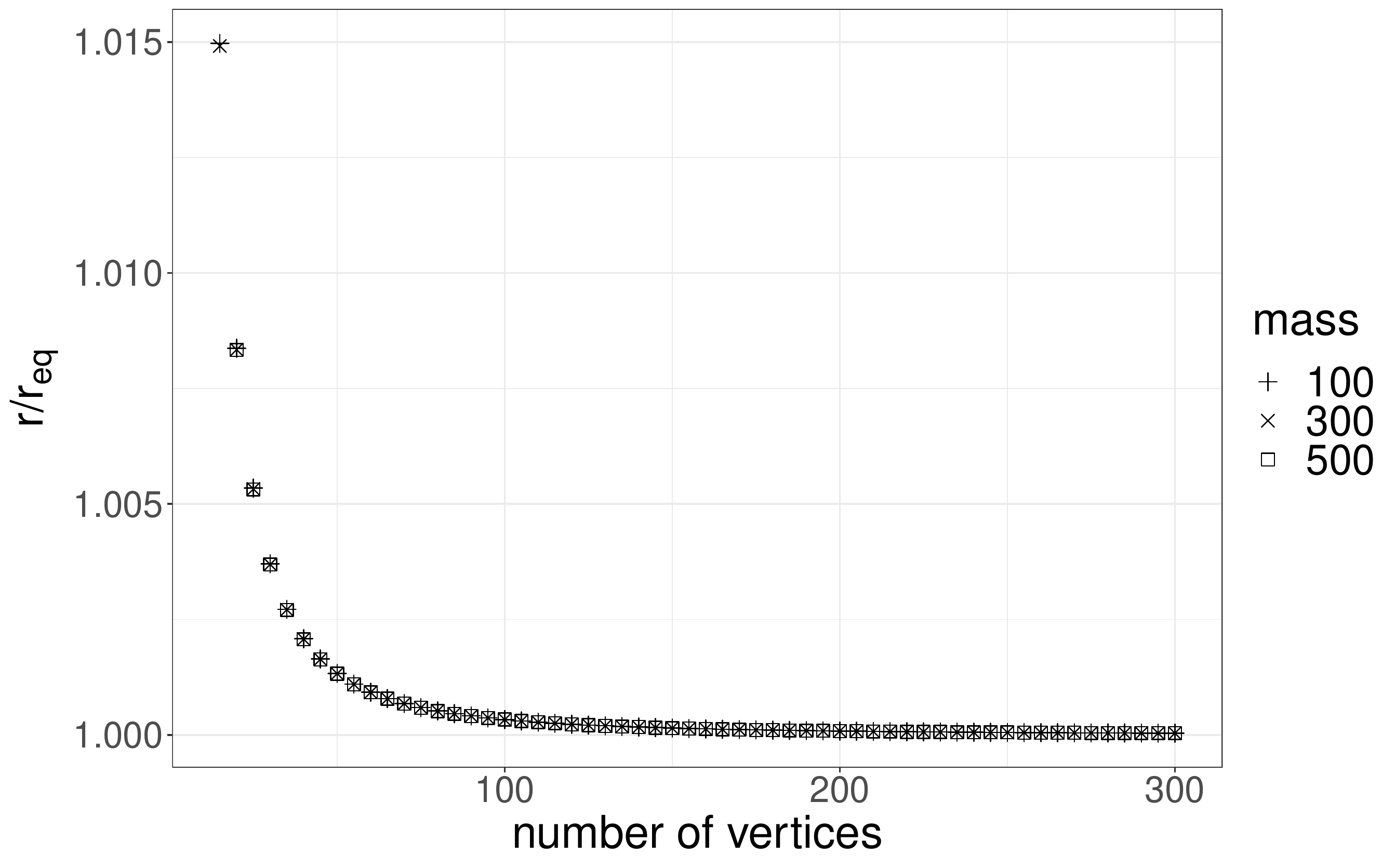}
    \caption{Impact of the number of vertices on the cell radius obtained after relaxation. The equilibrium radius corresponds to Eq.~(\ref{eq:radius}).}
    \label{fig:conv_radius}
\end{figure}

To find the optimal value for $n$, a circular cell is initialized with a radius $r\neq r_{eq}$~(\ref{eq:radius}), and it is let free to relax toward its equilibrium state. 
In that particular case, no mass is added to or removed from the system, thus, the cell can only change its radius to balance bending, spring and internal pressure forces.
It is expected that if $n$ is high enough, then we should end up with $r=r_{eq}$ at the end of the simulation. If not, $n$ must be increased to reduce numerical errors.
Corresponding results are reported in Fig.~\ref{fig:conv_radius} for a number of vertices that varies from 20 to 300, three different values of the cell mass $m$, while keeping other parameters fixed. The latter mass values correspond to three equilibrium radii, and a fortiori, three different cell states (see Fig.~\ref{fig:cell_density}). 

Interestingly, results show a rapid convergence of our model, and it is sufficient to rely on $n\approx 100$ vertices to discretize the cell membrane in an accurate manner. 
\rcol{It is worth noting that similar convergence studies were conducted for test cases shown hereafter, and they confirmed that converged results are achieved for $n=120$. 
Nevertheless, such a number of vertices might be too optimistic in case cells encounter large deformations. In that case, vertices should locally be added/removed, in order to accurately capture modifications of the membrane, by following rules described in Sec.~\ref{subsec:cell_description}.
In this work, deformations remain small, and consequently, it is not necessary to dynamically adjust the number of vertices unless cells are in a proliferating state. Hence, $n=120$ in Secs.~\ref{subsec:cell_compression} and~\ref{subsec:motility}, whereas it is dynamically adjusted for cell proliferation studies in Sec.~\ref{sec:Validation}.}

\subsection{Cell compression\label{subsec:cell_compression}}

Cells can generally be assimilated to incompressible, visco-elastic materials~\citep{JAMALI_PO_5_2010}. 
To test both properties, cell compression is performed for different compression force intensities, and the cell area is monitored over time. 

\begin{figure}[!btp]
    \centering
    \begin{subfigure}{0.19\textwidth}
        \includegraphics[width=\textwidth]{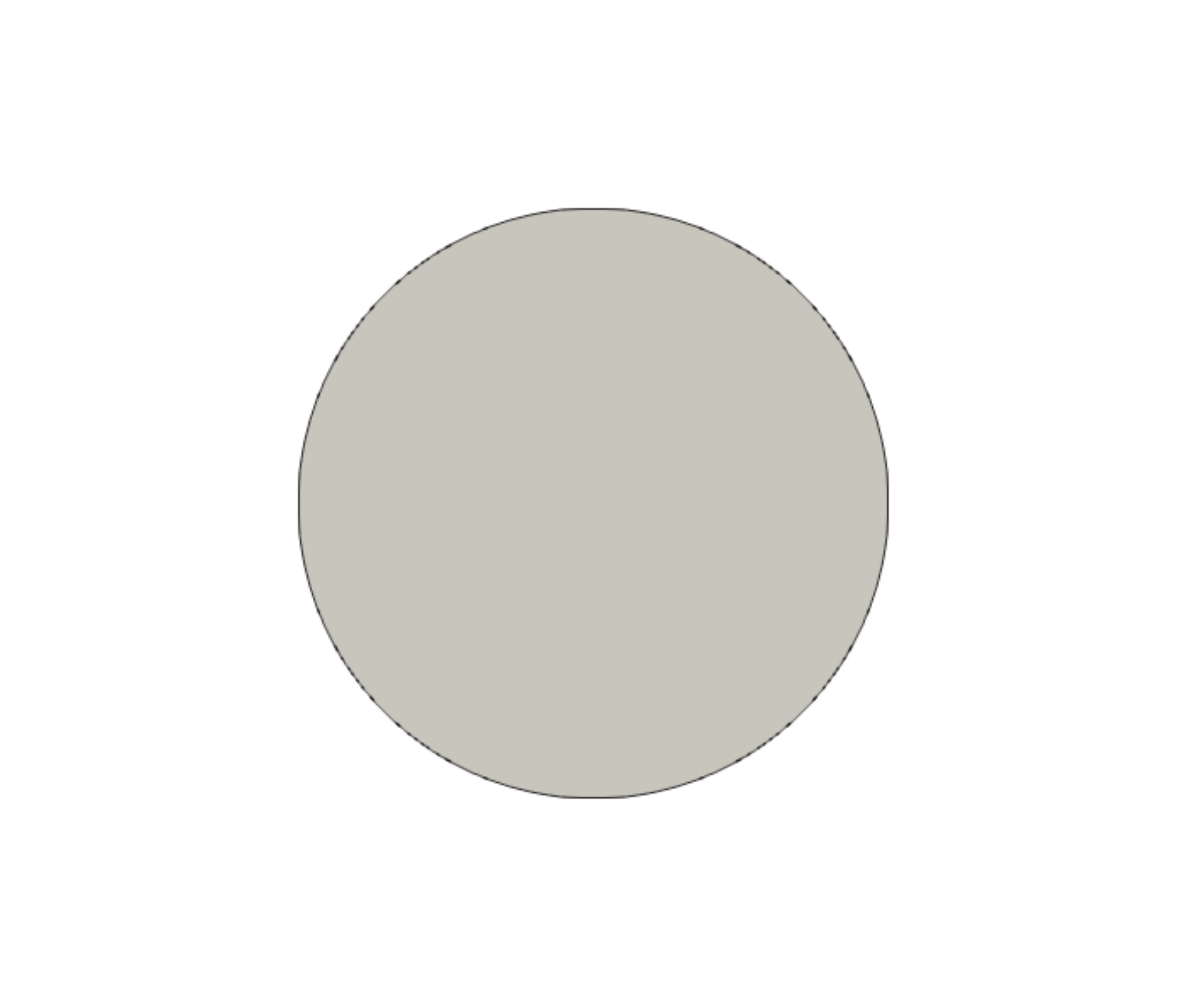}
        \caption{Initial state: t=0}
        \label{fig:compr_init}
    \end{subfigure}
    \begin{subfigure}{0.19\textwidth}
        \includegraphics[width=\textwidth]{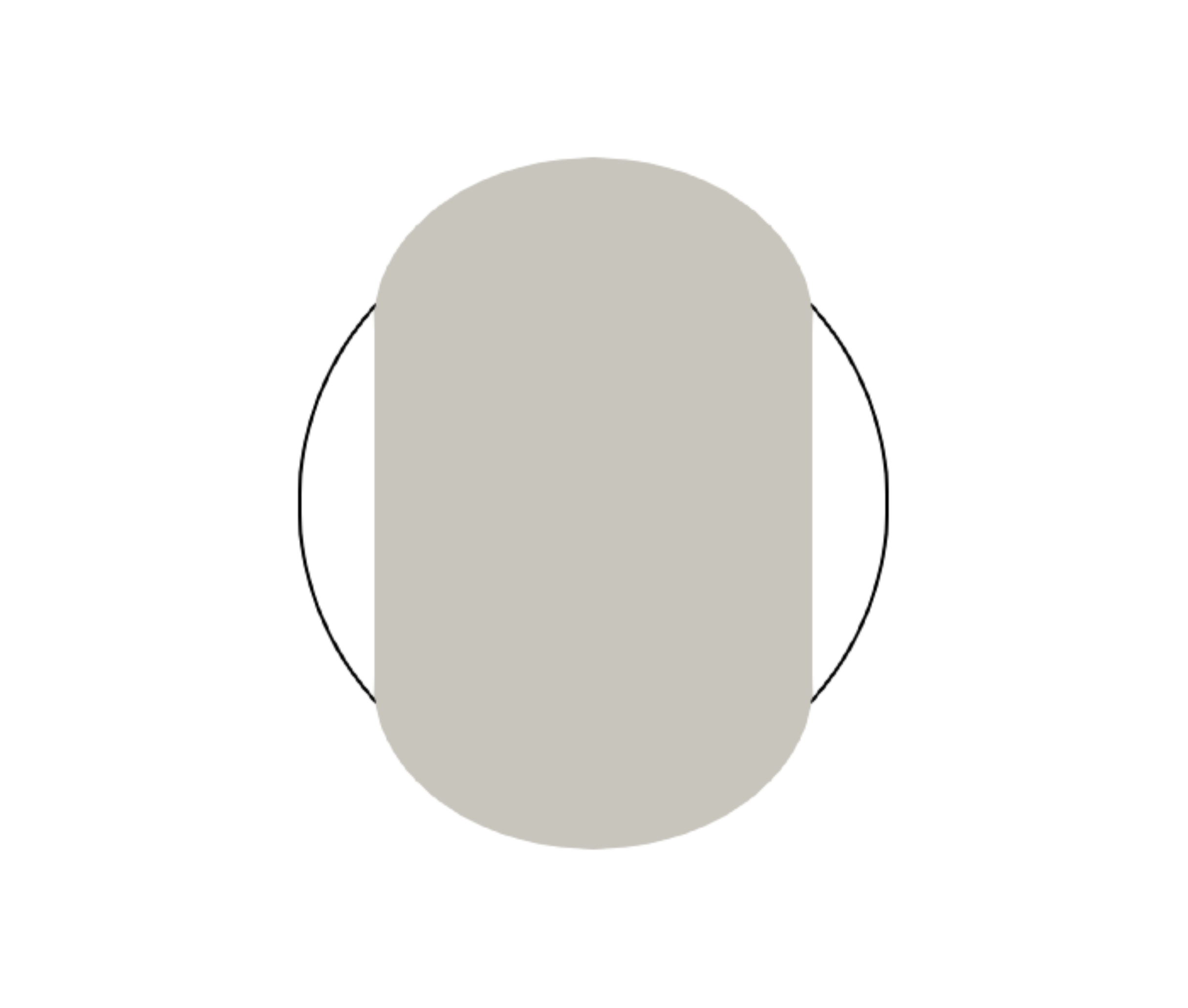}
        \caption{Compression: t=0.05}
        \label{fig:compr_compr}
    \end{subfigure}
    \begin{subfigure}{0.19\textwidth}
        \includegraphics[width=\textwidth]{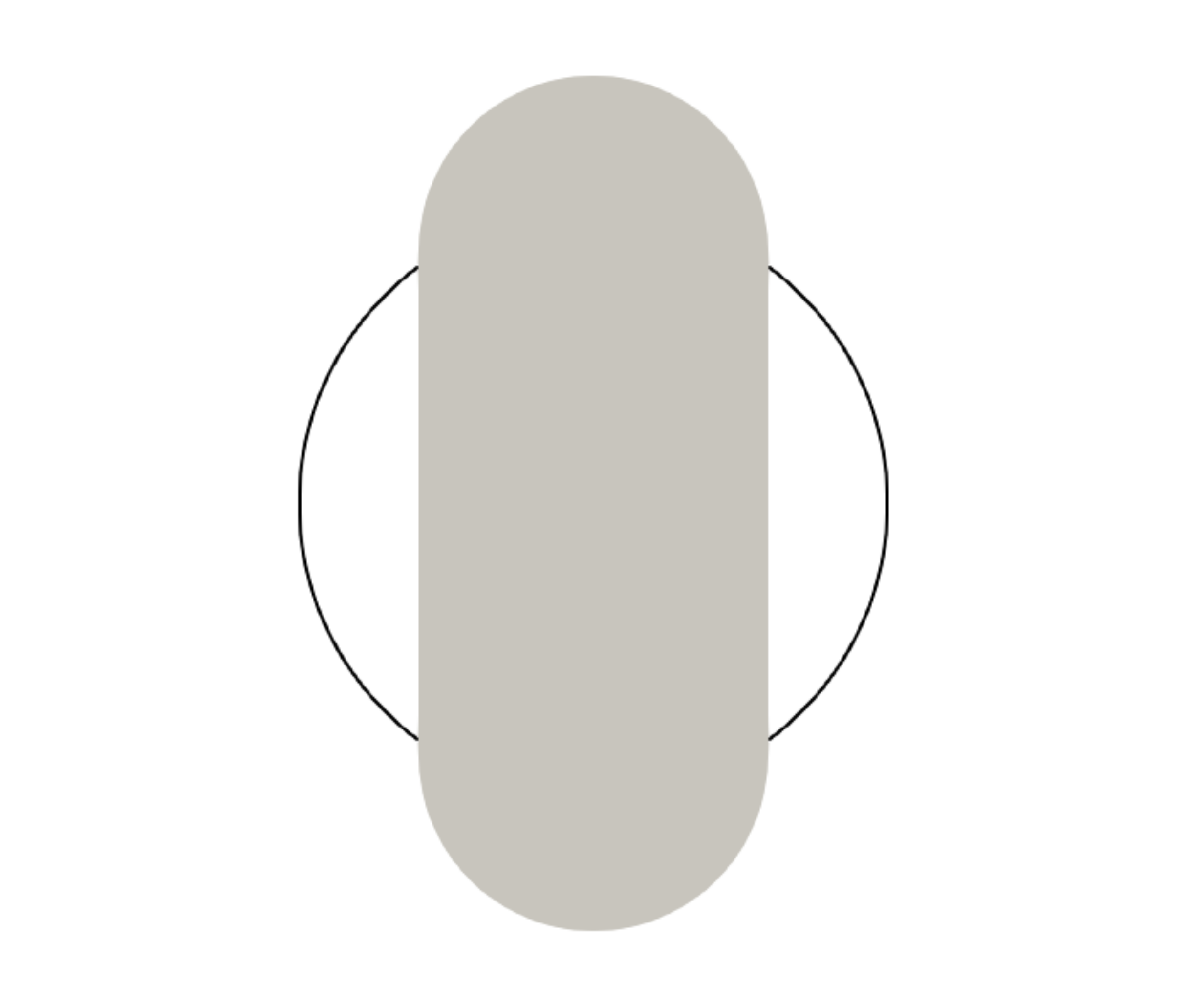}
        \caption{Max compression: t=1}
        \label{fig:compr_max}
    \end{subfigure}
    \begin{subfigure}{0.19\textwidth}
        \includegraphics[width=\textwidth]{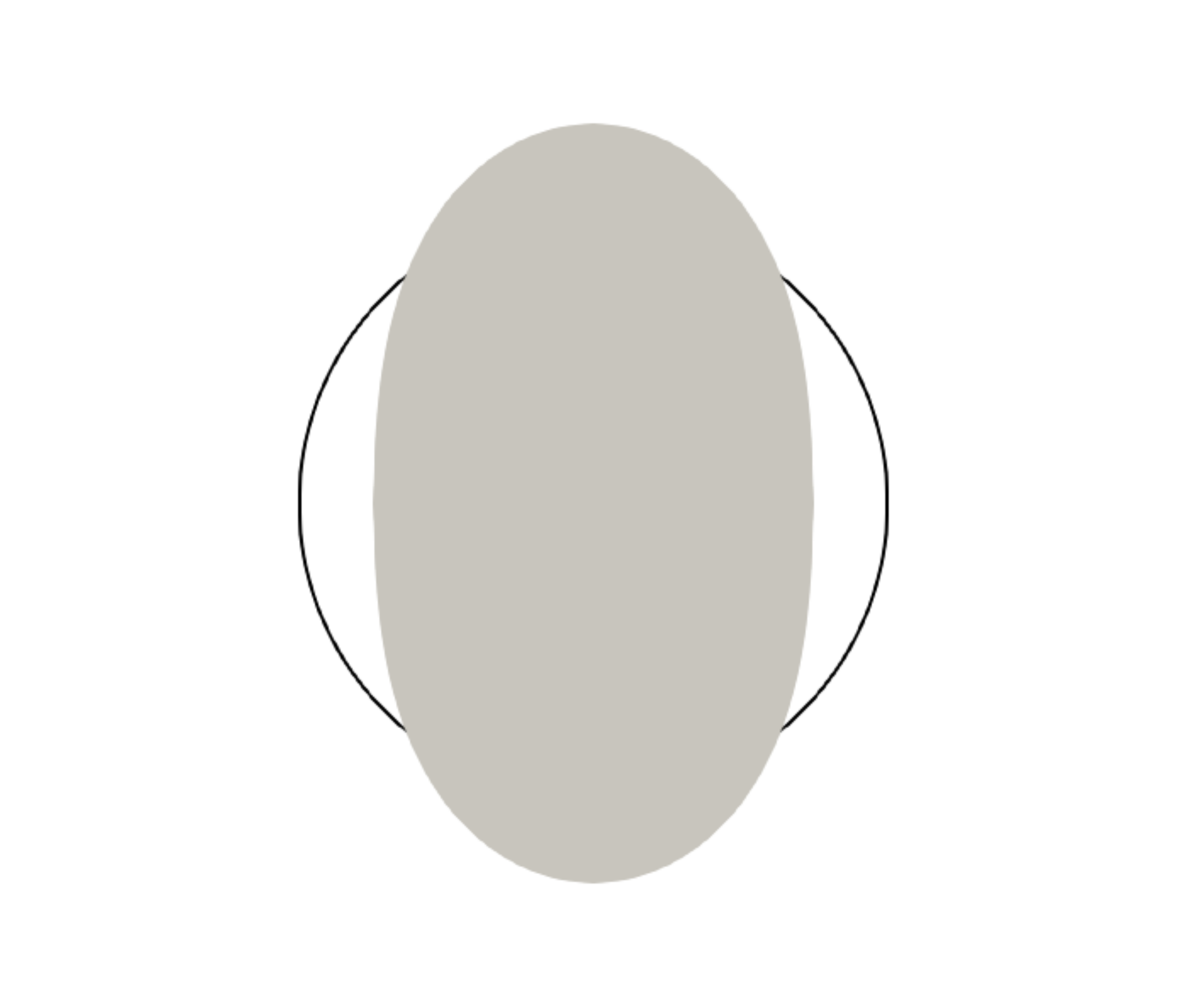}
        \caption{Relaxation: t=1.05}
        \label{fig:compr_relax}
    \end{subfigure}
    \begin{subfigure}{0.19\textwidth}
        \includegraphics[width=\textwidth]{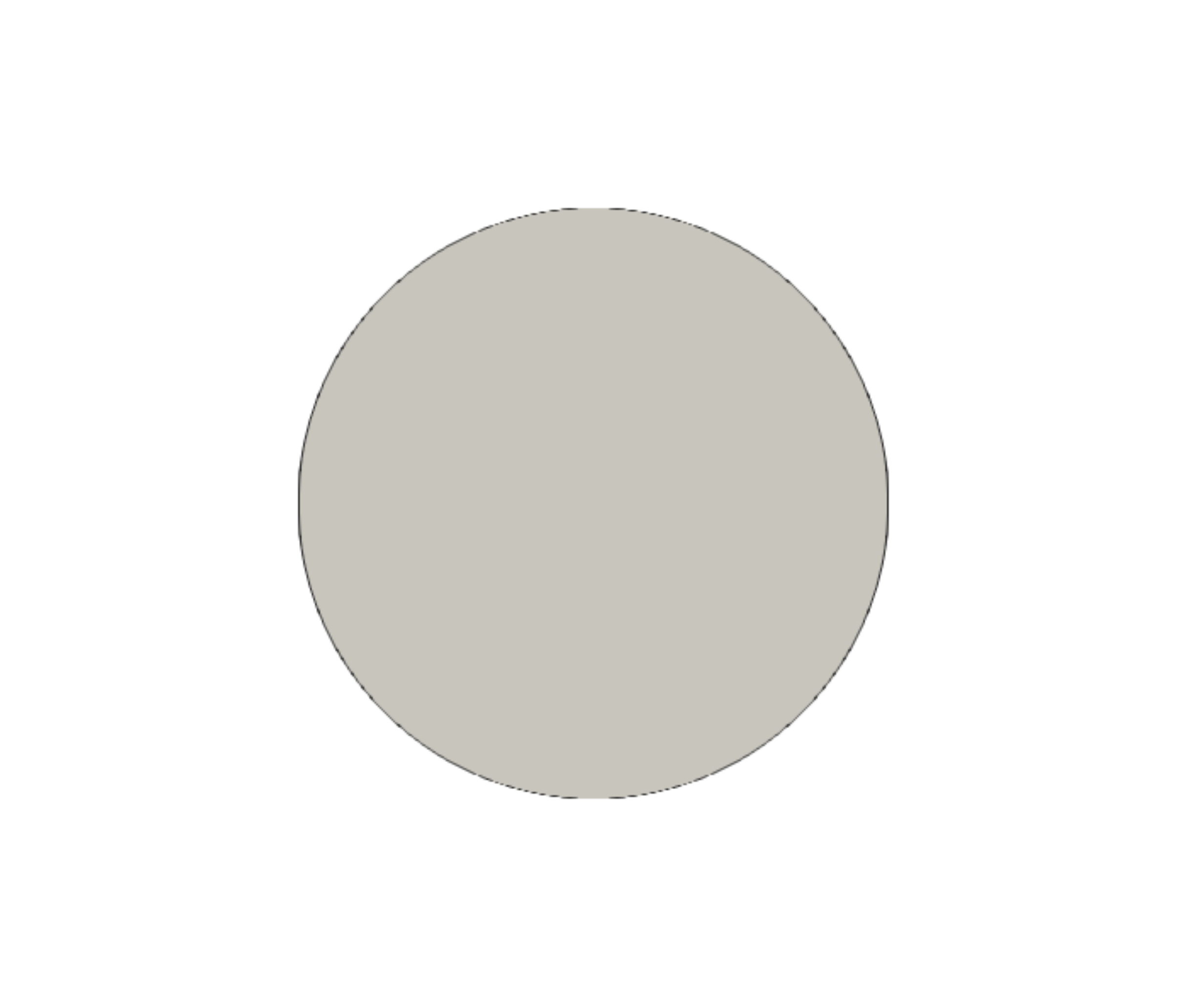}
        \caption{Final relaxation: t=2}
        \label{fig:compr_end}
    \end{subfigure}
    \caption{Time evolution of a cell compression/relaxation, starting from a circular state (black circle). The time scale is normalized by the number of iterations required to reach maximal compression.}
    \label{fig:compr_evo}
\end{figure}

Qualitatively speaking, the simulation is initialised with a cell at equilibrium (Fig.~\ref{fig:compr_init}). The cell is compressed between two walls moving along the horizontal axis (Fig.~\ref{fig:compr_compr}). The wall movement is imposed through a constant force, that is eventually balanced by the cell inner pressure force (Fig.~\ref{fig:compr_max}). The two walls are then removed, and the cell can relax toward is equilibrium state (Figs.~\ref{fig:compr_relax} and~\ref{fig:compr_end}).

\begin{figure}[!btp]
    \centering
    \hspace*{\fill}
    \begin{subfigure}{0.45\textwidth}
      \centering
    \includegraphics[width=\textwidth]{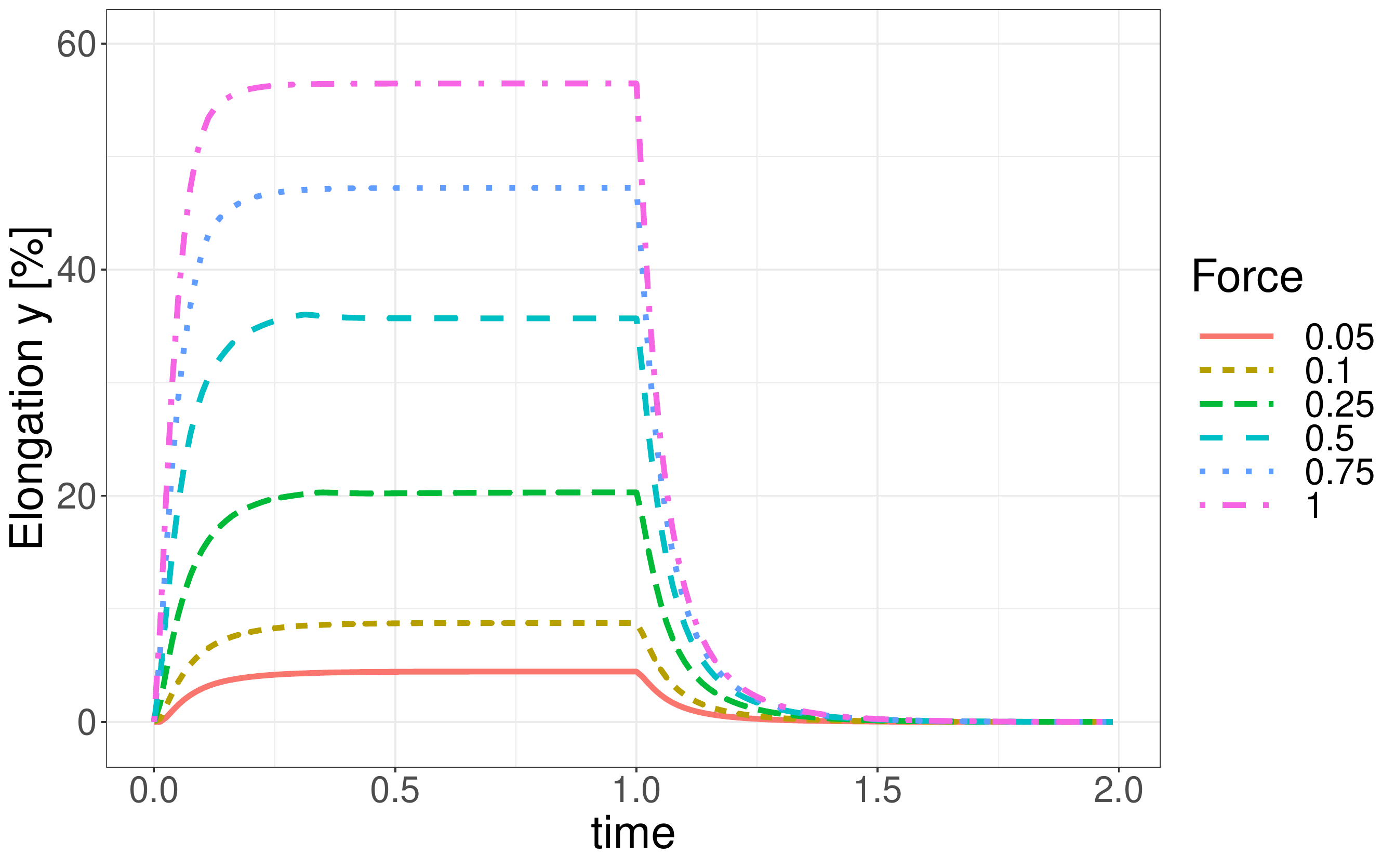}
    \caption{Evolution of the cell deformation along the $y$-axis during the compression/relaxation process.}
    \label{fig:deformation}
    \end{subfigure}
    \hfill
    \begin{subfigure}{0.45\textwidth}
      \centering
        \includegraphics[width=\textwidth]{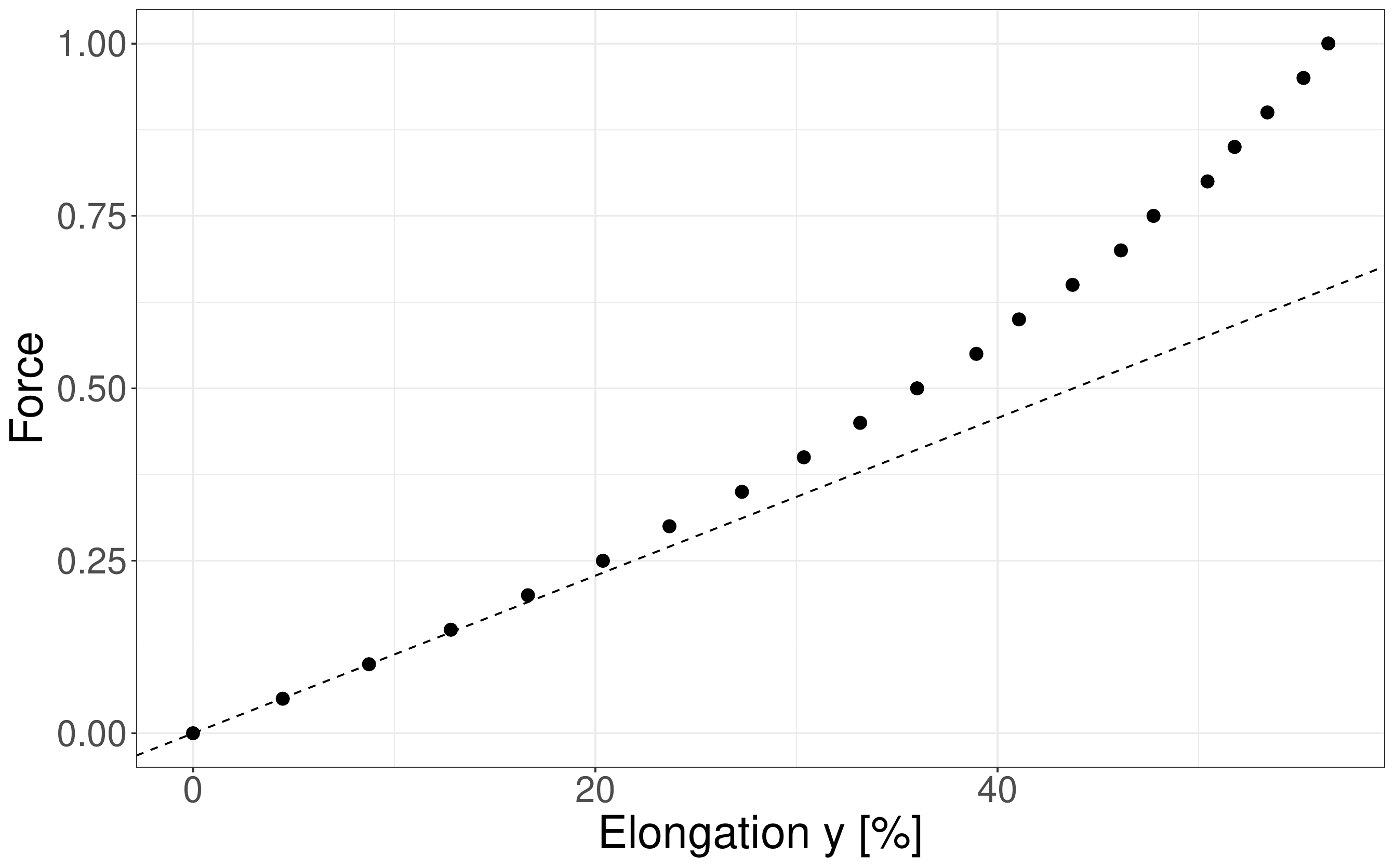}
        \caption{Membrane elongation along $y$-axis with respect to the applied compression force.}
        \label{fig:stressstrain}
    \end{subfigure}
    \hspace*{\fill}
    \caption{Compression of a single cell. Left: Time evolution for different compression force intensities (the time scale is normalized by the number of iterations required to reach maximal compression) Right: Maximal cell elongation evolution with respect to the applied force.}
    \label{fig:cell_compression}
\end{figure}

Looking more in-depth at the cell deformation, it is clear that the compression induces a non-linear behavior of the cell elongation (Fig.~\ref{fig:deformation}). In fact, the latter follows a linear slope in the early compression/relaxation state, which is then rapidly damped, hence confirming the visco-elastic behavior of the simulated cell. In addition, we do recover a linear behavior for low force amplitudes, whereas the cell starts showing some resistance to compression after $~20\%$ of elongation along the $y$-axis (Fig.~\ref{fig:stressstrain}).

It is interesting to note that similar tendencies were obtained in the context of cell monolayer compression/elongation, via both experiments and numerical simulations (based on vertex models), as presented by~\citep{HARRIS_PNAS_41_2012} and~\citep{MERZOUKI_SM_12_2016} respectively.

Moreover, the cell always recovers its initial shape at $t=2$, and during the whole simulation deviation from the initial area remain negligible, i.e., $\vert A(t=0)-A(t)\vert / A(t=0) \leq 10^{-5}$. This means the cell does have an incompressible behavior even if its inner structure (cytoskeleton, cytoplasm and nucleus) is not explicitly accounted for in our formalism.

\subsection{Simplified cell migration in a tissue \label{subsec:motility}}

In the context of vertex models, the simulation of active cell migration requires the inclusion of a mechanism that imposes topological (T1) transitions~\citep{FLETCHER_PBMB_113_2013,MERZOUKI_PhD_2018}.
Hereafter, we show that such a property can be simulated in a simple manner, because interstitial gaps are naturally accounted for in our formalism. While this test case does not reproduce a specific biological system, it is nevertheless the starting point to simulate ordered migration within epithelia~\citep{TRICHAS_PLOSB_10_2012,GAUQUELIN_SM_15_2019}.   

\begin{figure}[!ht]
    \centering
    \begin{subfigure}{0.3\textwidth}
        \centering
        \includegraphics[width=\textwidth]{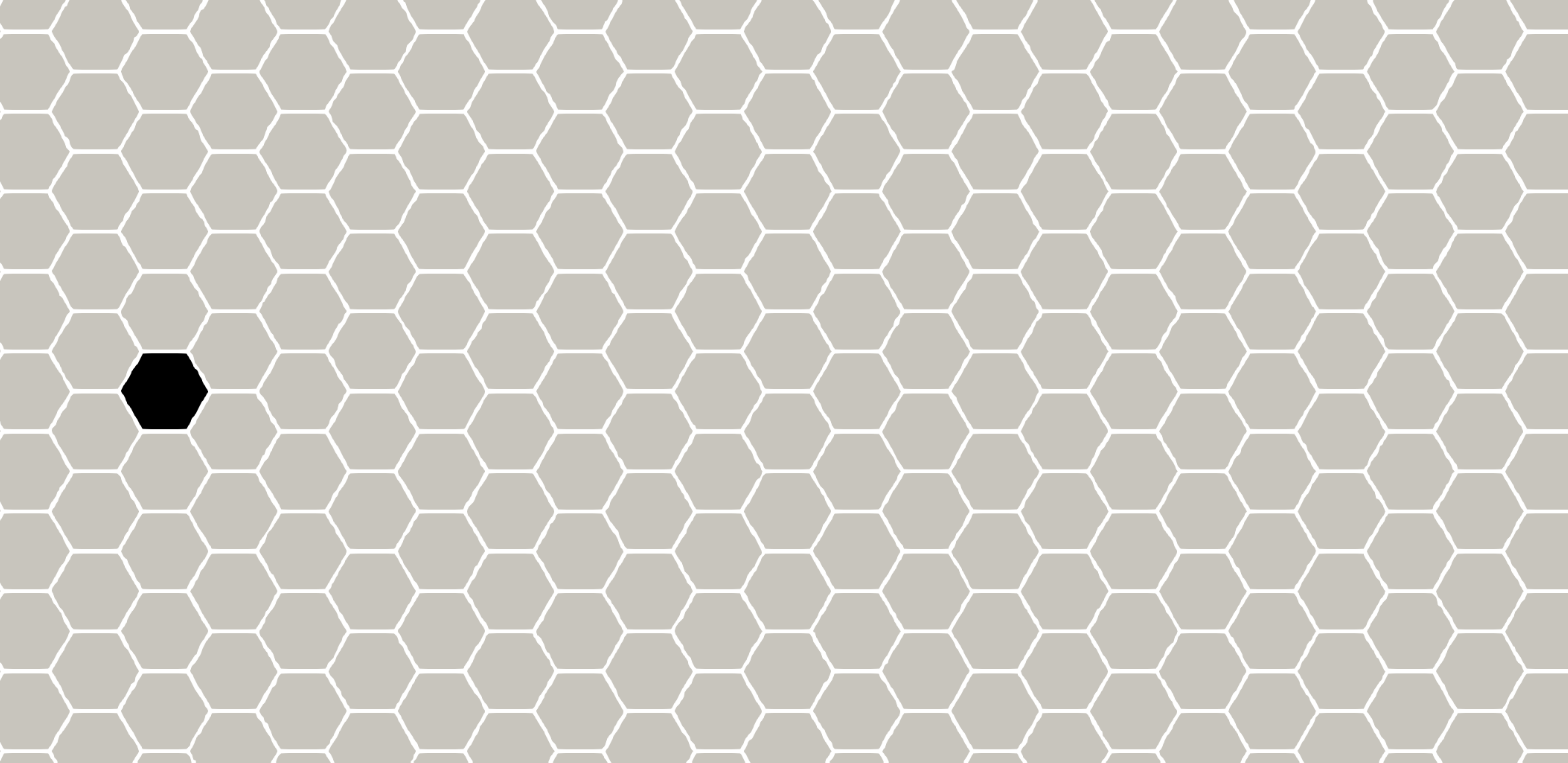}
    \end{subfigure}
    \begin{subfigure}{0.3\textwidth}
        \centering
        \includegraphics[width=\textwidth]{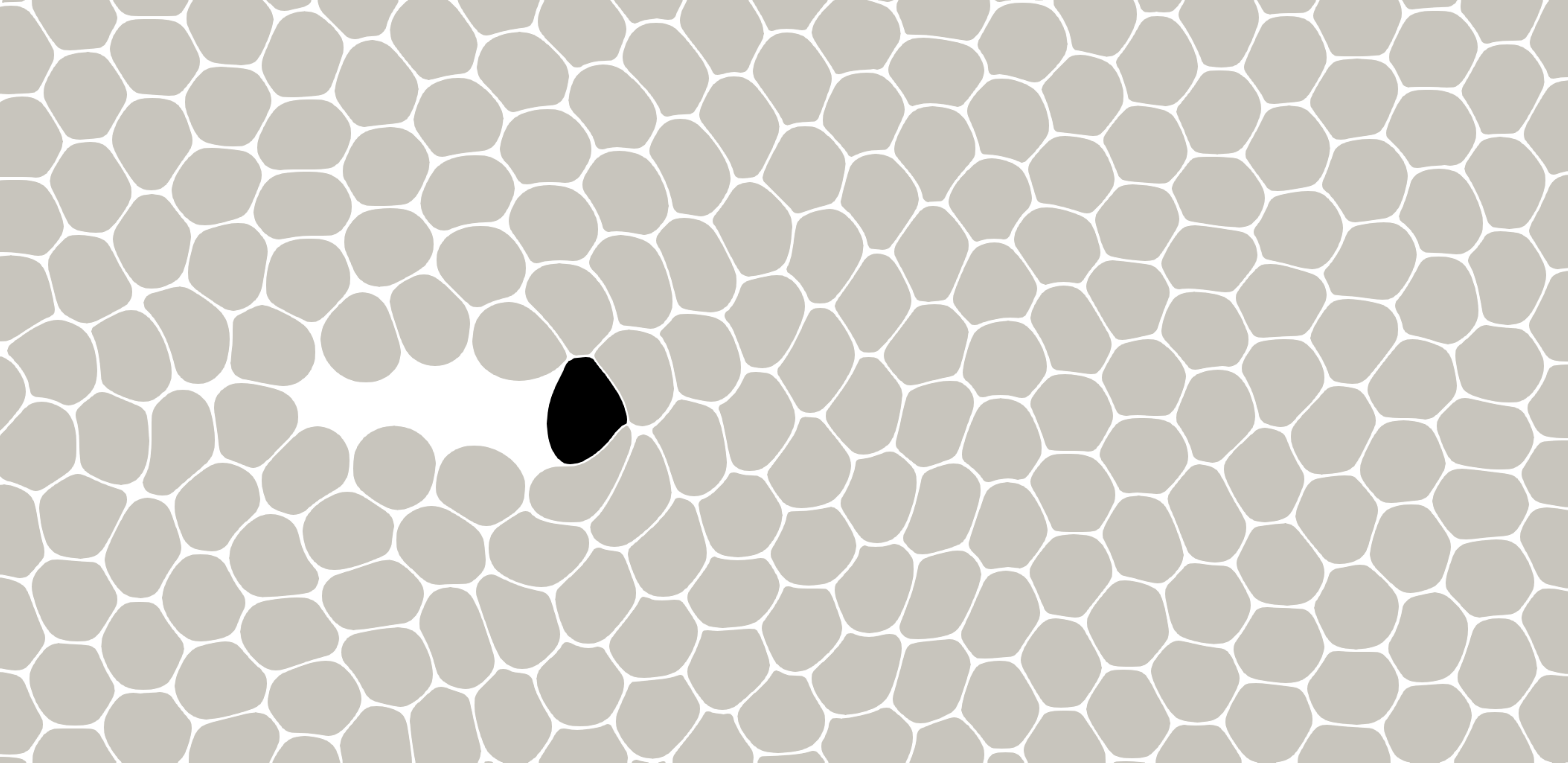}
    \end{subfigure}
    \begin{subfigure}{0.3\textwidth}
        \centering
        \includegraphics[width=\textwidth]{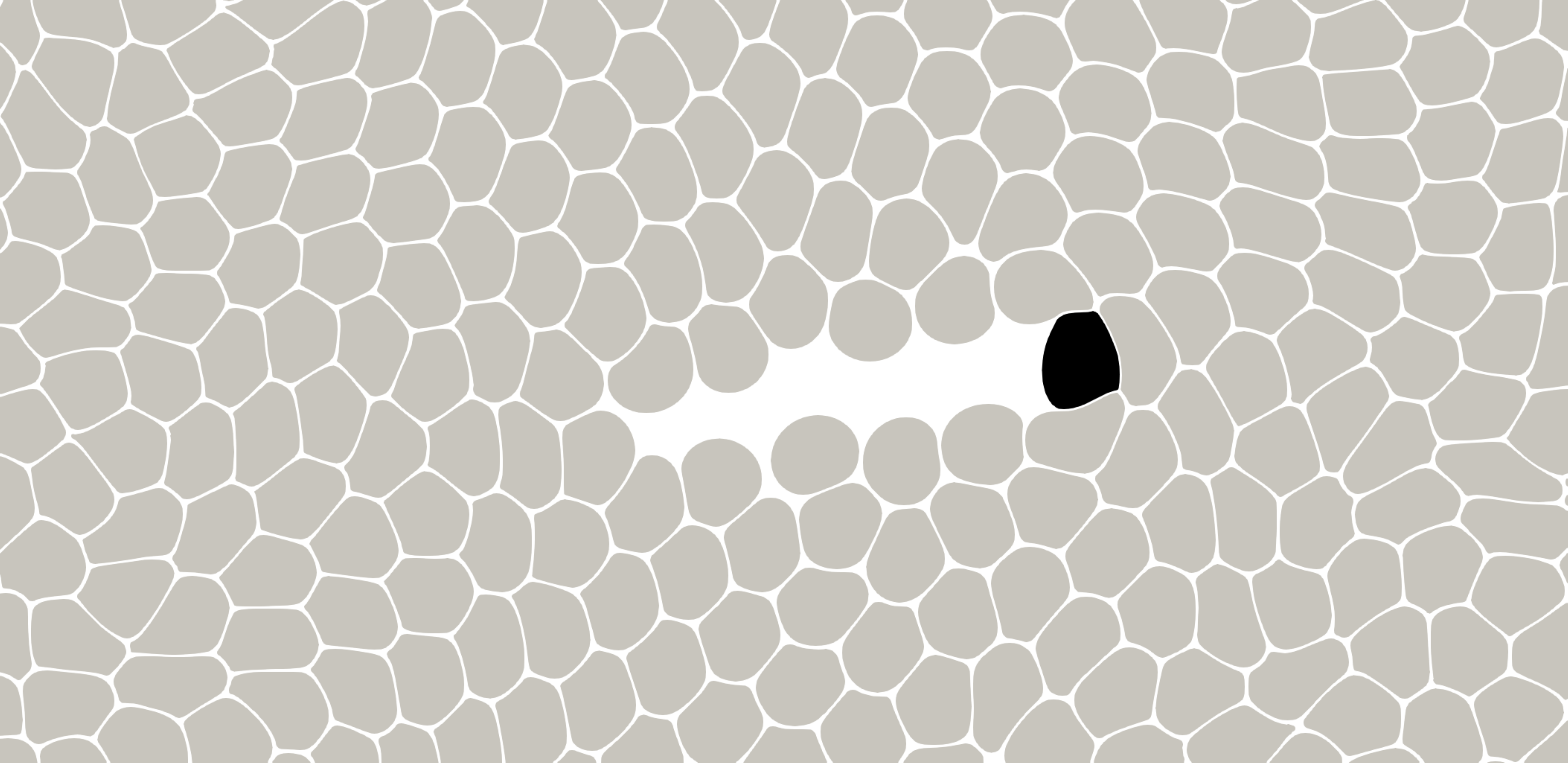}
    \end{subfigure}
    \caption{Cell migration inside an epithelial tissue. By applying an external force to the (black) cell, active migration can be triggered.}
    \label{fig:moving}
\end{figure}

To validate this mechanism, we reproduce the same experiment as in~\citep{FLETCHER_PBMB_113_2013}, starting from a tissue composed of hexagonal cells (Fig.~\ref{fig:moving}). The latter is an idealized description of a cell monolayer, as experimentally observed by~\citep{LEWIS_AR_39_1928}. From this, an external force is applied to a single (black) cell. This force points toward the right, and its amplitude is chosen so that it can balance the pressure exerted by surrounding cells. Interestingly, the active cell is able to migrate toward the right by sneaking in between other cells. While moving, it leaves a small trail in its wake that is rapidly occupied by other cells. In absence of cell proliferation, the behavior of surrounding cells is purely mechanical, and it aims at decreasing internal constraints of the tissue. 

This methodology is currently being extended to a full tissue of cells, in order to investigate how cell motility impacts the proliferation of epithelial monolayers in a confined environment, as experimentally investigated by~\citep{GAUQUELIN_SM_15_2019}. Corresponding results will be presented in a future work.

\section{Cell tissue behavior\label{sec:Validation}}

\subsection{Free and chemically controlled proliferation \label{subsec:growth_w_wo_sig}}

A common validation test consists in letting cells proliferate in an unbounded domain, while monitoring their number of neighbors. The latter quantity is well documented, notably, thanks to measurements from biological experiments \citep{LEWIS_AR_39_1928} as well as numerical approaches like the vertex model \citep{FARHADIFAR_CB_17_2007,MERZOUKI_SM_12_2016}. Two particular cases will be investigated, namely, free and chemically controlled proliferation. The latter requires the simulation of a chemical signal, as discussed in Section~\ref{subsec:signaling}.

\begin{figure}[!ht]
    \centering
    \begin{subfigure}{0.3\textwidth}
    \centering
        \includegraphics[width=\textwidth]{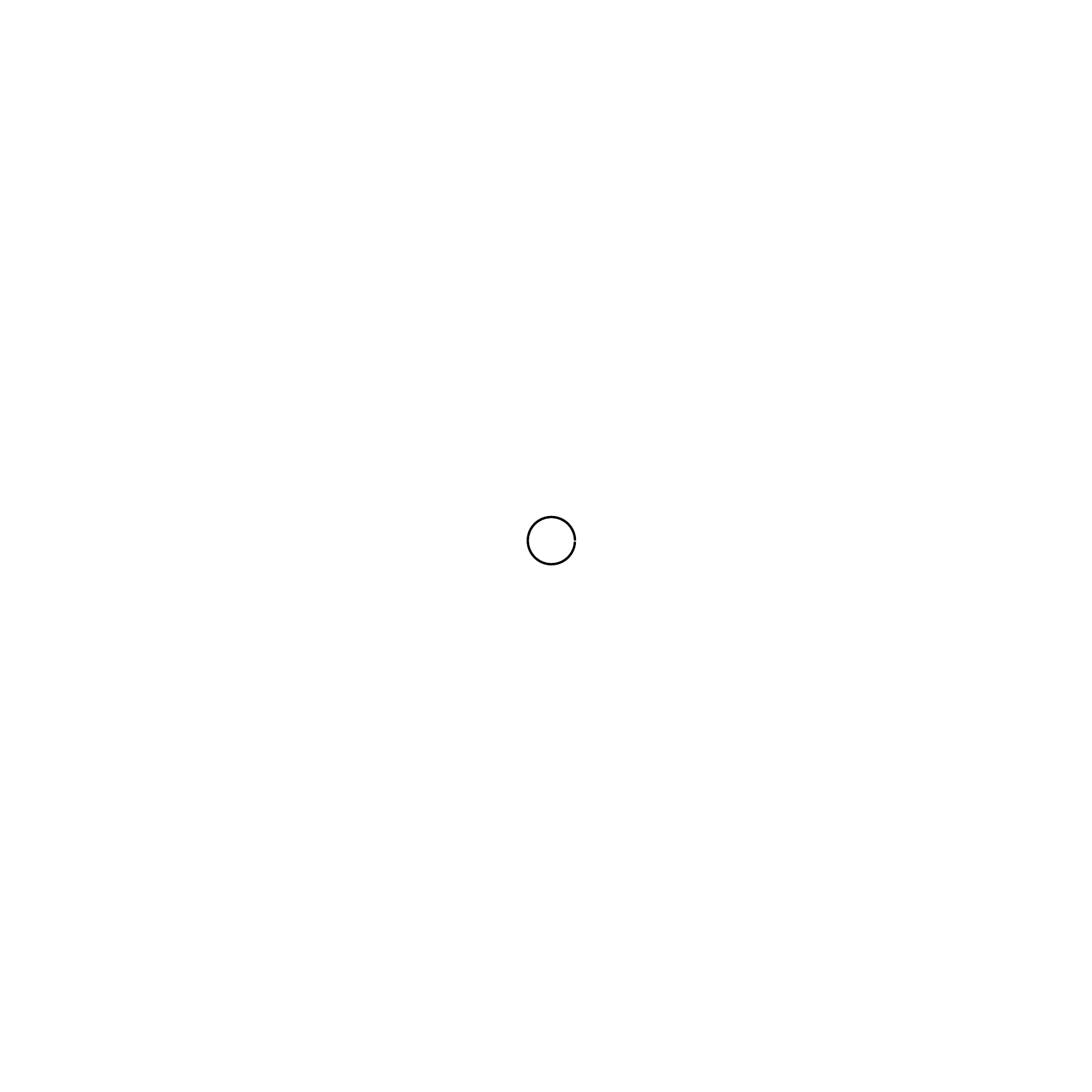}
        \caption{Initial state}
        \label{fig:tiss_prolif_start}
    \end{subfigure}
    \begin{subfigure}{0.3\textwidth}
    \centering
        \includegraphics[width=\textwidth]{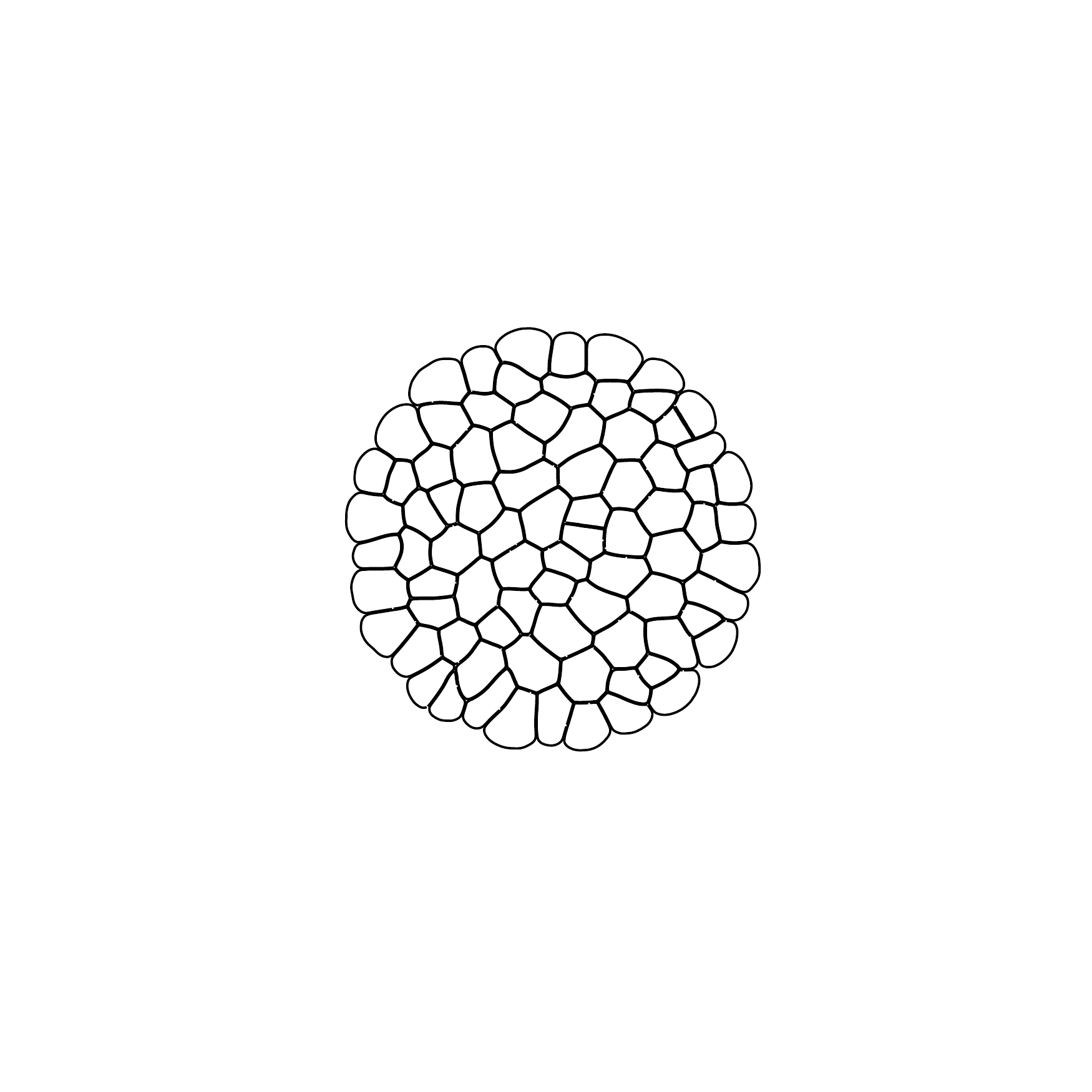}
        \caption{Intermediate state}
        \label{fig:tiss_prolif_mid}
    \end{subfigure}
    \begin{subfigure}{0.3\textwidth}
    \centering
        \includegraphics[width=\textwidth]{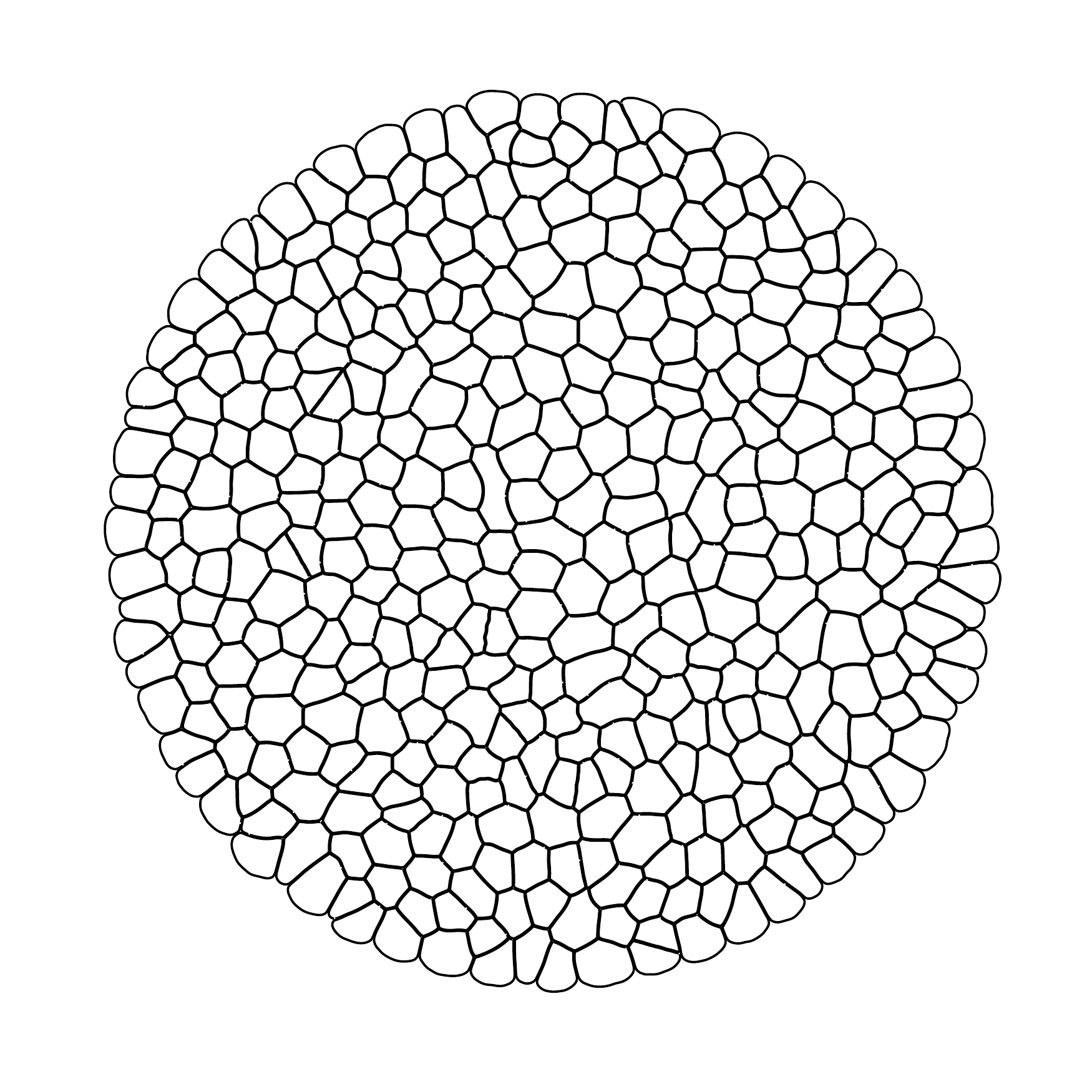}
        \caption{End of the simulation with $\sim 400$ cells}
        \label{fig:tiss_prolif_end}
    \end{subfigure}
    \caption{Unbounded tissue growth with cells in constant proliferating mode.}
    \label{fig:tiss_prolif}
\end{figure}

Let us start with the setup for the freely proliferating cell tissue. 
It consists of a single (circular) cell that is in a proliferating state, and which is located at the center of the simulation domain (Fig.~\ref{fig:tiss_prolif_start}). This cell is free to grow and divide since no external constraint is applied to it. Each daughter cell is in a proliferating state and continues to give birth to new cells (Fig.~\ref{fig:tiss_prolif_mid}). In fact, only the external pressure exerted by surrounding cells can reduce their growth rate, as explained in Section~\ref{subsec:growth}. As there is no way to break the symmetry of the system, the shape of the final tissue stays almost circular (Fig.~\ref{fig:tiss_prolif_end}).


\begin{figure}[!ht]
    \centering
    \begin{subfigure}{0.3\textwidth}
    \centering
        \includegraphics[width=\textwidth]{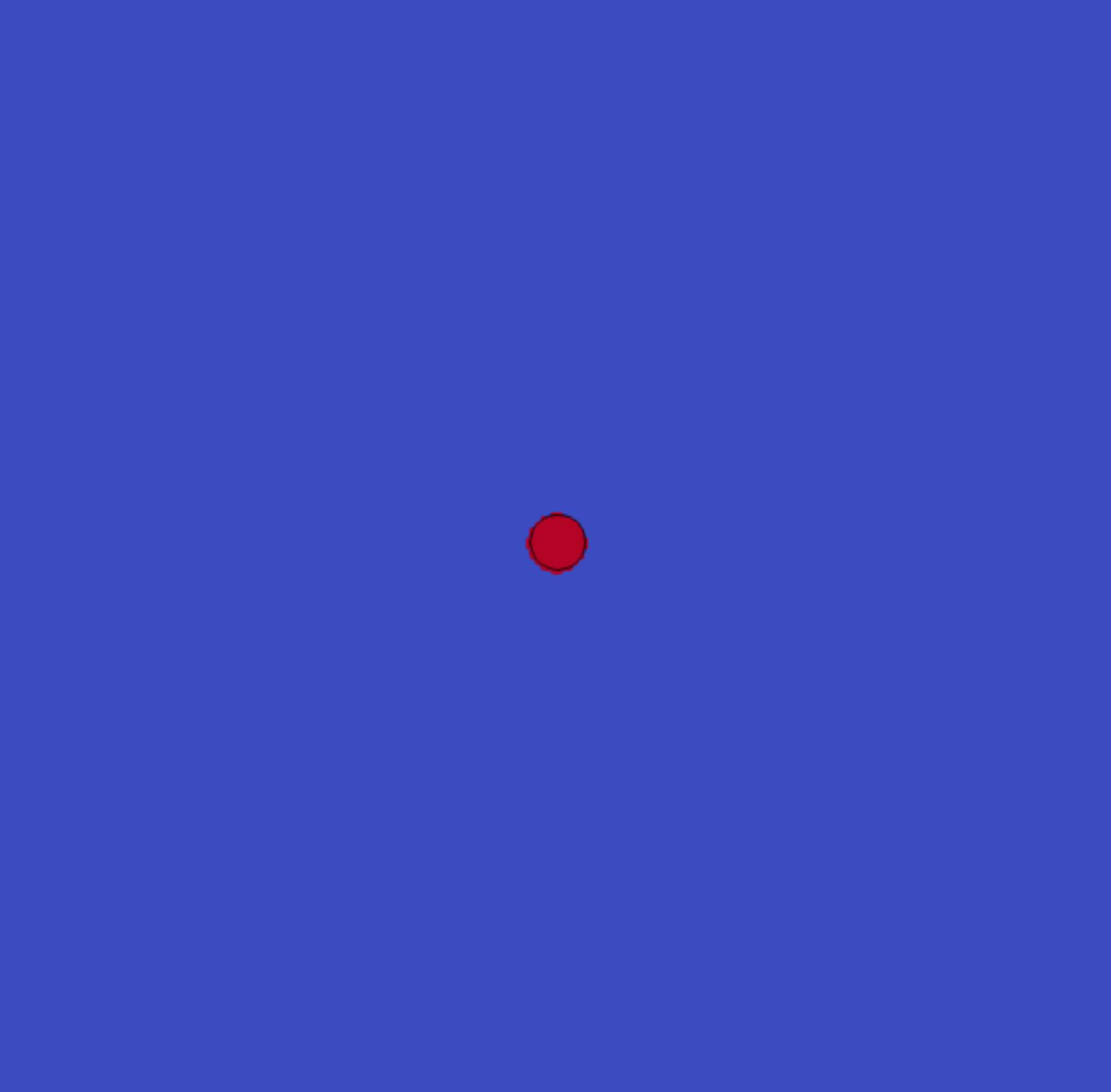}
        \caption{Initial state}
        \label{fig:tiss_prolif_chem_start}
    \end{subfigure}
    \begin{subfigure}{0.3\textwidth}
    \centering
        \includegraphics[width=\textwidth]{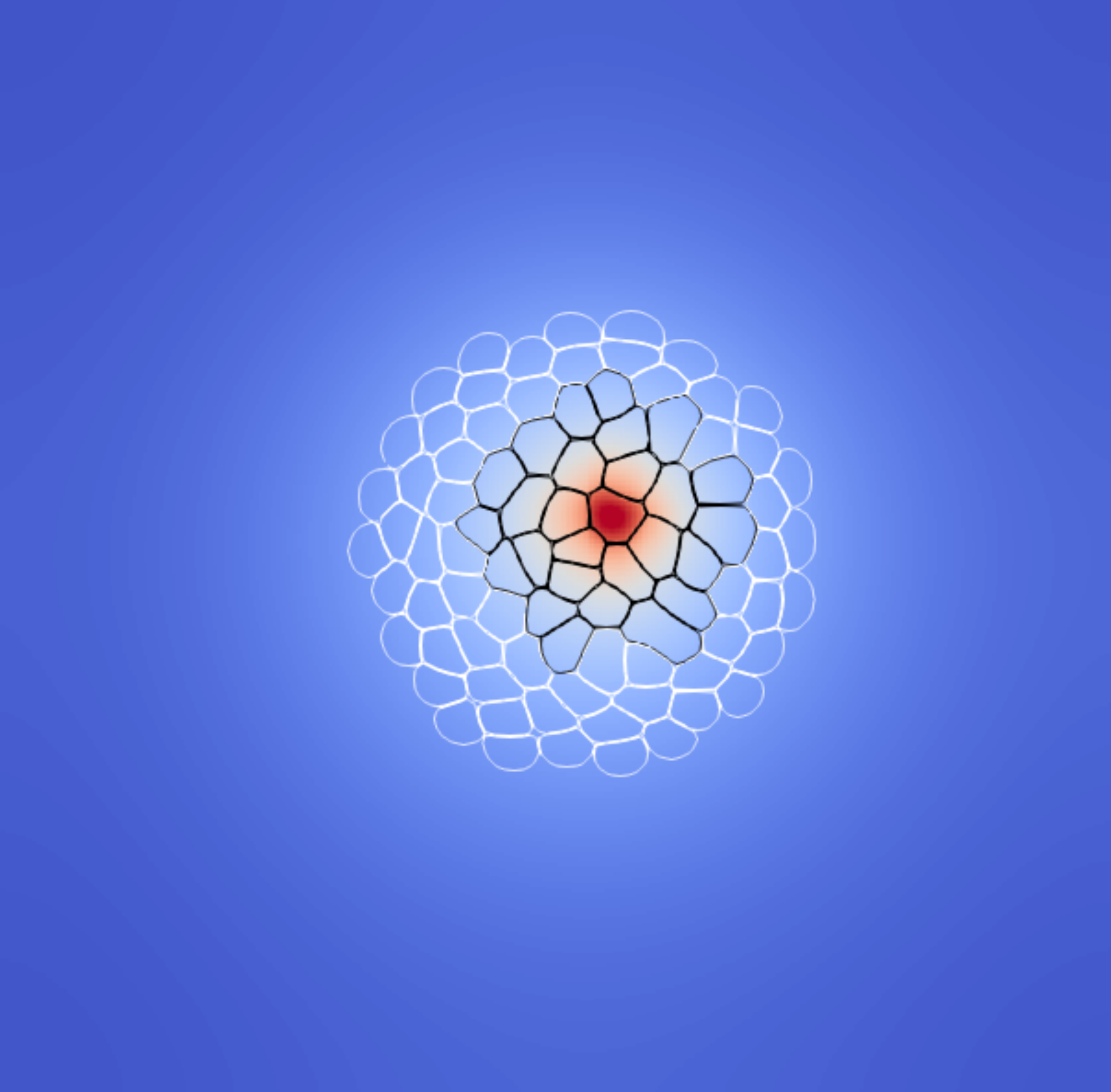}
        \caption{Intermediate state}
        \label{fig:tiss_prolif_chem_mid}
    \end{subfigure}
    \begin{subfigure}{0.3\textwidth}
    \centering
        \includegraphics[width=\textwidth]{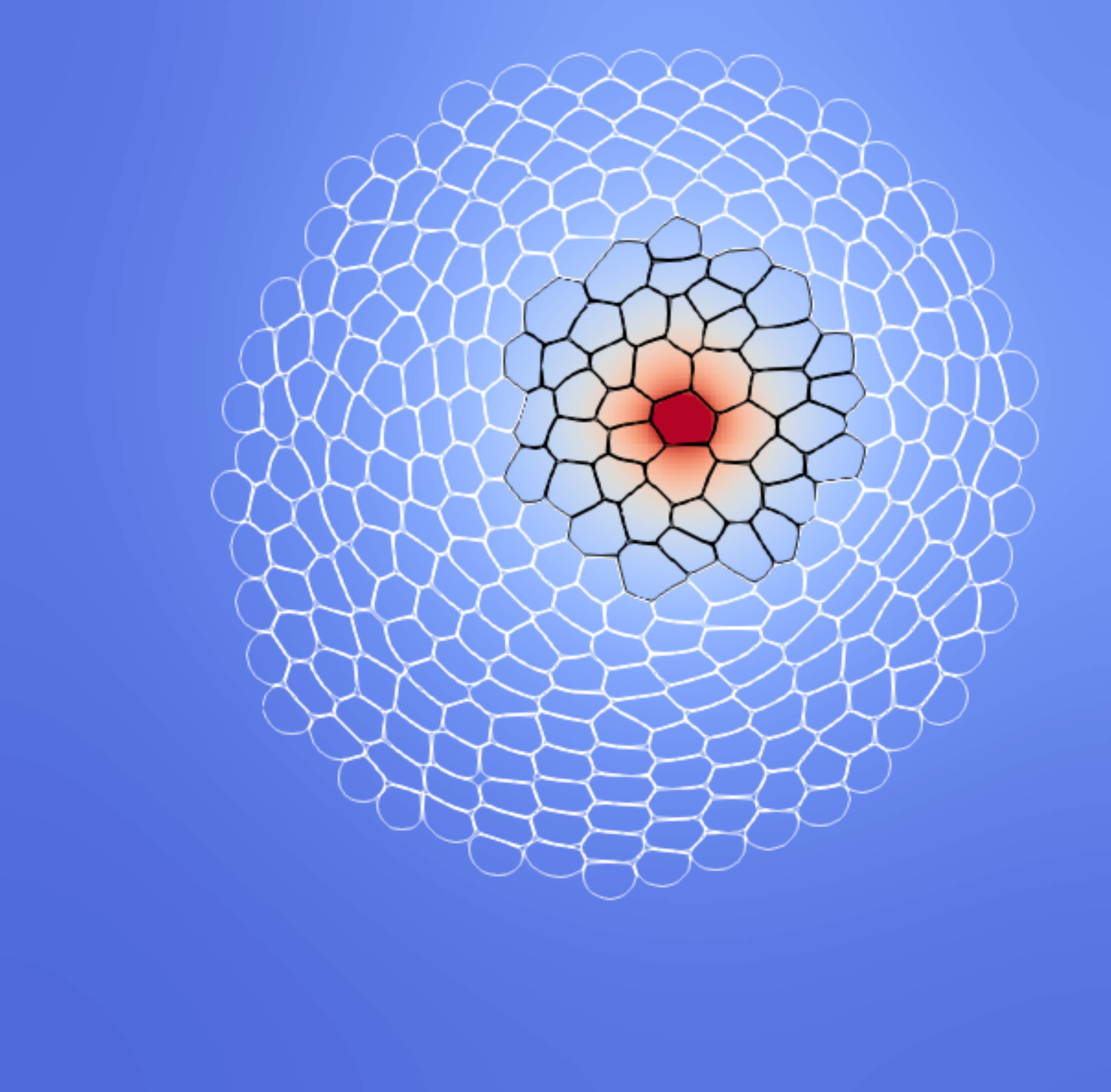}
        \caption{End of the simulation with 300 cells}
        \label{fig:tiss_prolif_chem_end}
    \end{subfigure}
    \caption{Tissue growth with signaling. Cells are either in proliferating (black) or resting state (white). Superimposed to the cell tissue is the density of the chemical signal created by the initial (red) cell.}
    \label{fig:tiss_prolif_chem}
\end{figure}

Continuing with chemically controlled cell proliferation, the setup is similar to the previous one, with the exception that the mother cell is now producing a chemical signal that diffuses over the domain (Fig.~\ref{fig:tiss_prolif_chem_start}). During the whole simulation, this initial cell will be the sole cell to produce a chemical signal that can trigger the proliferating mode of surrounding cells. As explained in Section~\ref{subsec:signaling}, if the signal density inside a cell is above the switching threshold, then the cell can proliferate, otherwise the cell is considered to be at rest.
As time is advancing, the cell producing the signal moves, and its motion depends on (1) the splitting axis used for mitosis, and (2) forces exerted by surrounding cells (Fig.~\ref{fig:tiss_prolif_chem_mid}). As a consequence, the tissue is not guaranteed to have a circular shape anymore (Fig.~\ref{fig:tiss_prolif_chem_end}).

Experiments conducted by~\citep{LEWIS_AR_39_1928} show that, in average, each cell have six neighboring cells, which corresponds to cells with a hexagonal shape. This is not surprising as cells inside a tissue tend to fill all the gaps, eventually leading to a \textit{regular} tiling of the Euclidean plane. Interestingly, simulations show very similar results despite the simplicity of our model (see Table~\ref{tab:comp}). In addition, the chemical signaling only have little impact on the averaged shape of cells, as already reported for immersed-boundary-based methods~\citep{TANAKA_PhD_2016}.
\begin{table}[!ht]
\centering
\begin{tabular}{|l|c|}
\hline 
 & mean number of neighboring cell \\
\hline
 Experiment Drosophila~\citep{LEWIS_AR_39_1928} & $5.99$\\
 \rcol{Experiment Allium~\citep{Korn_Spalding_1973}}	& $5.89$\\
 \rcol{Experiment Dryopteris~\citep{Korn_Spalding_1973}}	& $6.03$\\
 \rcol{Experiment Euonymus~\citep{Korn_Spalding_1973}}	& $6.04$\\
 Simulation with chemical signaling & $5.93$\\
 Simulation without signaling & $5.90$ \\ 
\hline 
\end{tabular}
\caption{Cell distributions obtained by several experiments~\citep{LEWIS_AR_39_1928,Korn_Spalding_1973} and our simulations.}
\label{tab:comp}
\end{table}

\begin{figure}[H]
\centering
\includegraphics[trim=0 0.5cm 0 1cm,width=0.9\textwidth]{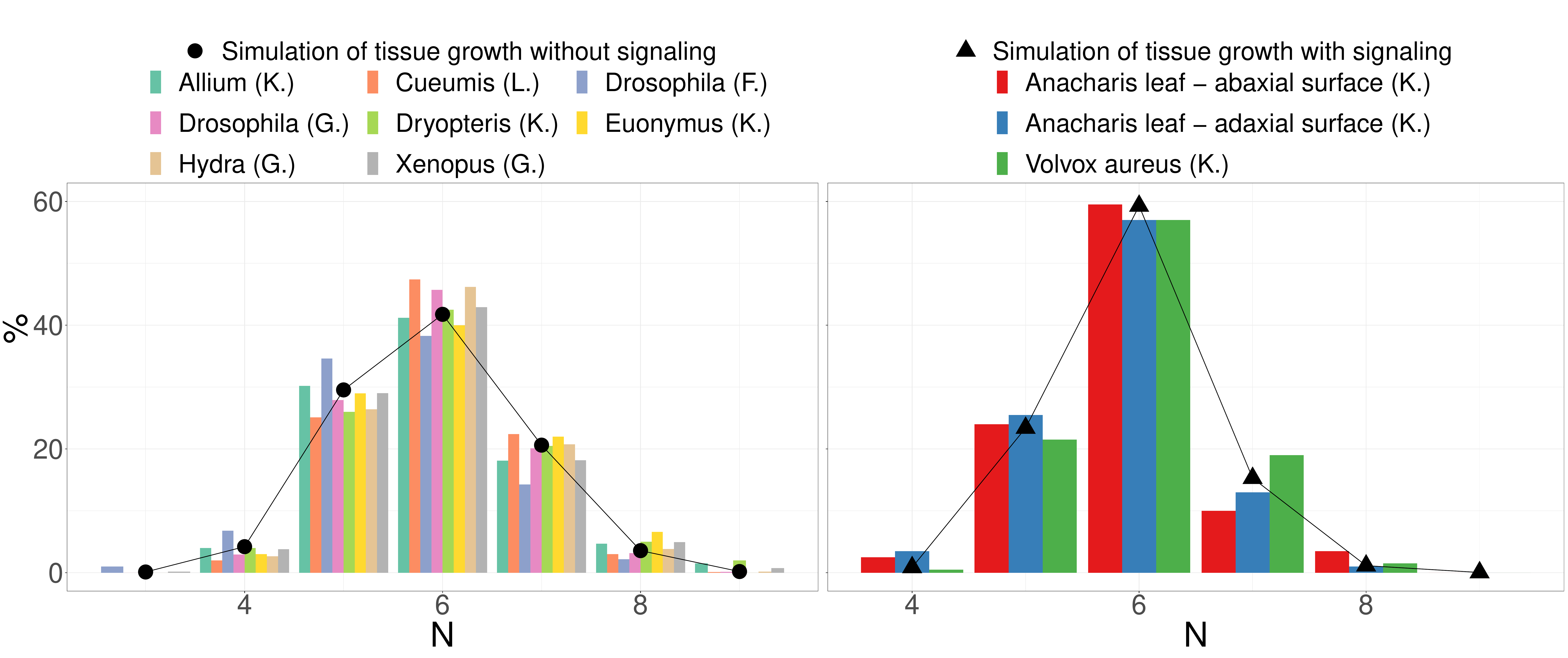}
\caption{Distribution of the number of neighbors $N$, obtained by experiments~\citep{LEWIS_AR_39_1928,Korn_Spalding_1973,Gibson_Patel_Nagpal_Perrimon_2006,FARHADIFAR_CB_17_2007} and our simulations.}
\label{fig:dist}
\end{figure}

Going into more details, experimental results highlight an asymmetric distribution of the number of neighboring cells~\citep{Korn_Spalding_1973,Gibson_Patel_Nagpal_Perrimon_2006,FARHADIFAR_CB_17_2007,Gibson_Gibson_2009}. Indeed, cells are more likely to have five neighbors instead of seven. This is also observed in our simulations with and without chemical signaling (see Fig.~\ref{fig:dist}). \rcol{It is interesting to note that, while the distribution obtained without chemical signaling fits a large set of experimental data, further accounting for chemical signaling only recovers results obtained experimentally for the Anacharis leaf and Volvox green algae~\citep{Korn_Spalding_1973}. One possible explanation might be that, for the latter experiments, an underlying mechanism is biasing the cell growth~\citep{Korn_Spalding_1973}), as it is the case for simulations with chemical signaling. Yet, this hypothesis remains to be confirmed by a more in-depth comparison based on a larger set of experimental data.}

\subsection{Growth under constraint \label{subsec:capsule}}
Instead of using a chemical signaling to drive the state of the cell, it is common to rely on a stochastic process to switch from the resting state to its proliferation counterpart. This probability to switch from one state to the other can be related to the external pressure applied to the cell, similarly to what was proposed in Section~\ref{subsec:growth} to reduce (or even stop) the cell growth. Here it reads as

\begin{equation}\label{eq:probSwtich}
    P_{switch} = \begin{cases}
    a_{prolif} \left(1 - \frac{p_{ext}}{p_{max}} \right) & \text{if } p_{ext}\leq p_{max} \\
    0 & \text{otherwise}    
    \end{cases}
\end{equation}
with $a_{prolif}$ a parameter used to better control the probability $P_{switch}$~(\ref{eq:probSwtich}). Hence, if $p_{ext}\leq p_{max}$ then there is a probability $P_{switch}>0$ for the cell to start proliferating, but if $p_{ext}> p_{max}$, the cell will stop proliferating. In practice, this probability to switch between states is tested at each iteration, with $a_{prolif}=10^{-3}$.

\begin{figure}[!ht]
    \centering
    \begin{subfigure}{0.3\textwidth}
        \includegraphics[width=\textwidth]{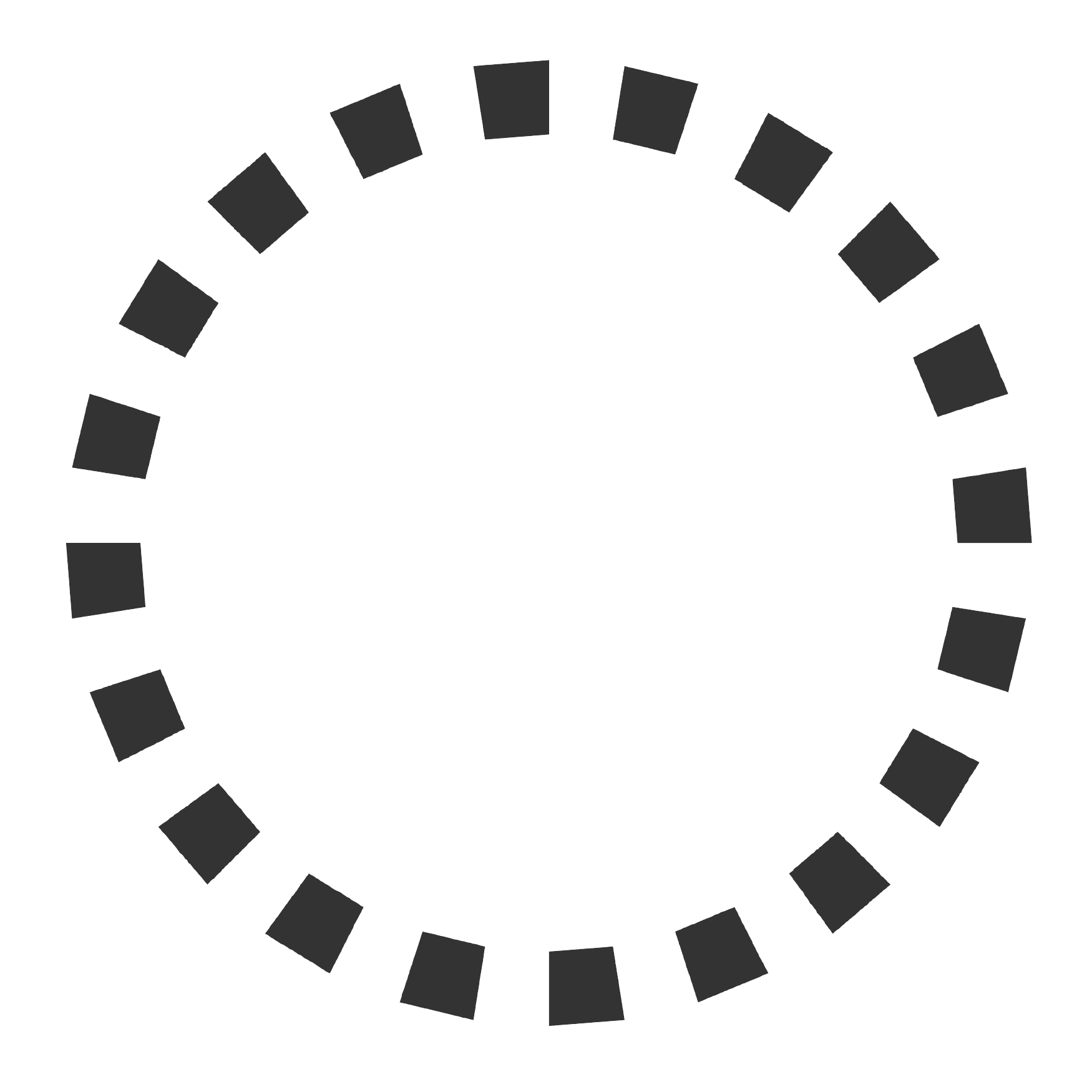}
        \caption{t = 0}
        \label{fig:stopGrowth_0}
    \end{subfigure}
    \begin{subfigure}{0.3\textwidth}
        \includegraphics[width=\textwidth]{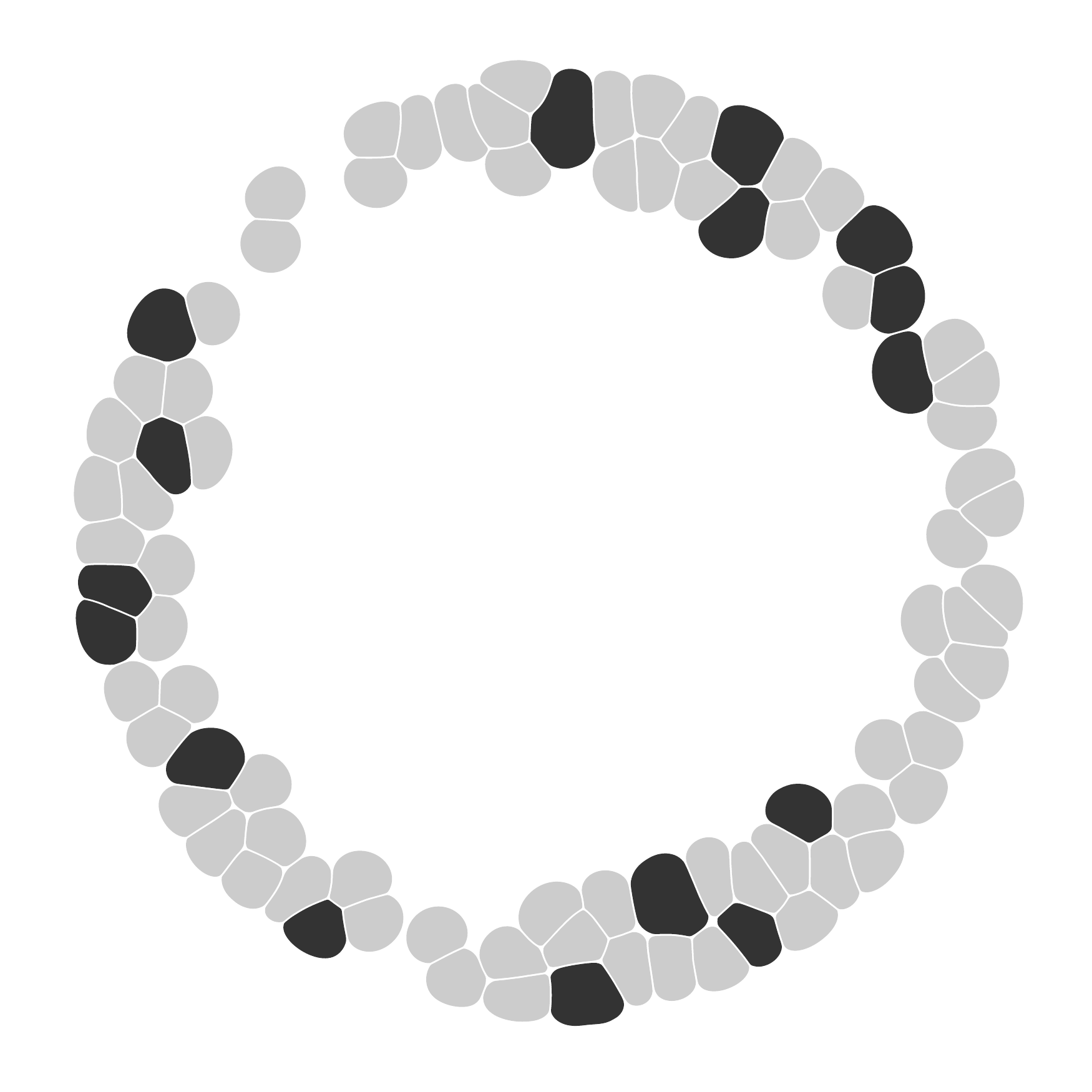}
        \caption{t = 1}
        \label{fig:stopGrowth_1}
    \end{subfigure}
    \begin{subfigure}{0.3\textwidth}
        \includegraphics[width=\textwidth]{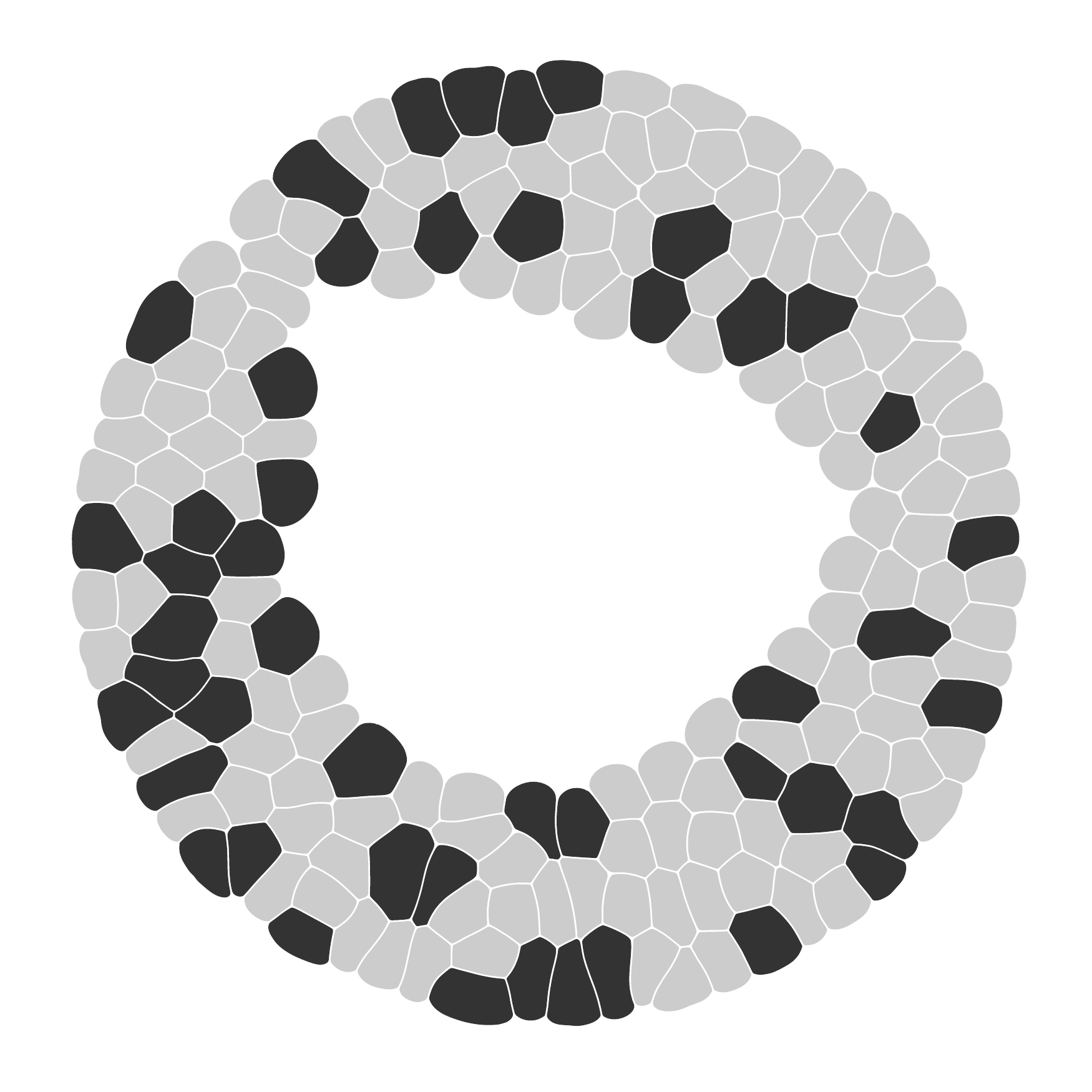}
        \caption{t = 2}
        \label{fig:stopGrowth_2}
    \end{subfigure}
    \begin{subfigure}{0.3\textwidth}
        \includegraphics[width=\textwidth]{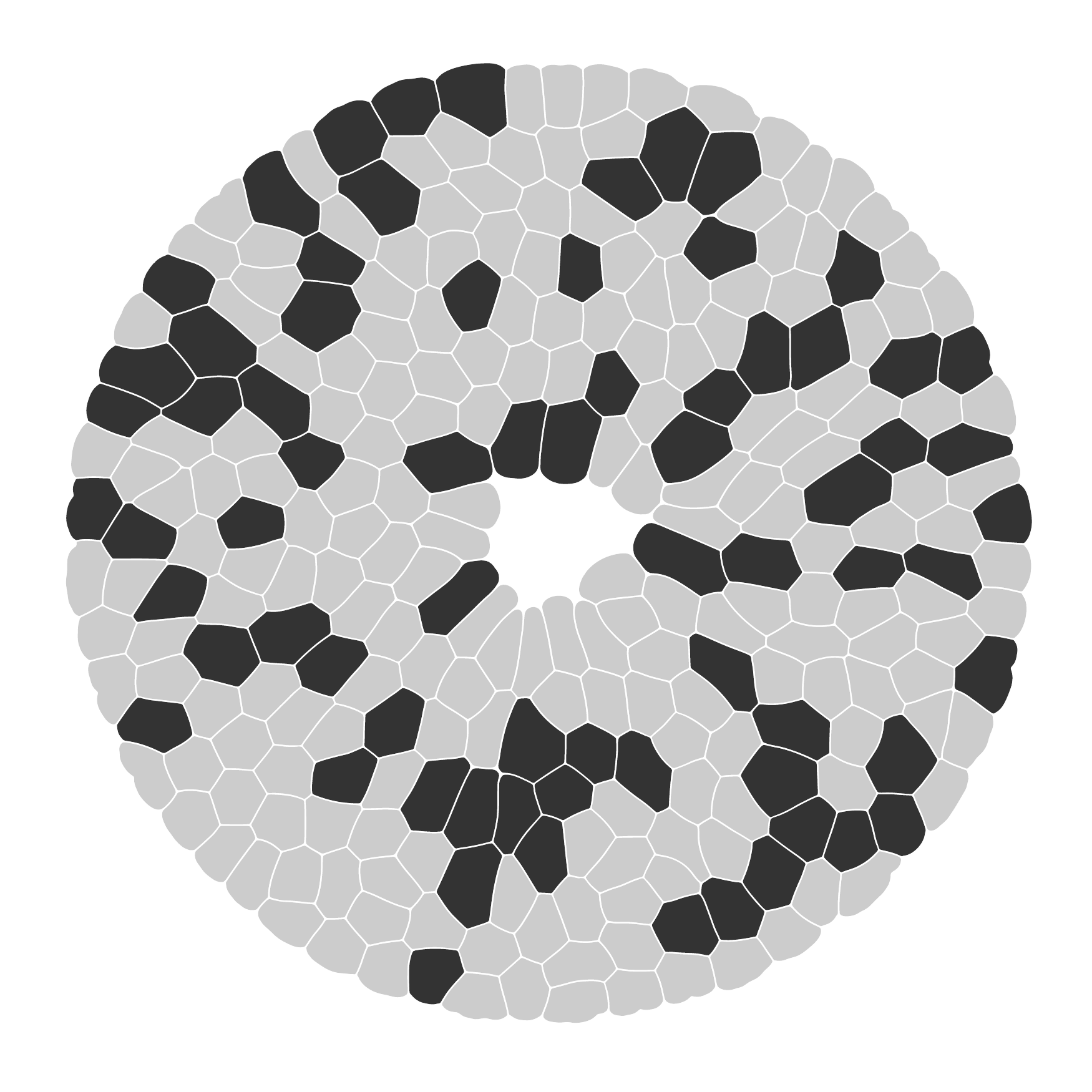}
        \caption{t = 3}
        \label{fig:stopGrowth_3}
    \end{subfigure}
    \begin{subfigure}{0.3\textwidth}
        \includegraphics[width=\textwidth]{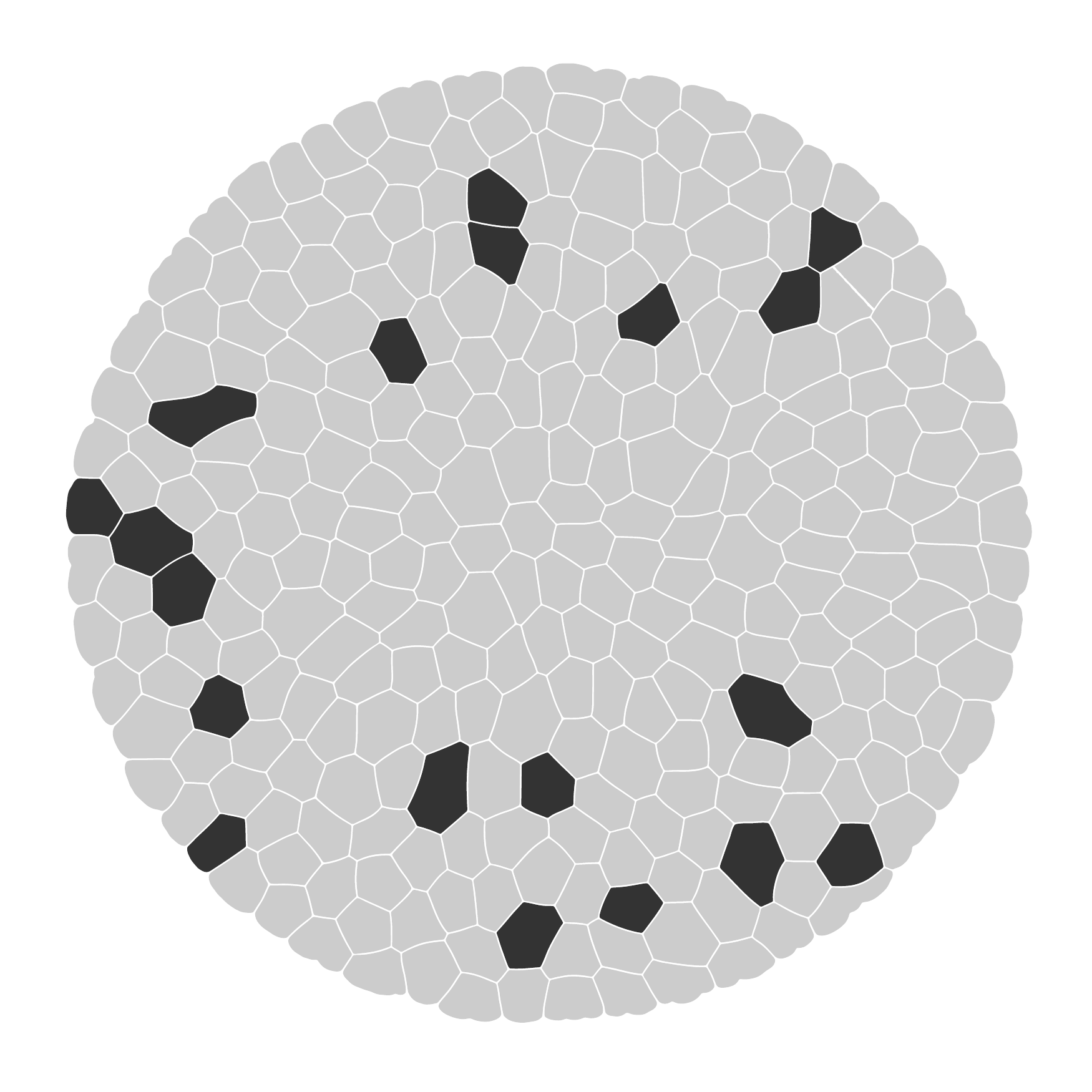}
        \caption{t = 4}
        \label{fig:stopGrowth_4}
    \end{subfigure}
    \begin{subfigure}{0.3\textwidth}
        \includegraphics[width=\textwidth]{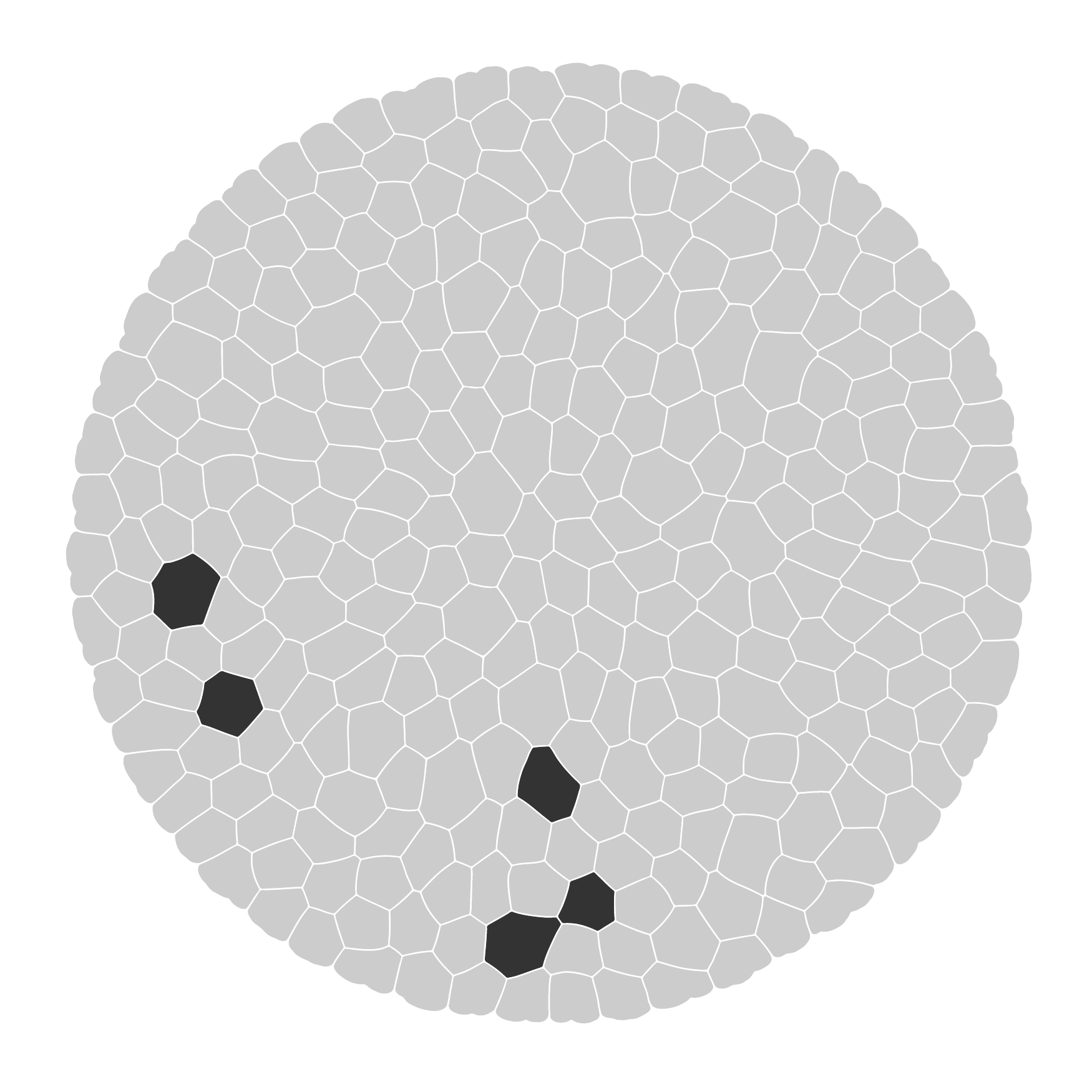}
        \caption{t = 5}
        \label{fig:stopGrowth_5}
    \end{subfigure}
    \caption{Simulation of the cells growth inside a capsule. Black cells are in proliferation mode, whereas grey cells are at rest.}
    \label{fig:capsule_stop}
\end{figure}

As a preliminary validation of such a mechanism, it is proposed to investigate the proliferation of cells inside a circular capsule. In practice, the capsule acts as an elastic soft wall, which can be deformed. The force required to deform is assumed to increase with the capsule radius. Hence, cells can push the capsule when proliferating, but they are more likely to invade the empty space at the center of the domain, due to the force applied on cells by the capsule. 
This is numerically investigated, starting from an initial configuration of twenty cells that are uniformly distributed over the inner part of the capsule (Fig.~\ref{fig:stopGrowth_0}). The capsule is modeled using an external force that acts on the vertices of the cell, and whose amplitude depends on the distance from the capsule center:
\begin{equation}
    \bm{F}_{ext}(\bm{x}_i) = \begin{cases}
    - k_{caps} \frac{\bm{x}_i - \bm{c}_{caps}}{\left \lVert \bm{x}_i - \bm{c}_{caps} \right \rVert}\left(\left \lVert \bm{x}_i - \bm{c}_{caps} \right \rVert - r_{caps}\right) & \text{if } \left \lVert \bm{x}_i - \bm{c}_{caps} \right \rVert \geq r_{caps}\\
    0 & \text{otherwise}    
    \end{cases}
\end{equation}
with $r_{caps}$ the capsule radius, $k_{caps}$ the capsule spring constant, $\bm{c}_{caps}$ the center of the capsule.

\begin{figure}[!ht]
    \centering
    \begin{subfigure}{0.4\textwidth}
        \centering
        \includegraphics[scale=0.5]{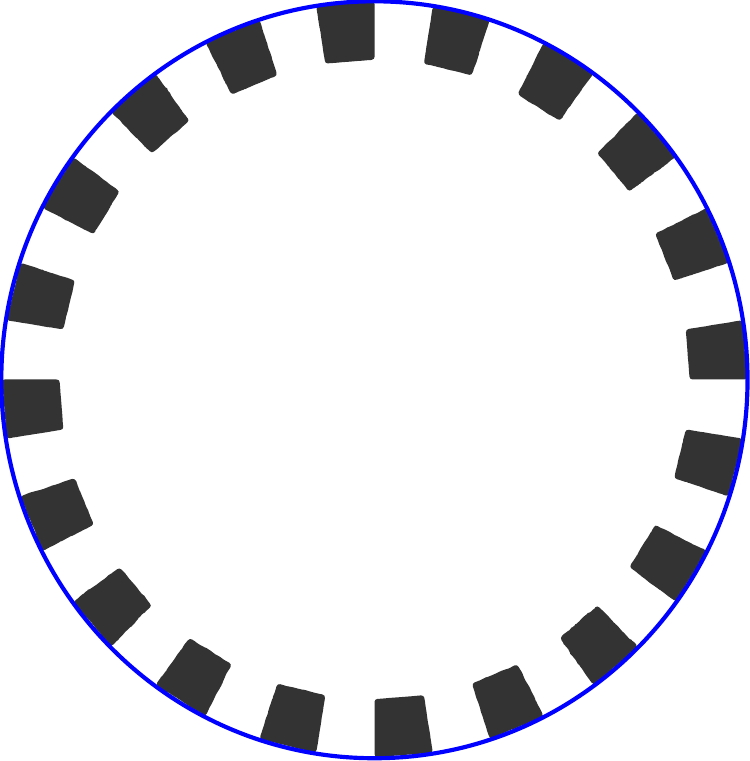}
        \caption{Initial state of the capsule}
    \end{subfigure}
    \begin{subfigure}{0.4\textwidth}
        \centering
        \includegraphics[scale=0.5]{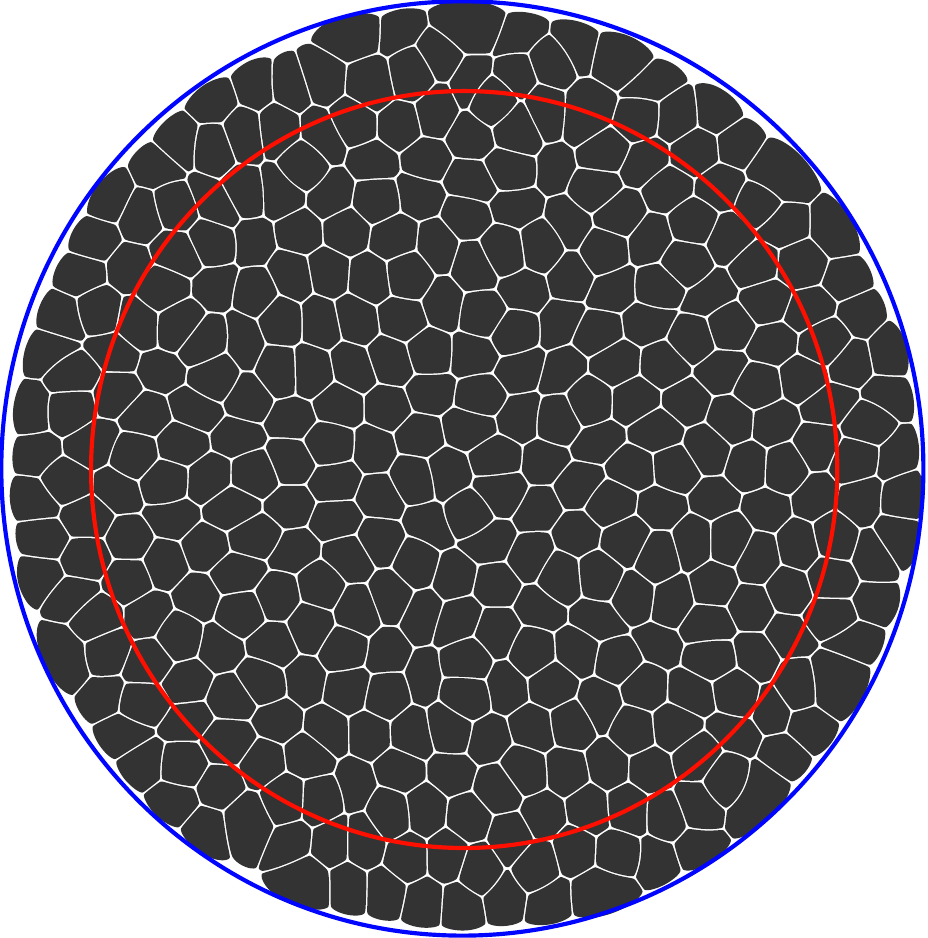}
        \caption{Final state of the capsule}
    \end{subfigure}
    \caption{Evolution of the capsule size, with the actual size in blue and the initial size in red}
    \label{fig:capsule}
\end{figure}

Corresponding results are compiled in Fig.~\ref{fig:capsule_stop}. As anticipated, once cells have populated the inner surface of the capsule, they tend to invade the center part of the simulation domain. While proliferating toward the capsule center, the pressure inside the cell tissue rises. Soon, more and more cells can no longer withstand it, and their proliferation ends. A comparison of the initial and final capsule size is proposed in Fig.~\ref{fig:capsule}.     

Ongoing research shows that with this kind of configuration, the fraction of cells that are proliferating is assumed to follow two distinct steps~\citep{ROUX_PRIVATE_2020}. At the beginning of the simulation, a constant ratio of proliferating cells is observed. This corresponds to the phase where cells are progressively invading the center part of the capsule without encountering any trouble in terms of external pressure. Once a large number of cells have been created, the pressure inside the tissue increases which decreases the number of cells that can continue to proliferate. At some point, the pressure is so high that no cell can proliferate anymore. This obviously depends on the parameter $p_{max}$, but also on the force applied by the capsule on cells. To properly understand the impact of each term, comparisons with experimental data are being performed. Corresponding results will be presented in a future work.


\section{Conclusion\label{sec:Conclusion}}

This paper proposes a new framework for detailed tissue morphogenesis. It relies on the numerical discretization of the cell membrane through approximately one hundred vertices, whose space and time evolution is governed by Newton's law of motion. A number of forces can be accounted for in order to reproduce the mechanical properties of a cell (elasticity, bending, incompressibility). Interestingly, the present formulation can also simulate active processes such as mitosis, apoptosis, and (simplified) cell migration even though the inner structure of the cell (cyoplasm, cytoskeleton and nucleus) is not explicitly modelled. All of this is presented in details, and preliminary studies are conducted to validate the proposed approach. This includes benchmark tests at a single-cell scale (compression and migration), as well as, at full tissue scale (free, chemically-driven and geometrically constrained proliferation). First results are encouraging, and they will be consolidated through more comparisons with experimental data in a future work. \rcol{Eventually, the proposed methodology only requires a discretization of the cell membrane in terms of vertices. Hence, it can easily be extended to simulate 3D configurations by considering a 3D tessellation of the cell membrane, and then applying forces to the corresponding vertices. Such a 3D formulation will be presented in a future work by~\citep{RUNSEN_UnderPreparation_2021}.}

\section*{Acknowledgement}
Fruitful discussions with D. Iber and her team, as well as, A. Roux and F. Raynaud are gratefully acknowledged. This work was supported by the Swiss National Science Fund SNF through the Sinergia Project No. 170930, ``A 3D Cell-Based Simulation Framework for Morphogenetic Problems''.

\appendix
\section{Derivation of the cell radius formula assuming an equilibrium state\label{app:appendix}}

\begin{figure}[ht]
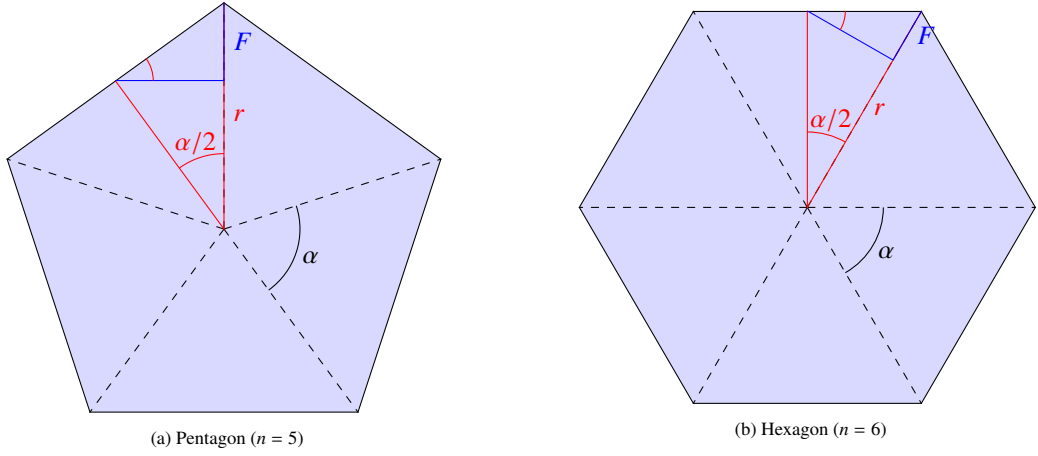

\hspace*{\fill}
\begin{subfigure}{.4\textwidth}
  \centering
  \poly{5}{3}{r}{F}
  \caption{Pentagon ($n=5$)}
  \label{fig:sub-first}
\end{subfigure}
\hfill
\begin{subfigure}{.4\textwidth}
  \centering
  \poly{6}{3}{r}{F}
  \caption{Hexagon ($n=6$)}
  \label{fig:sub-second}
\end{subfigure}
\hspace*{\fill}
\caption{Illustrations of $n$-polygonal cells with either an odd or even number of sides.}
\label{fig:fig}
\end{figure}

Let us consider an individual cell at equilibrium, whose shape is described as an $n$-polygon, where $n$ is the number of vertices of the polygon. At equilibrium, the membrane and pressure forces balance each other, hence, imposing a constant cell area. For an $n$-polygon, this area reads as
\begin{equation}
A = \frac{1}{2} n r^2 \sin\left(\frac{2\pi}{n}\right),
\end{equation}
with the internal angle $\alpha = \frac{2\pi}{n}$, and $r$ the circumradius. In addition, the edge length $l_0$ is defined as
\begin{equation}
l_0 = 2r\sin\left(\frac{\pi}{n}\right).
\end{equation}
In practice, only the value of $r$ is unknown. In order to derive its expression at equilibrium ($r_{eq}$), let us start from the force balance at vertex $i$. In the present model, forces tha apply at vertex $i$ are
\begin{align}
\bm{F}_{B}(\bm{x}_i) &= \rcol{k_b} \left(\frac{\bm{x}_{i+1} - \bm{x}_i}{\lVert \bm{x}_{i+1} - \bm{x}_i \rVert} + \frac{\bm{x}_{i-1} - \bm{x}_i}{\lVert \bm{x}_{i-1} - \bm{x}_i \rVert} \right), \\[0.1cm]
\bm{F}_{S}(\bm{x}_i) &= k_s \left[ (\bm{x}_{i+1} - \bm{x}_i) + (\bm{x}_{i-1} - \bm{x}_i) \right], \\[0.1cm]
\bm{F}_{IP}(\bm{x}_i) &= p \frac{ \lVert \bm{x}_{i+1} - \bm{x}_i \rVert + \lVert \bm{x}_{i+1} - \bm{x}_i \rVert}{2}  \hat{n}_i,
\end{align}
with $i-1$ and $i+1$ the previous and the next vertices, $\rcol{k_b}$ the bending constant, $k_s$ the spring constant, $p$ the internal pressure, and $\bm{\hat{n}}_i$ the normal vector at vertex $i$. The local equilibrium imposes the following force balance:
\begin{equation}
\bm{F}_{B}(\bm{x}_i) + \bm{F}_{S}(\bm{x}_i) + \bm{F}_{IP}(\bm{x}_i) = 0.
\label{eq:equi}
\end{equation}
The latter can further be decomposed into a normal and a tangential force balance. Interestingly, tangential components cancel each other out by symmetry, and only normal components remain. This is explained by the fact that the cell is a regular polygon, hence, $l = \lVert \bm{x}_{i+1} - \bm{x}_i \rVert = \lVert \bm{x}_{i-1} - \bm{x}_i \rVert$. 
Furthermore, the normal projection of a membrane force $\bm{F}$ is $F = \lVert \bm{F} \rVert \sin(\pi/2)$, with $\bm{F}=\bm{F}_S$ or $\bm{F}_B$. 
As the membrane force is in the direction opposite to the normal, the force balance~(\ref{eq:equi}) becomes:
\begin{align}
-F_{B}(\bm{x}_i) - F_{S}(\bm{x}_i) + F_{IP}(\bm{x}_i) &= 0, \\[0.1cm]
\eta \left(\frac{m}{A} - \rho_0 \right) l &= 2 \rcol{k_b} \sin\left(\frac{\pi}{n}\right) + 2 k_s l \sin\left(\frac{\pi}{n}\right),\\[0.1cm]
\eta \left(\frac{2m}{n r^2 \sin \left(\frac{2\pi}{n} \right)} - \rho_0 \right) 2 r \sin\left(\frac{\pi}{n}\right) &= 2 \rcol{k_b} \sin\left(\frac{\pi}{n}\right) + 4 k_s r \sin\left(\frac{\pi}{n}\right)^2.
\end{align}
Assuming that the number of vertices $n$ is large enough, then $\sin(\pi/n) \approx \pi/n$, which leads to the following simplifications:
\begin{align}
\eta \left(\frac{2m}{n r^2 \frac{2\pi}{n}} - \rho_0 \right) 2 r \frac{\pi}{n} &= 2 \rcol{k_b} \frac{\pi}{n} + 4 k_s r \left(\frac{\pi}{n}\right)^2, \\[0.1cm]
\eta \left(\frac{2m}{r^2 2\pi} - \rho_0 \right) r &= \rcol{k_b} + 2 \pi r \frac{k_s}{n}, \\[0.1cm]
\frac{2m\eta}{2r\pi} - \eta \rho_0 r &= \rcol{k_b} + 2\pi r \frac{k_s}{n}, \\[0.1cm]
2 \eta m &= 2 r \pi \rcol{k_b} + 4 \pi^2 r^2 \frac{k_s}{n} + 2 \pi \eta \rho_0 r^2,\\[0.1cm]
0 &= r^2 \left(4\pi^2 \frac{k_s}{n} + 2\pi \eta \rho_0 \right) + 2 \pi \rcol{k_b} r - 2\eta m.
\end{align}
By solving the above quadratic equation, one ends up with
\begin{equation}
r = \frac{-2\pi \rcol{k_b} \pm \sqrt{\left(2 \pi \rcol{k_b}\right)^2 + 4 \left(4\pi^2 \frac{k_s}{n} + 2\pi \eta \rho_0 \right) 2\eta m}}{2\left(4\pi^2 \frac{k_s}{n} + 2\pi \eta \rho_0 \right)}.
\end{equation}
Keeping the only solution that leads to $r>0$, one eventually obtains:
\begin{align}
r_{eq} &= \pi \rcol{k_b} \frac{\sqrt{1 + \frac{2\eta m}{\pi^2 \rcol{k_b}^2}\left(4\pi^2 \frac{k_s}{n} + 2\pi \eta \rho_0 \right)}-1}{4\pi^2 \frac{k_s}{n} + 2\pi \eta \rho_0}.
\end{align}

As the radius of the cell needs to be independent of the cell discretization, $r$ should be independent of $n$. This condition is reached when 
\begin{equation}\label{eq:kCondition}
k_s/n = \mathrm{cst}.
\end{equation}
In our formulation, the membrane is considered as a unique spring with a global constant $K_s$, and the subdivision of the membrane into $n$ vertices amounts to replacing the global spring by $n$ local springs with identical constant $k_s$. As the springs are in series, the equivalent spring reads as
\begin{equation}
\frac{1}{K_s} = \sum_{i=1}^{n} \frac{1}{k_s} = \frac{n}{k_s}.
\end{equation} 
The condition~(\ref{eq:kCondition}) is then automatically verified, hence, leading to an equilibrium cell radius $r_{eq}$ that is independent of the membrane discretization.

\section{Simulation parameters\label{app:param}}
This appendix summarizes all parameter used in this work. They are grouped by in several categories depending on their physical meaning. While Tab.~\ref{tab:paramCell} gathers all parameters related to the cell mechanical and biological properties, those related to signaling and the growth under constraint (Sec.~\ref{subsec:signaling} and~\ref{subsec:capsule} respectively) are compiled in Tab.~\ref{tab:paramSigCaps}. As a sidenote, the same parameter $p_{max}$ is used to control (i) the growth/relaxation of cells~(\ref{eq:growthRelax}), and (ii) the probability to switch from the cell state from proliferating to rest~(\ref{eq:probSwtich}).

\begin{table}[h!]
    \centering
    \begin{tabular}{c c c c c c c c c c c c c c c}
\toprule
\toprule
  \multicolumn{1}{c}{Cell} & & \multicolumn{2}{c}{Membrane} & &  \multicolumn{3}{c}{Adherence} & & \multicolumn{6}{c}{Growth/Relaxation} \\
  \cmidrule{0-0} \cmidrule{3-4} \cmidrule{6-8} \cmidrule{10-15}
     $n_0$ & & $\rcol{k_b}$ & $k_s$ & &  $d_0$ & $d_{max}$ & $k_a$ & & $\rho_0$ & $\eta$ & $A_0$ & $\nu_{relax}$ & $\nu$  & $p_{max}$\\ 
     120   & & 0.1   & 0.002 & &  0.5   & 1.0       & 0.1   & & 1.05     & 1.0    & 200   & 0.01          & 0.0025 & 0.05\\
\toprule
\toprule
    \end{tabular}
    \caption{Simulation parameters related to the cell discretization, as well as, its mechanical and biological properties.}
    \label{tab:paramCell}
\end{table}

\begin{table}[h!]
    \centering
    \begin{tabular}{c c c c c c c c}
\toprule
\toprule
  \multicolumn{3}{c}{signaling} & & \multicolumn{4}{c}{Capsule} \\
  \cmidrule{1-3}\cmidrule{5-8}
     $\tau_p$ & $\tau_d$ & $\theta_{sig}$ & & $p_{max}$ & $a_{prolif}$ & $k_{caps}$ & $r_{caps}$\\ 
      0.001 & 0.0001 & 0.025 & & 0.05 & 0.001 & 0.01 & 130\\
\toprule
\toprule
    \end{tabular}
    \caption{Simulation parameters related to the signaling mechanism, and the external forcing used in the case of the growth inside a capsule.}
    \label{tab:paramSigCaps}
\end{table}

\end{document}